\documentclass[journal]{IEEEtran}
\usepackage[nosort]{cite}
\usepackage{overpic}
\usepackage[normalem]{ulem}
\usepackage[caption=false,font=scriptsize,labelfont=sf,textfont=sf]{subfig}
\usepackage{amsmath,graphicx}
\usepackage[linesnumbered,ruled]{algorithm2e}
\usepackage{multirow}
\usepackage{hhline}
\usepackage{enumitem}
\usepackage{pifont}
\usepackage{bm}
\usepackage[dvipsnames,table]{xcolor}
\usepackage{amsmath}
\usepackage{url}
\usepackage{amsfonts}
\usepackage{soul}
\usepackage{booktabs}
\usepackage{soul}

\newcommand{\purple}[1]{{\textcolor{purple}{#1}}}
\ifCLASSINFOpdf
\else
\fi

\begin{document}
\bstctlcite{IEEEexample:BSTcontrol}

\title{Towards Effective and Efficient Non-autoregressive decoders for Conformer and LLM-based ASR using Block-based Attention Mask}

\author{Tianzi Wang, Xurong Xie, Zengrui Jin, Mengzhe Geng, Jiajun Deng, Zhaoqing Li, Shoukang Hu, Shujie Hu, Guinan Li, Mingyu Cui, Helen Meng,~\IEEEmembership{Fellow,~IEEE}, Xunying Liu,~\IEEEmembership{Member,~IEEE}
\thanks{Tianzi Wang, Jiajun Deng, Zhaoqing Li, Shujie Hu, Guinan Li, Mingyu Cui, Helen Meng, and Xunying Liu are with the Chinese University of Hong Kong, China (email: \{twang, jjdeng, zqli, sjhu, gnli, mycui, hmmeng, xyliu\}@se.cuhk.edu.hk)}
\thanks{Xurong Xie is with the 
Institute of Software, Chinese Academy of Sciences, China (email: xurong@iscas.ac.cn)}
\thanks{Zengrui Jin is with Tsinghua University, China (email:~zrjin@tsinghua.edu.cn)}
\thanks{Mengzhe Geng is with National Research Council Canada, Canada (email: Mengzhe.Geng@nrc-cnrc.gc.ca)}
\thanks{Shoukang Hu is with Nanyang Technological University, Singapore (email: shoukang.hu@ntu.edu.sg)}
\thanks{Xunying Liu and Xurong Xie are the corresponding authors.}}

\markboth{IEEE Transactions on Audio, Speech and Language Processing}%
{Bare Demo of IEEEtran.cls for IEEE Journals}

\maketitle
\begin{abstract}
Automatic speech recognition (ASR) systems often rely on autoregressive (AR) Transformer decoder architectures, which limit efficient inference parallelization due to their sequential nature. To this end, non-autoregressive (NAR) approaches aim primarily to achieve significant decoding speedup while the maintaining recognition accuracy that is comparable to AR baselines. This paper proposes a novel NAR block-based attention mask decoder (AMD) that effectively improves decoding efficiency while maintaining ASR accuracy, and also offering flexibility in balancing the performance-efficiency trade-off on both Conformer and large language model (LLM)-based ASR systems. The proposed AMD performs parallel inference within contiguous blocks of output labels while maintaining monotonic left-to-right prediction between blocks. A one-pass beam search algorithm is designed to dynamically fuse Connectionist Temporal Classification (CTC), AR decoder, and AMD probabilities. Experiments are conducted on normal speech LS960 and DBank elderly speech across: a) The Conformer encoder-decoder ASR system with filterbank input features; b) its integration with WavLM features; and c) further advancement by integrating an LLM-based decoder. On the LS960 task, the proposed AMD empowered tripartite decoder achieves decoding speedup ratios of up to 1.44x, 1.55x, and 2.31x under the three model configurations over the CTC + AR baselines, without statistically significant WER increases. When operating with real-time factors (RTFs) comparable to the baselines, the tripartite decoder produces statistically significant WER reductions of 0.19\%, 0.62\% and 0.13\% absolute (4.3\%, 16.3\%, and 3.8\% relative). Similar improvements are also obtained on the DBank task, where the tripartite decoder accelerates decoding by up to 1.38x, 1.64x and 1.61x without statistically significant WER increase, and yields statistically significant WER reductions of 0.46\%, 0.38\% and 0.41\% absolute (1.8\%, 1.8\%, and 2.0\% relative) when operating with RTFs comparable to the CTC + AR baselines.
\end{abstract}
\IEEEpeerreviewmaketitle
\vspace{-5mm}
\section{Introduction}
\label{introduction}

\IEEEPARstart{S}{tate-of-the-art} automatic speech recognition (ASR) systems, in particular Conformer encoder-decoder models with joint Connectionist Temporal Classification (CTC) and attention cost ~\cite{watanabe2017hybrid, guo2021recent, kim2017joint, sainath2019two, yao2021wenet}, are often based on an autoregressive (AR) Transformer decoder architectures. Under this paradigm, AR model inference is conditioned only on the left-context, enforcing a strictly monotonic, left-to-right decoding process in which each output token is predicted one at a time. AR Transformer decoders have been successfully adopted by current ASR systems, and in a wider context, mainstream large language models (LLMs)~\cite{radford2023robust, barrault2023seamless,mann2020language, du2021glm, hoffmann2022training, chowdhery2023palm, chiang2023vicuna, chung2024scaling,achiam2023gpt,touvron2023llama,bai2023qwen,yang2024qwen2,zhang2022opt,henighan2020scaling,kojima2022large,huang2023large,rae2021scaling,jiang2023mistral7b} that have been increasingly integrated into these systems ~\cite{chu2023qwen,du2023lauragpt,hu2024wavllm,shu2023llasm,wang2023blsp,tang2023salmonn,rubenstein2023audiopalm,chen2023x,huang2024audiogpt,li2023prompting,yu2024connecting,fathullah2024prompting,wu2023decoder,das2024speechverse} and sparked a paradigm shift in speech technology research. However, the sequential nature of AR Transformer decoders limits their potential for model inference parallelization for practical application scenarios that are not only performance-critical but also efficiency-sensitive. 
One general solution to address the above issue is to adopt non-autoregressive (NAR) Transformer based decoder designs. NAR Transformers provide more powerful parallelization than their AR counterparts to improve inference speed. Their efficient designs have been widely exploited across a wide range of applications including, but not limited to, machine translation ~\cite{gu2017non,wang2018semi,guo2019non,qian2021glancing,huang2022directed, ma2019flowseq, stern2019insertion, gu2019levenshtein, libovicky2018end, kaiser2018fast, lee2018deterministic, gu2020fully}, dialogue systems ~\cite{han2020non, zou2021thinking, huang2021sarg,liang2023open} and speech translation ~\cite{chuang2021investigating,inaguma2021orthros,inaguma2021non,inaguma2021fast}. 
\vspace{-2mm}
\subsection{Non-autoregressive Approaches for ASR}
\label{intro_approach}
However, the development of NAR Transformer decoder architectures that can flexibly balance the performance-efficiency trade-off for ASR systems remains a notoriously challenging “zero sum game", and has attracted increasingly research attention in recent years. To this end, prior researches include, but not limited to, the following categories: \textbf{a) mask-based NAR }~\cite{higuchi2020mask,higuchi2021improved, song2021non,chen2020non,sunder2025non, higuchi2023mask} learn to fill the randomly masked training label tokens conditioned on the unmasked ones. Among these, conditional masked language models (CMLMs) were first introduced in ~\cite{chen2020non}, before being further used by Mask-CTC and its variants ~\cite{higuchi2020mask,higuchi2021improved,song2021non} to refine CTC predictions. \textbf{ b) alignment-refinement based approaches} ~\cite{chi2020align, chen2021align,fan2021cass,fan2021improved,fan2023ctc} that aim to refine the initial CTC produced alignment by injecting noisy labels extracted from, for example, auto-encoders in the Align-Refine method ~\cite{chi2020align}, or noise perturbed encoder alignment posteriors in the Align-Denoise approach ~\cite{chen2021align}. Further developments in this category include CASS-NAT and its variants ~\cite{fan2021cass,fan2021improved,fan2023ctc}, which employ CTC alignments to extract token-level acoustic embeddings to allow parallel decoding using bidirectional attention-based decoders. \textbf{c) integrate-and-fire (IF) based approaches} ~\cite{dong2020cif,yu2021boundary,gao2022paraformer,zou2024paraformer}. Continuous IF (CIF) ~\cite{dong2020cif} alignment modules and their variants ~\cite{yu2021boundary} are utilized by Paraformer ~\cite{gao2022paraformer} and E-Paraformer ~\cite{zou2024paraformer} models to perform parallel inference over output tokens. \textbf{d) hybrid AR+NAR decoders} that exploit their complementarity in combination ~\cite{sudo20224d,tian2022hybrid,wang2022deliberation}. The 4D ASR approach ~\cite{sudo20224d} jointly trains CTC~\cite{graves2006connectionist}, attention, RNN-T and mask-prediction decoders, while using only mask-prediction decoder for NAR inference. ~\cite{tian2022hybrid} adopts NAR for initial prediction followed by AR rescoring. Deliberation-based approaches ~\cite{wang2022deliberation} refine streaming RNN-T hypotheses using Align-Refine NAR. These hybrid approaches either operate with an expensive multi-pass decoding strategy, or rely solely on a single NAR decoder to “correct” the errors in the initial recognition hypotheses. 
In contrast, only limited researches have been conducted on LLM oriented NAR decoding techniques. Existing researches in this direction largely focus on using multi-pass rescoring based approaches. For example, PaLM 2 language model ~\cite{anil2023palm} was used in ~\cite{huang2024multilingual} to rescore the CTC decoding outputs produced by the first decoding pass Universal Speech Model (USM) ~\cite{zhang2023google}. Beyond these LLM rescoring based shallow fusion approaches, there is a notable lack of exploration on effective and efficient NAR decoding techniques when LLMs serve directly as the decoders in a single pass recognition framework ~\cite{chu2023qwen,du2023lauragpt,hu2024wavllm,shu2023llasm,wang2023blsp,tang2023salmonn,rubenstein2023audiopalm,chen2023x,huang2024audiogpt,li2023prompting,yu2024connecting,fathullah2024prompting,wu2023decoder,das2024speechverse,saon2025granite}.
\subsection{Key Challenges in Non-autoregressive ASR}
\label{intro_challenges}
Efforts on developing high-performance and low-latency NAR-based ASR models require several key challenges to be addressed. 

\textbf{a)} NAR models’ assumption of conditional independence among output tokens results in an intrinsic lack of monotonic sequence modeling constraints, which leads to their large performance gap against state-of-the-art ASR systems based on AR designs.
The individual output tokens predicted by the NAR decoder are assumed to be temporally independent against each other. Despite recent attempts to mitigate such modeling deficiency of NAR ASR systems using, for example, Mask-CTC and its improved variants ~\cite{higuchi2020mask, higuchi2021improved}, or iterative alignment-refinement approaches ~\cite{chi2020align, chen2021align}, their performance gap against AR counterparts still exists. 

\textbf{b)} \textbf{Lack of effective and efficient one-pass decoding algorithm} that is purpose-designed for NAR decoders and their further integration with CTC and AR ones. Prior research in this direction either deployed standalone NAR decoders in a later rescoring pass within a multi-pass decoding framework ~\cite{sudo20224d}, while the first pass recurrent neural network transducer (RNN-T) decoding serves to produce an initial set of hypotheses ~\cite{wang2022deliberation}, or vice versa when the NAR decoders are used in the initial N-best generation before AR rescoring ~\cite{tian2022hybrid}. 

\textbf{c) Lack of efficient and effective NAR decoders in LLM-based speech and audio models} that are mainly based on
AR decoder architectures~\cite{chu2023qwen,du2023lauragpt,hu2024wavllm,shu2023llasm,wang2023blsp,tang2023salmonn,rubenstein2023audiopalm,chen2023x,huang2024audiogpt,li2023prompting,yu2024connecting,fathullah2024prompting,wu2023decoder,das2024speechverse}. Research on LLM-oriented NAR decoding techniques remains limited to date. Existing efforts predominantly following multi-pass rescoring-based paradigms~\cite{huang2024multilingual}, rather than employing LLMs directly as NAR decoders in a one-pass recognition framework.

\textbf{d) Lack of application to atypical speech domains}. Prior research on NAR-based ASR  have been mainly conducted on typical speech where large quantities of domain data are often available ~\cite{higuchi2020mask, higuchi2021improved, song2021non, chen2020non, chi2020align, chen2021align, fan2021cass, fan2021improved, fan2023ctc, dong2020cif, yu2021boundary, gao2022paraformer, zou2024paraformer, sudo20224d, tian2022hybrid, wang2022deliberation,ng2021pushing,higuchi2021comparative}. In contrast, their efficacy when being applied to medical and healthcare domain data, for example, the highly scarce, disfluent, inarticulate and diverse elderly speech recorded during neurocognitive impairment assessment interviews, remains under-explored to date. 
\subsection{The Proposed Approach}
\label{sec:1.3}
To this end, building upon~\cite{wang2024towards}, this paper proposes a novel non-autoregressive block-based attention mask decoder (AMD) that flexibly balances performance-efficiency trade-offs for both Conformer~\cite{gulati2020conformer} encoder-decoder and \textbf{LLM-based} ASR systems. The AMD leverages both parallel NAR inference and monotonic left-to-right AR prediction.  A beam search algorithm is designed to leverage a dynamic fusion of CTC, AR decoder, and AMD probabilities. In addition to fixed-size attention-masking blocks during NAR inference, mixed-size blocks were also explored to facilitate cold start monotonic inference (block size = 1) for the initial N labels of each speech segment, before switching to parallel label prediction (block size $>$ 1) for the remaining labels. The effectiveness of our AMD is demonstrated on both WavLM ~\cite{chen2022wavlm} features empowered Conformer encoder-decoder models and LLM (Llama-3.2-1B-Instruct ~\cite{dubey2024llama}) based ASR systems across both typical and atypical, elderly speech domain data. 

Experiments are conducted on the benchmark LibriSpeech 960-hour (LS960) normal speech dataset~\cite{panayotov2015librispeech} and DementiaBank Pitt~\cite{becker1994natural} (DBank) elderly speech corpus across three model configurations: {\bf a)} The Conformer encoder-decoder ASR system with filterbank (FBank) input features; {\bf b)} its integration with WavLM features; and {\bf c)} further advancement by replacing the original decoder with an LLM-based decoder.

On the LS960 task, the proposed AMD empowered tripartite decoder respectively achieves speedup ratios of up to {\bf 1.44x}, {\bf 1.55x}, and {\bf 2.31x} under the three model configurations, without statistically significant WER increase over the CTC + AR baselines. When operating with RTFs comparable to the CTC + AR baselines, the tripartite decoder produces statistically significant absolute WER reductions of {\bf 0.19\%}, {\bf 0.62\%} and {\bf 0.13\%} ({\bf 4.3\%}, {\bf 16.3\%}, and {\bf 3.8\%} relative) across the three configurations over their respective CTC + AR baselines. Similar trends are observed on the DBank task.


The main contributions of this paper are as follows:

\textbf{a)} This paper presents a novel block-based attention mask decoder (AMD) that flexibly balances performance-efficiency trade-offs for Conformer ASR systems in recognition time. For the first time, this NAR decoder allows: 1) decoding time speedup via non-autoregressive, parallel inference without increasing ASR WERs; and 2) statistically significant WER reductions over AR decoders when operating with the same decoding real-time factors. In contrast, large performance degradation was often observed in prior research when alternative forms of NAR decoders ~\cite{higuchi2020mask,higuchi2021improved,chen2021align}, e.g. Mask-CTC, were used, or failed to provide the WERs at equivalent real- time factors for fair comparison ~\cite{gao2022paraformer}. 

\textbf{b)} A novel one-pass beam search algorithm is designed to leverage a dynamic fusion of CTC, AR decoder, and AMD probabilities. In contrast, prior research have largely deployed NAR decoders in a more time-consuming multi-pass decoding framework ~\cite{sudo20224d, wang2022deliberation, song2021non, pmlr-v139-qi21a, liang-etal-2022-janus}.

\textbf{c)} In contrast to prior LLM-based ASR systems that largely operate with AR decoders, this paper presents a novel NAR LLM-based decoder using our AMD. Experimental results show that compared to the CTC + AR baseline, the proposed tripartite decoder achieves significant decoding speedups without a statistically significant WER increase. And when operating at an RTF comparable to the baseline, the tripartite decoder yields statistically significant WER reductions.
In contrast, prior research either uses only LLM-based AR decoders~\cite{chu2023qwen,du2023lauragpt,hu2024wavllm,shu2023llasm,wang2023blsp,tang2023salmonn,rubenstein2023audiopalm,chen2023x,huang2024audiogpt,li2023prompting,yu2024connecting,fathullah2024prompting,wu2023decoder} or employs LLM-oriented NAR decoding techniques in multi-pass rescoring
~\cite{huang2024multilingual}.

\textbf{d)} This paper demonstrates that the ASR performance gains and efficiency enhancement from AMD tripartite decoders on typical speech (LS960) are preserved when they are applied to highly scarce, disfluent, inarticulate and diverse elderly speech. These consistent improvements across typical and atypical speech domains highlight the robustness of our approach. In contrast, prior NAR-based ASR research primarily focused on typical speech domains~\cite{higuchi2020mask, higuchi2021improved, song2021non, chen2020non, chi2020align, chen2021align, fan2021cass, fan2021improved, fan2023ctc, dong2020cif, yu2021boundary, gao2022paraformer, zou2024paraformer, sudo20224d, tian2022hybrid, wang2022deliberation} with abundant training data.

Compared to our prior conference paper \cite{wang2024towards}, this paper represents a large extension include: 1) novel algorithmic enhancements through the integration of our AMD framework with LLMs, featuring a parameter-efficient shared-backbone design; and 2) extensive experimental validation, scaling results to the full LS960 dataset and a new domain of atypical elderly speech (DBank).

The rest of the paper is organized as follows. Section \ref{sec:2} reviews the hybrid CTC + AR encoder-decoder based Conformer ASR system. Section \ref{sec:3} describes the proposed AMD-empowered tripartite decoder for Conformer ASR, as well as their further integration with WavLM features and LLM-based decoders. Section \ref{sec:4} presents the one-pass beam search algorithm and mixed block size decoding approaches for the AMD tripartite decoder. Section \ref{sec:exp_setup} details the experimental setup, Section \ref{sec:implementation_details} describes the ablation studies that inform our design choices, and Section \ref{sec:exp} presents the final comparative results on the normal speech LS960 and DBank elderly speech corpora. Conclusion is drawn in Section \ref{sec:conclusion} together with discussion of possible future work.
\begin{figure*}[t]
    \centering
    \begin{minipage}{0.47\linewidth}
        \centering
        \includegraphics[width=\linewidth]{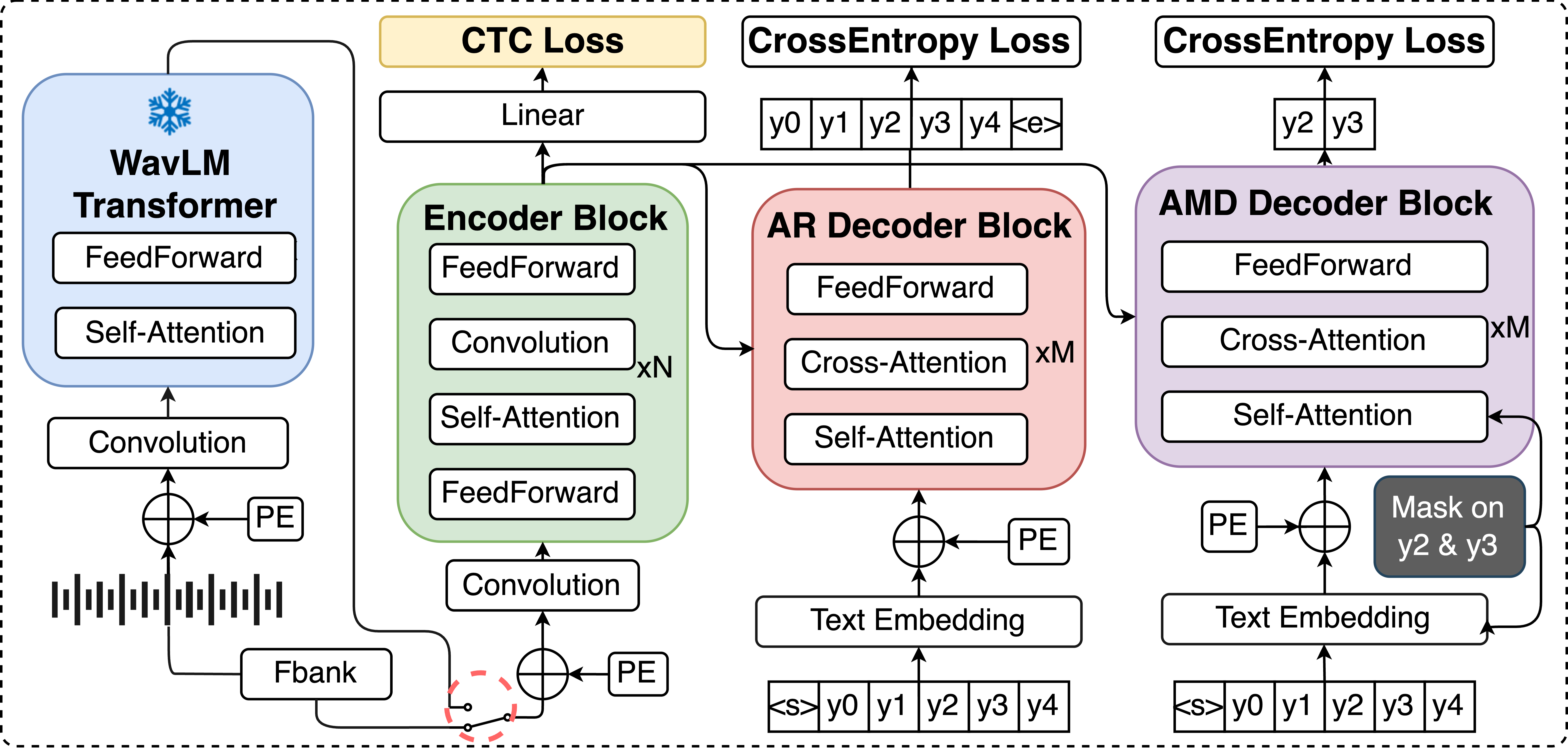}
        \caption{The proposed ASR system architecture illustrating two configurations: using FBank features directly as input (lower one in the red dotted circle); or using frozen pre-trained WavLM (blue) as a feature extractor (upper one in the red dotted circle). Both configurations utilize the Conformer encoder followed by a tripartite decoder that includes the proposed non-autoregressive block-based attention mask decoder (AMD) (purple) in addition to the CTC module (yellow) and attention-based AR decoder (red).}
        \label{fig:wavlm_main}
    \end{minipage}
    \hfill
    \begin{minipage}{0.48\linewidth}
        \centering
        \includegraphics[width=\linewidth]{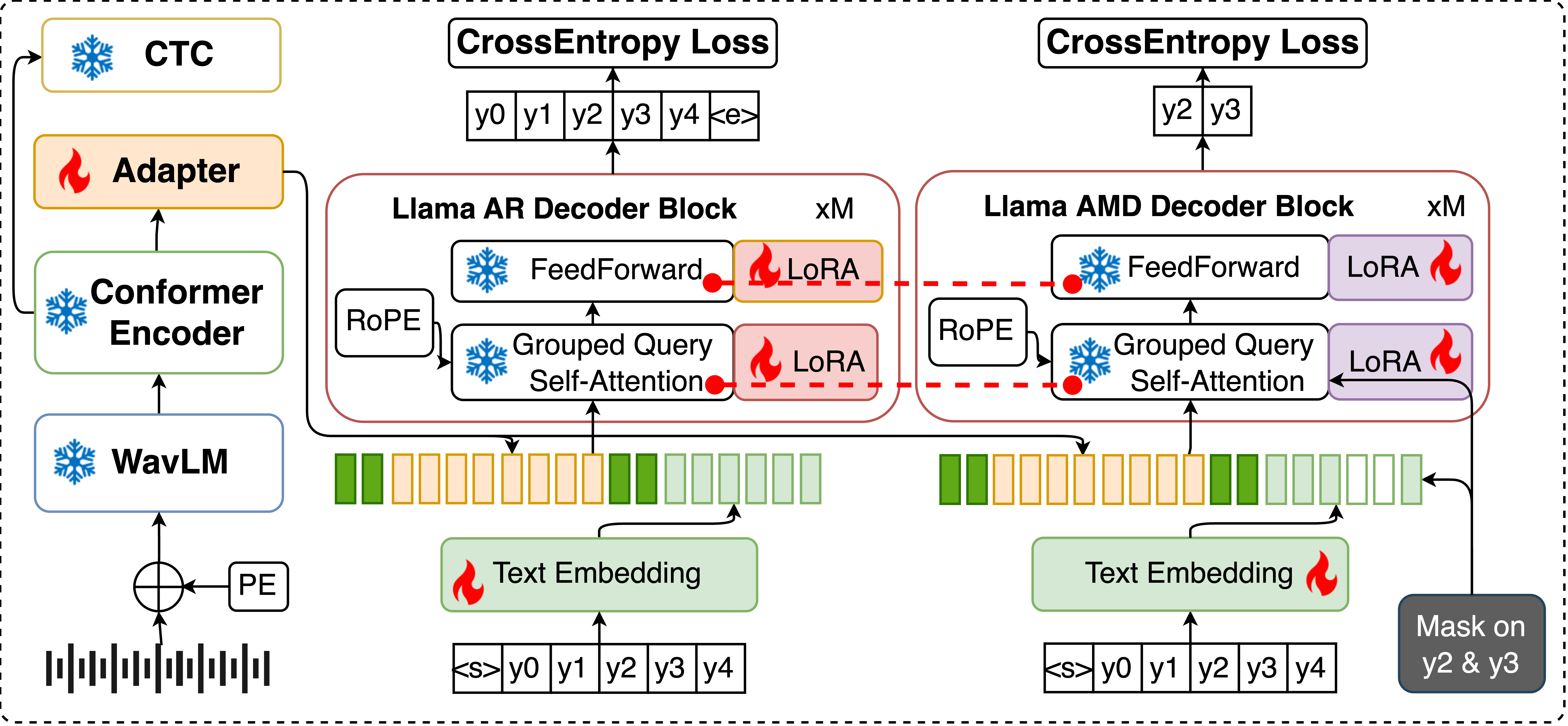}
        \caption{The enhanced ASR architecture incorporating frozen pre-trained WavLM feature extractor, Conformer encoder and CTC decoder, integrated with Llama-based AR and AMD decoders. While the Llama backbone remains frozen, only the adapters (left, yellow) between encoder and decoder and decoder-specific LoRA (red and purple) parameters are trained, along with text embeddings (green). The adapted encoder outputs are concatenated with text embeddings before feeding into the decoders. Dotted red lines denote shared weights between components.}
        \label{fig:main-llama}
    \end{minipage}
    \vspace{-3mm}
\end{figure*}
\begin{figure}[t]
    \centering
    \includegraphics[width=0.9\linewidth]{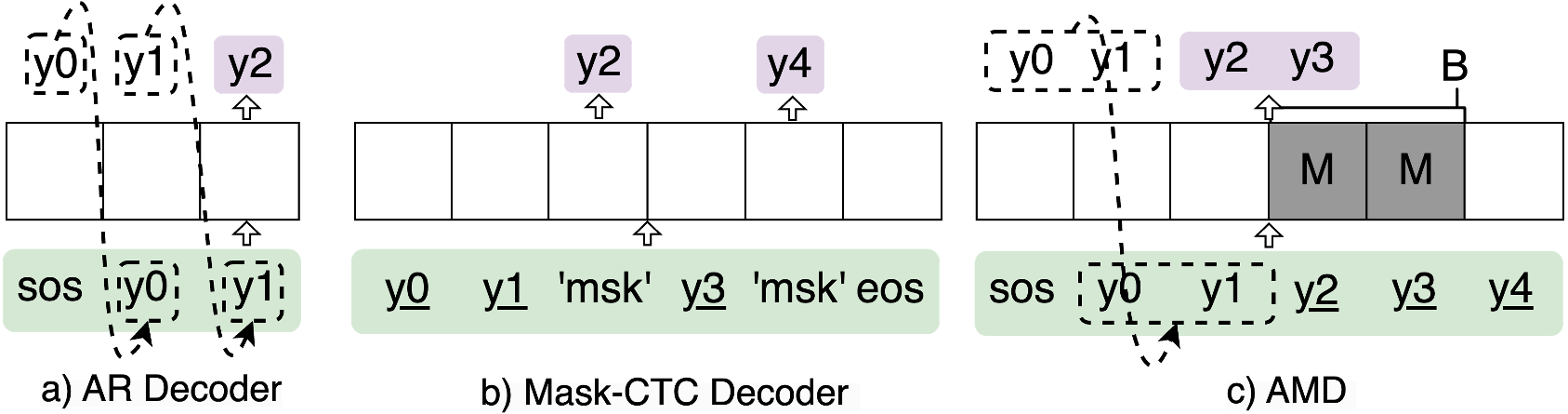}
    \vspace{-2mm} 
    \caption{Inference using: a) an AR decoder, b) a Mask Prediction decoder;  c) the proposed AMD. ``msk'' refers to input mask tokens, \text{\rm\colorbox{gray}{M}} is the attention mask within contiguous blocks for parallel inference via AMD, `B' refers to the block size. Tokens (green) denote decoder inputs at the current inference step. Tokens (purple) denote predicted tokens at current step. Tokens in dashed boxes in a) and c) represent those from the previous inference step. Tokens with ``\_" denote those obtained from CTC prediction. }
    \label{fig:attn}
    \vspace{-5mm}
\end{figure}

\section{Hybrid CTC + AR Encoder-Decoder Based ASR}
\label{sec:2}
This paper employs a hybrid CTC-attention encoder-decoder (AED) based ASR system ~\cite{watanabe2017hybrid} comprising three key components: A shared Conformer-based encoder, a CTC decoder and an attention-based AR decoder. The combined CTC plus AR loss function is used for model training, 
\vspace{-1mm}
\begin{equation}
\vspace{-1mm}
\mathcal{L}_{\text{C-A}}=\gamma_{1}\mathcal{L}_{\text{CTC}} + \gamma_{2}\mathcal{L}_{\text{AR}},
\label{eq:hybrid2}
\end{equation}
where $\gamma_1$ and $\gamma_2$ denote the weights 
applied to the CTC and AR decoder objectives during training, respectively.

This ASR system employs a label synchronous autoregressive beam search strategy ~\cite{watanabe2017hybrid} during inference, incorporating both CTC and AR decoder probabilities to generate the final hypothesis. For a partial hypothesis $\boldsymbol{h}{\leq i}$ at the $i$-th decoding step, its score is computed as
\vspace{-1mm}
\begin{equation}
\vspace{-1mm}
\alpha_{\text{C-A}}(\boldsymbol{h}_{\leq i})=\lambda_{1}\alpha_{\text{CTC}}(\boldsymbol{h}_{\leq i}) + \lambda_{2}\alpha_{\text{AR}}(\boldsymbol{h}_{\leq i}),
\label{eq:c-a}
\end{equation}
where $\lambda_1$ and $\lambda_2$ represent the respective weights for the CTC and AR decoder scores during the decoding process.

The CTC score $\alpha_{\text{CTC}}$ represents the log probability of all possible token sequences sharing the common prefix $\boldsymbol{h}_{\leq i}$,
\vspace{-1mm}
\begin{equation}
\vspace{-1mm}
\alpha_{\text{CTC}}(\boldsymbol{h}_{\leq i})=\text{log}P_{\text{CTC}}(\boldsymbol{h}_{\leq i}, \cdots\mid\boldsymbol{\mathcal{X}}),
\label{eq:ctc_p}
\end{equation}
where $\boldsymbol{\mathcal{X}}$ denote the encoder outputs. 

The AR decoder generates token probabilities conditioned on both the previous tokens $\boldsymbol{h}_{<i}$ and encoder outputs $\boldsymbol{\mathcal{X}}$, expressed as $P_{AR}(y_i\mid\boldsymbol{h}_{<i},\boldsymbol{\mathcal{X}})$. The AR decoder score is
\vspace{-2mm}
\begin{equation}
\alpha_{\text{AR}}(\boldsymbol{h}_{\leq i})=\text{log}\prod_{j=1}^iP_{\text{AR}}(y_{j}\mid\boldsymbol{h}_{<j},\boldsymbol{\mathcal{X}}).
\label{eq:ar_p}
\vspace{-3mm}
\end{equation}
However, the sequential nature of the AR decoder limits its potential for parallelization, making it a bottleneck in efficiency-sensitive applications.

\section{AMD Empowered Tripartite Decoder}

\label{sec:3}
To address this limitation, we propose an AMD-empowered tripartite decoder designed to flexibly balance the performance-efficiency trade-off. As detailed in Section \ref{sec:3.1}, this architecture integrates a novel non-autoregressive block-based attention mask decoder (AMD) (Figure \ref{fig:wavlm_main}, purple) alongside the standard CTC module (Figure \ref{fig:wavlm_main}, yellow) and attention-based AR decoder (Figure \ref{fig:wavlm_main}, red). Further integration with both WavLM features (Figure \ref{fig:wavlm_main}, blue) and LLM-based decoder (Figure \ref{fig:main-llama}) are introduced in Section \ref{sec:3.2} and Section \ref{sec:3.3} respectively.

\subsection{Block-Based Attention-Mask Decoder}
\label{sec:3.1}
Figure \ref{fig:attn} shows three decoding approaches respectively using: a) an AR decoder; b) a Mask Prediction decoder; and c) our proposed AMD.

As illustrated in Figure \ref{fig:attn} (a), the AR decoder generates output tokens autoregressively, with each token's probability conditioned on all previously predicted tokens in a strictly left-to-right manner. In contrast, the NAR Mask-CTC decoder ~\cite{higuchi2020mask} enables concurrent prediction of multiple tokens by replacing selected input positions with special ``msk'' symbols (e.g., ${y_2}$ and $y_4$, Figure \ref{fig:attn} (b)), thereby relaxing the sequential dependencies between masked positions. To address potential performance degradation associated with predicting all tokens simultaneously, the Mask-CTC decoder can be restricted to selectively predict only a small subset of tokens, with these positions determined by CTC scores at inference time.

However, the embedding of the ``msk'' token can still receive non-zero attention from other positions. In contrast, the proposed AMD enforces a hard constraint by directly modifying the attention scores to completely prevent the masked positions from contributing to the context vectors of other tokens. As illustrated in Figure \ref{fig:attn} (c), AMD uniquely combines parallel NAR inference and AR prediction. It employs block-based attention masks ``M'' to perform parallel inference within contiguous token blocks (e.g., ${y_2}$ and ${y_3}$), while maintaining sequential AR prediction between blocks. To prevent information leakage, AMD sets both the attention weights and token embedding outputs at masked positions to zeros. Unlike Mask-CTC, AMD performs prediction over all tokens without subset selection.

For a given input sequence $\boldsymbol{h}$ and encoder outputs $\boldsymbol{\mathcal{X}}$, the AMD probability for the $j$-th token within an attention-masked block spanning positions $[i, i+B-1]$ is defined as
\vspace{-1mm}
\begin{equation}
P_{\text{AMD}}(y_{j}\mid\boldsymbol{h}, \boldsymbol{\mathcal{X}})=P_{\text{AMD}}(y_j\mid\boldsymbol{h}_{<i},\boldsymbol{h}_{>i+B-1}, \boldsymbol{\mathcal{X}})
\label{eq:amd_p}
\vspace{-1mm}
\end{equation}
where $B$ represents the block size.
During training, AMD utilizes ground truth labels as input. We employ a specific training strategy to enhance the AMD model's generalization to all tokens in the training data, and its adaptability to varying block sizes during inference. For each training sentence, we perform four forward passes over all tokens. Each pass uses attention-mask blocks of a different size. The size is randomly sampled from the range [1, $L$], where $L$ is the sentence length. 
The AMD loss function $\mathcal{L}_{\text{AMD}}$ is formulated as
\begin{equation}
\label{eq:3}
\mathcal{L}_{\text{AMD}}=-\sum_{n=1}^4\text{log}\prod_{j=1}^L P_{\text{AMD}}(y_{j}\mid\boldsymbol{h}_{<i},\boldsymbol{h}_{>i+B_n-1}, \boldsymbol{\mathcal{X}})
\end{equation}

In this work, we investigate two distinct random masking strategies in our proposed AMD for implementing the AMD loss function $\mathcal{L}_{\text{AMD}}$,

\textbf{a)} Uniform attention-mask block size sampling (UNI.) maintains consistent block sizes within each of the 4 attention masked forward passes ($n=1,\dots,4$) performed over each utterance. 
For a given utterance of $L$ tokens, each forward pass $n$ randomly samples and uses {\bf a single block size} $B_n \in [1,L]$ that remains constant for all the token positions $1 \leq j \leq L$ within this utterance.

\textbf{b)} Variable attention-mask block size sampling (VAR.) introduces varying block sizes within each of the 4 forward passes performed for a training data utterance. 
For a given utterance of $L$ tokens, each forward pass $n$ randomly samples and uses {\bf a locally varying block size} $B_n \in [1,L]$ that remains constant for all the token positions $j$ within the same attention masked region ($i \leq j \leq i+B_n-1$) within this utterance.

The AMD framework is designed to be model-agnostic and can be integrated into various AR-based architectures. This is demonstrated in the following sections, where it is applied to a Conformer-based ASR system (enhanced with pre-trained speech features) and an LLM-based ASR system.

\subsection{Integration with Pre-trained Speech Model}
\label{sec:3.2}

A common approach of integrating pre-trained speech models~\cite{baevski2020wav2vec,hsu2021hubert,chen2022wavlm,babu2021xls} into ASR systems is to use their features. These domain invariant features are learned during  pre-training using large quantities of diverse speech data.
To accomplish this, the constructed ASR system employs WavLM ~\cite{chen2022wavlm} as the feature extractor, which is then passed through a Conformer-based encoder, followed by the proposed tripartite decoder.
WavLM is particularly well-suited for this role as it comprises CNN downsampling layers and Transformer blocks that jointly learn masked speech token prediction and denoising during pre-training. 

As shown in Figure \ref{fig:wavlm_main} (left, blue), let $\boldsymbol{\mathcal{M}}_{\text{WL}}$ denote the WavLM-model that produces contextualized representations $\boldsymbol{\mathcal{X}}_{\text{WL}} \in \mathbb{R}^{T' \times D}$ for a given input waveform $\boldsymbol{x} \in \mathbb{R}^T$, 
\vspace{-1mm}
\begin{equation}
\boldsymbol{\mathcal{X}}_{\text{WL}} = \boldsymbol{\mathcal{M}}_{\text{WL}}(\boldsymbol{x} \mid \bar{\boldsymbol{W}}_{\text{WL}}),
\label{eq:wl}
\vspace{-1mm}
\end{equation}
where $T$ and $T'$ denote the temporal dimensions before and after downsampling is performed by WavLM's CNN encoder, respectively. 
$D$ is the WavLM features dimensionality, and $\bar{\boldsymbol{W}}_{\text{WL}}$ represents the frozen WavLM parameters.
The extracted representations $\boldsymbol{\mathcal{X}}_{\text{WL}}$ are first transformed by a linear projection layer to match the Conformer encoder input dimension ($D \rightarrow D'$, where $D'$ is the input dimensionality of the Conformer encoder), then processed by the Conformer encoder $\boldsymbol{\mathcal{M}}_{\text{CF}}$, before producing the final {\it \bf tandem} encoder outputs $\boldsymbol{\mathcal{X}} \in \mathbb{R}^{T'' \times D''}$, as shown in Figure \ref{fig:wavlm_main} (the upper choice of Conformer input features in the red dotted circle),
\vspace{-1mm}
\begin{equation}
\boldsymbol{\mathcal{X}} = \boldsymbol{\mathcal{M}}_{\text{CF}}(\text{Linear}(\boldsymbol{\mathcal{X}}_{\text{WL}})\mid \boldsymbol{W}_{\text{CF}}),
\label{eq:cf}
\vspace{-1mm}
\end{equation}
where $T''$ is the temporal dimension after Conformer's downsampling, $D''$ and $\boldsymbol{W}_{\text{CF}}$ denote the output dimensionality and parameters of the Conformer encoder, respectively.

\subsection{Integration with Large Language Model based Decoder}
\label{sec:3.3}
Integrating a pre-trained LLM as an ASR decoder presents several design challenges, such as how to efficiently fuse modalities between the speech encoder and the text-based LLM, and how to adapt the fused LLM to the ASR task in a parameter-efficient manner. Integrating the proposed AMD framework introduces the additional challenge of supporting two distinct decoding objectives (AR and AMD) on a single backbone without doubling the model size. To address these, a novel, parameter-efficient architecture is proposed, as illustrated in Figure \ref{fig:main-llama}. In this architecture, a pre-trained Llama-3.2-1B\footnote{\url{https://huggingface.co/meta-llama/Llama-3.2-1B}} LLM is adopted as a shared, frozen backbone for both the AR and AMD decoders. The parameter increase is small, consisting of the AMD-specific text embedding and output layers, as well as the separate lightweight LoRA modules for each of the AR decoder and AMD.
A trainable adapter $\boldsymbol{\mathcal{M}}_{\text{ADP}}$ is introduced between the pre-trained speech encoder and the LLM-based decoder for modality fusion, to produce modality adapted features $\boldsymbol{\mathcal{X}}_{\text{ADP}}$ as 
\vspace{-1mm}
\begin{equation}
\boldsymbol{\mathcal{X}}_{\text{ADP}} = \boldsymbol{\mathcal{M}}_{\text{ADP}}(\boldsymbol{\mathcal{X}}\mid\boldsymbol{W}_{\text{ADP}}),
\label{eq:adp}
\end{equation}
\begin{equation}
\boldsymbol{\mathcal{M}}_{\text{ADP}}(\boldsymbol{\cdot}) = \operatorname{Linear}(\operatorname{ReLU}(\operatorname{Linear}(\operatorname{Attn}(\boldsymbol{\cdot})))),
\label{eq:adp_arch}
\end{equation}
where $\operatorname{Attn}(\cdot)$ denotes a self-attention mechanism, followed by two sequential linear transformations $\operatorname{Linear}(\cdot)$ with an intermediate ReLU activation $\operatorname{ReLU}(\cdot)$ applied in between these two.
All the trainable parameters of these adapter internal modules are collectively denoted as $\boldsymbol{W}_{\text{ADP}}$.

The adapted representations $\boldsymbol{\mathcal{X}}_{\text{ADP}}$ are concatenated with the embedded instruction prompts and text tokens $\boldsymbol{h}$, where the text embedding layer $E(\cdot)$ is applied to all text inputs, 
\begin{equation}
\boldsymbol{\mathcal{H}} = E(\boldsymbol{s_1}) \oplus \boldsymbol{\mathcal{X}}_{\text{ADP}} \oplus E(\boldsymbol{s_2}) \oplus E(\boldsymbol{h}),
\label{eqn:llm-speech-adapt-input}
\vspace{-1mm}
\end{equation}
where $\oplus$ denotes concatenation along temporal dimensions. The two instruction prompts $\boldsymbol{s_1}$ and $\boldsymbol{s_2}$ are ``\texttt{THE SPEECH IS:}'' and  ``\texttt{THE TRANSCRIPT IS:}'', respectively.

Let $\bar{\boldsymbol{W}} \in \mathbb{R}^{d_{out} \times d_{in}}$ stand for one of the frozen weight matrices in the feed-forward networks and group query attention layers (query, key, value, and output projections) of the pre-trained Llama model. 
LoRA ~\cite{hu2021lora} introduces a low-rank offset, bias parameter matrix that is target domain fine-tuned, before being added to the frozen Llama model parameters .
\vspace{-1mm}
\begin{equation}
\boldsymbol{W}=\bar{\boldsymbol{W}} + \Delta=\bar{\boldsymbol{W}}+\boldsymbol{BA},
\label{eq:lora}
\vspace{-1mm}
\end{equation}
where $\boldsymbol{B} \in \mathbb{R}^{d_{out} \times r}$ and $\boldsymbol{A} \in \mathbb{R}^{r \times d_{in}}$ represent the trainable low-rank adapter parameter matrices,  and $r$ is the rank of the decomposition, with $r \ll \min(d_{in}, d_{out})$.
Both the AR and AMD decoders share the same base parameters $\bar{\boldsymbol{W}}$, while maintaining their respective LoRA parameters $\Delta_{\text{AR}}$ and $\Delta_{\text{AMD}}$. 

Given the LoRA adapted speech encoder features that are augmented with instruction prompts in Eqn. (\ref{eqn:llm-speech-adapt-input}), 
 $\boldsymbol{\mathcal{H}}$, the respective AR and AMD probabilities assigned the $j$-th token in the predicted sequence $\boldsymbol{h}$ are defined as
\begin{equation}
P_{\text{AR}}(y_j\mid\boldsymbol{\mathcal{H}}) = P_{\text{AR}}(y_j\mid\boldsymbol{\mathcal{H}}_{<j};\Delta_{\text{AR}}, \bar{\boldsymbol{W}}),
\label{eq:llm-p-ar}
\end{equation}
\vspace{-1em}
\begin{equation}
\resizebox{0.43\textwidth}{!}{$P_{\text{AMD}}(y_j\mid\boldsymbol{\mathcal{H}}) = P_{\text{AMD}}(y_j\mid\boldsymbol{\mathcal{H}}_{<i},\boldsymbol{\mathcal{H}}_{>i+B-1};\Delta_{\text{AMD}}, \bar{\boldsymbol{W}})$},
\label{eq:llm-p-amd}
\end{equation}
where $j$ and $i$ denote the $j$-th token and the AMD attention mask block's starting position $i$ within the full composite sequence $\boldsymbol{\mathcal{H}}$, respectively.
\section{Decoding Utilizing Tripartite Decoder}
\label{sec:4}
\begin{algorithm}[t]
\caption{\footnotesize NAR Beam Search with Tripartite Decoder}
\scriptsize
\label{alg:beamsearch}

\SetCommentSty{green}  
\DontPrintSemicolon
\SetKwComment{Comment}{$\#$\ }{} 
\SetKwFor{ParFor}{for}{do in parallel}{end}
\let\oldnl\nl
\newcommand{\nonl}{\renewcommand{\nl}{\let\nl\oldnl}}

\nonl $L_{\text{max}}$: Maximum allowed hypothesis length (e.g. 512)\;
\nonl $B$: AMD NAR parallel inference block size (e.g. 2)\;
\nonl $K_{\text{main}}$: main top-K hypotheses beam for tripartite decoder (e.g. 4)\;
\nonl $K_{\text{1}}$: local top-K hypotheses beam in AMD search (e.g. 5)\;
\nonl $K_{\text{2}}$: local top-K hypotheses beam in CTC + AMD search (e.g. 5)\;
\nonl ${\cal H}^\text{main}$: Top $K_{\text{main}}$ hypotheses by CTC + AR + AMD tripartite decoder\;
\nonl ${\cal H}_j^\text{C-M}$: Top $K_{\text{1}}$ hypotheses by CTC + AMD decoding up to slot j\;
\nonl $\boldsymbol{h}^{\text{CTC}}$: CTC greedy search 1-best hypothesis\;

\hrule 
\vspace{1pt} 

\Comment*[l]{Each fixed size attention-mask block of $B$ label slots}
\For{$i = 1 \text{ to } L_\text{max}$ \textbf{with step} $B$}{
    \ParFor{$\boldsymbol{h}_{< i} \in {\cal H}^\text{main}$}{
        \Comment*[l]{\purple{AMD decoding on each of $B$ in-block label slots}}
        \ParFor{$j = i \text{ to } i+B-1$}{
 \Comment*[l]{Top $K_{\text{1}}$ labels predicted for each of $B$ in-block slots, $\alpha_{\text{AMD}}$ refers to the AMD score as defined in Eqn. \ref{eq:alpha_amd}}
 $\cal{Y}$$_j = \{y_j | y_j \in \text{Topk($\alpha_{{\text{AMD}}}(\boldsymbol{h}_{\leq j}), K_{\text{1}}$)}\}$
 
 \Comment*[l]{Union with $j^{th}$ label of CTC hypothesis $\boldsymbol{h}^{\text{CTC}}$}
 $\cal{Y}$$_j=\cal{Y}$$_j \cup \boldsymbol{h}^{\text{CTC}}_{j}$
        }
        \Comment*[l]{\purple{CTC + AMD decoding on each of $B$ in-block label slots}}
        \For{$j = i \text{ to } i+B-1$}{
 \Comment*[l]{Connecting ($\circ$) previous paths ${\cal H}_{j-1}^{\text{C-M}}$ up to label slot $j-1$ located immediately before slot $j$ with the $K_{\text{1}}$+1 in-block label predictions ${\cal Y}_{j}$ at position $j$}
 ${\cal H}_j = \{\boldsymbol{h}_{\leq j-1}\circ y_i|\boldsymbol{h}_{\leq j-1} \in {\cal H}_{j-1}^{\text{C-M}}$ and $y_i \in \cal{Y}$$_j\}$\;
 \Comment*[l]{CTC + AMD scores for expanded paths up to slot $j$}
 \For{$\boldsymbol{h}_{\leq j} \in {\cal H}_j$}{
 $\alpha_{\text{C-M}}(\boldsymbol{h}_{\leq j}) = \lambda_{1}\alpha_{{\text{CTC}}}(\boldsymbol{h}_{\leq j})+\lambda_{3}\alpha_{{\text{AMD}}}(\boldsymbol{h}_{\leq j})$\;
 }  
 \Comment*[l]{Pruning to top $K_{\text{2}}$ partial hypotheses up to slot $j$}
 ${\cal H}_{j}^{\text{C-M}}=\mathop{\text{Topk}}\limits_{\boldsymbol{h}_{\leq j}\in {\cal H}_{j}}(\alpha_{\text{C-M}}(\boldsymbol{h}_{\leq j}), K_{\text{2}})$\;
        }
    }
    \Comment*[l]{\purple{CTC + AR + AMD Tripartite re-ranking by adding AR scores}}
    \ParFor{$\boldsymbol{h}_{\leq i+B-1} \in {\cal H}_{i+B-1}^{\text{C-M}}$}{
        $\alpha_{\text{C-M-A}}(\boldsymbol{h}_{\leq i+B-1}) = \alpha_{\text{C-M}}(\boldsymbol{h}_{\leq i+B-1})+\lambda_{2}\alpha_{{\text{AR}}}(\boldsymbol{h}_{\leq i+B-1})$\;
    }
    \Comment*[l]{Pruning to top $K_{\text{main}}$ hypotheses leading up to last in-block slot}
    ${\cal H}^\text{main}=\mathop{\text{Topk}}\limits_{\boldsymbol{h}_{\leq i+B-1}\in {\cal H}_{i+B-1}^{\text{C-M}}}(\alpha_{\text{C-M-A}}(\boldsymbol{h}_{\leq i+B-1}), K_{\text{main}})$\;
}
\Return{${\cal H}^\text{main}$}
\end{algorithm}
Unlike the training procedure, which utilizes ground-truth input labels for block attention masking, the AMD decoding process relies on label hypotheses generated by a preliminary CTC greedy search. We found this reliance on the CTC greedy search to be robust; an ablation study on LS960 test sets showed no statistically significant WER difference between using the CTC hypothesis and the oracle ground truth as future context (3.22\% vs. 3.21\% average WER).
The AMD probabilities are dynamically fused with the CTC and AR decoder scores in a novel beam search algorithm (Algorithm \ref{alg:beamsearch}) tailored for AMD in this paper. 

\noindent The algorithm encompasses several crucial components beyond its NAR-based parallel processing within each block (line 3-6, Algorithm \ref{alg:beamsearch}):

\noindent\textbf{(1)} The left-to-right sequential nature of AR prediction necessitates maintaining contextual continuity between adjacent blocks. This is achieved by establishing connections between existing partial hypotheses up to the $(j{-}1)$-th position and the top-K candidate labels for position $j$ (line 7-8). 

\noindent\textbf{(2)} To improve the hypothesis diversity, the algorithm incorporates the CTC greedy search best-path output alongside AMD decoded results (line 5).

\noindent\textbf{(3)} 
The search efficiency is optimized through a three-stage pruning strategy. The initial top $K$ pruning is applied to individual token outputs within blocks using AMD scores (line 4, via $K_\text{1}$). This is followed by the intermediate top-$K$ pruning that targets expanded partial sequences using a weighted combination of CTC and AMD scores with their respective weights $\lambda_1$ and $\lambda_3$ (line 9-12, via $K_\text{2}$). The final top-$K$ pruning step using the tripartite CTC + AR + AMD decoder scores re-ranked hypotheses, where $\lambda_2$ is the weight for the AR decoder (line 15-18, via $K_\text{main}$).
The CTC score can be calculated efficiently following ~\cite{watanabe2017hybrid},  while the AMD score for a partial hypothesis sequence $\boldsymbol{h}_{\leq i}$ is calculated as:
\vspace{-2mm}
\begin{equation}
\alpha_{\text{AMD}}(\boldsymbol{h}_{\leq i})=\text{log}\prod_{j=1}^iP_{\text{AMD}}(y_j\mid\cdot),
\label{eq:alpha_amd}
\vspace{-2mm}
\end{equation}
where $P_{\text{AMD}}(y_j\mid\cdot)$ follows the same formulation as training in Eqn. (\ref{eq:amd_p}) and Eqn. (\ref{eq:llm-p-amd}), except that during decoding, the future context is approximated using CTC greedy search hypotheses, i.e., $P_{\text{AMD}}(y_j\mid\boldsymbol{h}_{<i},\boldsymbol{h}^{\text{\tiny CTC}}_{>i+B-1}, \boldsymbol{\mathcal{X}})$ for standard AMD and $P_{\text{AMD}}(y_j\mid\boldsymbol{\mathcal{H}}_{<i},\boldsymbol{\mathcal{H}}^{\text{\tiny CTC}}_{>i+B-1};\Delta_{\text{AMD}}, \bar{\boldsymbol{W}})$ for LoRA-adapted LLM-based AMD, respectively.
\vspace{-2mm}
\subsection{AMD Decoding Using Mixed Block Sizes}
\label{sec:4.1}
\begin{figure}
    \centering
    \includegraphics[width=0.5\linewidth]{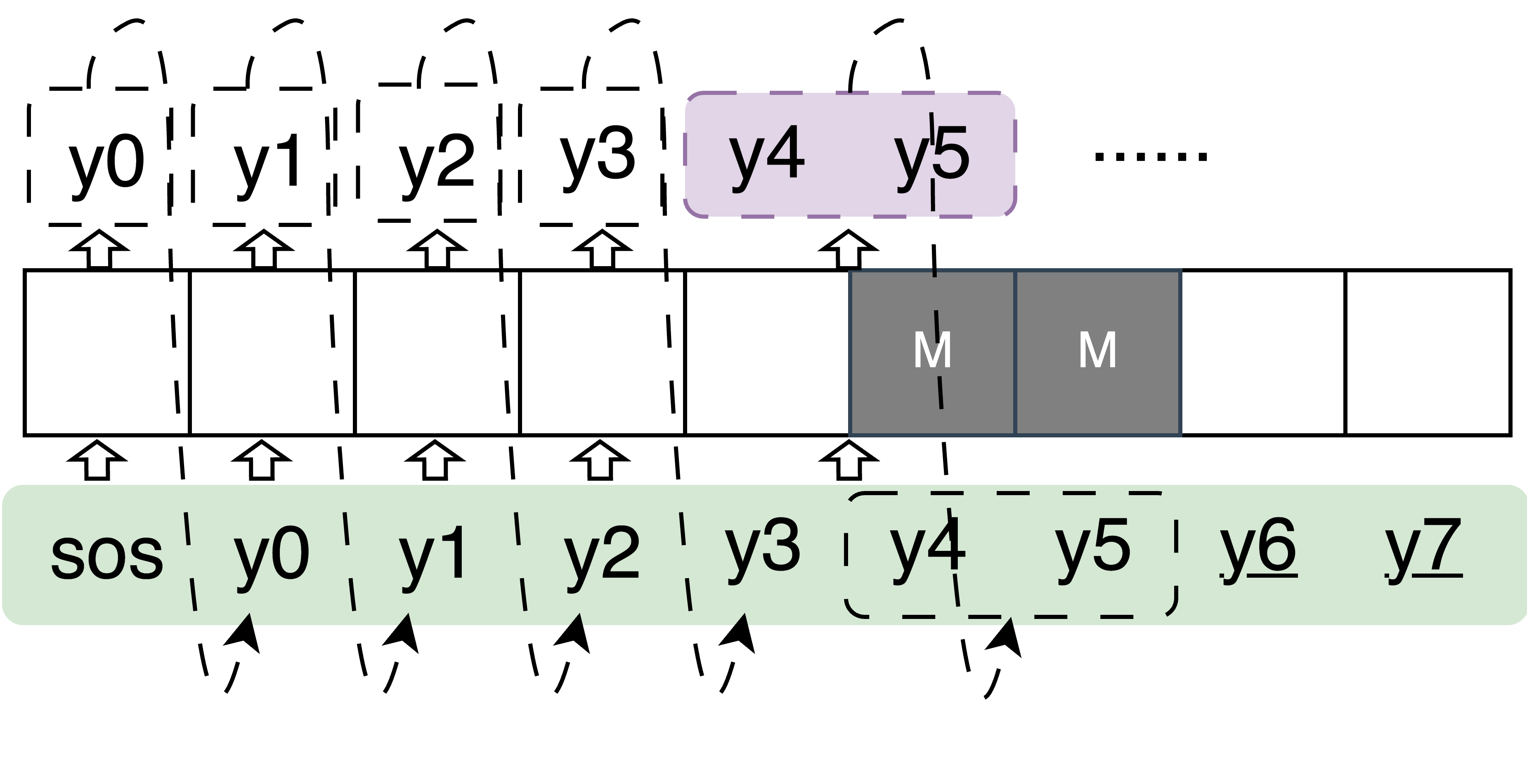}
    \vspace{-5mm}
    \caption{Inference using mixed size block sizes. Begins with cold start monotonic inference ($B=1$) for initial tokens y0-y3, then switches to larger ($B=2$) block sizes for parallel inference (y4, y5, pink)}
    \label{fig:mix-decoding}
    \vspace{-5mm}
\end{figure}

To achieve more flexible balance between performance and efficiency, we further propose AMD decoding using  mixed block sizes, as illustrated in Figure \ref{fig:mix-decoding}. In addition to using fixed-size attention-masking blocks during NAR inference, cold start monotonic inference (block size $B=1$) are employed for the initial N tokens' prediction. This enables a more steady, token by token build-up of history context in an AR fashion, before switching to faster parallel prediction ($B>1$) for the remaining tokens in the same utterance. This approach  flexibly balances the need of more precise history context modeling for initial few tokens, and the inference speed up brought by parallel decoding the remaining ones.   

\section{Experimental Setup}
\label{sec:exp_setup}
In this section, we detail the experimental setup used to evaluate the baseline and the proposed AMD-empowered tripartite decoder. The evaluation is conducted on the widely used LS960~\cite{panayotov2015librispeech} dataset and the DBank~\cite{becker1994natural} elderly speech corpus. Three model configurations are examined on both corpora:

{\it {Config. 1}}: The Conformer encoder-decoder ASR system with filterbank (FBank) input features, where the attention-based decoder is of approximately 30M parameters;

{\it {Config. 2}}: Integration of {\it {Config. 1}} with WavLM features as described in Section \ref{sec:3.2};

{\it {Config. 3}}: Further advances {\it {Config. 2}} by replacing the original decoder with a 1B-parameter LLM-based decoder as detailed in Section \ref{sec:3.3}.

Table~\ref{tab:model_params} lists the number of parameters for the baseline and proposed AMD-empowered ASR systems under {\it Config.~1-3}.

\begin{table}[h]
\vspace{-2mm}
\centering
\caption{Number of parameters for the baseline and proposed AMD-empowered ASR systems under {\it Config.~1-3}. ``M'' and ``B'' denote millions and billions of parameters, respectively.}
\label{tab:model_params}
\begin{tabular}{c l c}
\toprule
{\it Config.} & Decoder & \# Params \\
\midrule
\multirow{2}{*}{1} & CTC+AR & 116.15 M \\
 & CTC+AR+AMD & 146.50 M \\
\midrule
\multirow{2}{*}{2} & CTC+AR & 431.68 M \\
 & CTC+AR+AMD & 462.03 M \\
\midrule
\multirow{2}{*}{3} & CTC+AR$_\text{LLM}$ & 1.39 B \\
 & CTC+AR$_\text{LLM}$+AMD$_\text{LLM}$ & 1.42 B \\
\bottomrule
\end{tabular}
\end{table}

For training, models under {\it {Config. 1}} are trained on a single NVIDIA A40 GPU, models under {\it {Config. 2}} on 1$\times$A800 GPU, and models under {\it {Config. 3}} on 2$\times$A800 GPUs.
For RTF measurements, models under {\it {Config. 1}} and {\it \textbf{2}} are evaluated on a single NVIDIA A40-48GB GPU, while models under {\it {Config. 3}} are evaluated on 1$\times$A800-80GB GPU.
All experiments are conducted without external language models. Statistical significance testing is performed using matched pairs sentence-segment word error (MAPSSWE)~\cite{gillick1989some} at a significance level $\alpha$= 0.05.

\subsection{Experimental Setup for the LS960}
\subsubsection{Dataset}
The LS960~\cite{panayotov2015librispeech} corpus comprises 960 hours of English read speech from audiobooks for training, along with a 5.4-hour ``test-clean'' set featuring 40 speakers and a 5.1-hour ``test-other'' set featuring 33 speakers.

\subsubsection{Experimental Setup}
\label{sec:ls_expsetup}
\textbf{The Baseline Systems:}
The Conformer ASR \textbf{baseline system under {\it \textbf{Config. 1}}} using CTC + AR decoder 
is constructed following the ESPnet~\cite{watanabe2018espnet} recipe\footnote{\url{https://github.com/espnet/espnet/tree/master/egs2/librispeech/asr1#conformer-hop_length160}} with 12 Conformer blocks as encoder and 6 Transformer-based AR blocks as decoder\footnote{\# attention heads = 8; attention dim = 512; feed forward dim = 2048; convolution kernel size = 31.}.
The Conformer encoder achieves a 25Hz frame rate by applying a fourfold downsampling through its convolutional subsampling module to the input 80-dim FBank features extracted at 10ms intervals.
The AR decoder produces Byte-Pair-Encoding (BPE) ~\cite{gage1994new} tokens with a vocabulary size of 5000.
Speed perturbation ~\cite{ko2015audio} and SpecAugment ~\cite{park2019specaugment} are also performed for data augmentation. The model is trained for 50 epochs using the Adam~\cite{kingma2015adam} optimizer with a learning rate of 0.0025, 40,000 warmup steps, and a weight decay of 1e-6, using a combined CTC + AR loss function where weights are empirically set to $\gamma_1=0.3$ and $\gamma_2=0.7$ as specified in Eqn. (\ref{eq:hybrid2}). The final model is an average of the top 10 checkpoints based on dev set accuracy.
The \textbf{baseline system under {\it \textbf{Config. 2}} }is constructed following the ESPnet recipe\footnote{\url{https://github.com/espnet/espnet/tree/master/egs2/librispeech/asr1#self-supervised-learning-features-wavlm_large-conformer-utt_mvn-with-transformer-lm}}, with the WavLM features ($D=1024$) linearly projected to match the input dimension of Conformer encoder ($D'=80$). 
The other setups are the same as those in the baseline system under {\it {Config. 1}}.
For the \textbf{baseline system under {\it \textbf{Config. 3}}}, the WavLM feature extractor and Conformer encoder are identical to those used in the baseline model under {\it {Config. 2}}, with their weights inherited and kept frozen during training. A Llama-3.2-1B-Instruct model serves as the frozen backbone for AR$_{\rm LLM}$ decoder and is parameter-efficiently fine-tuned using the AdamW optimizer for 10 epochs (learning rate=0.001, warmup=1,000). The final model is obtained by element-wise averaging the LoRA weights from the top 5 checkpoints, selected based on dev set accuracy. The training costs for these baselines are as follows: the public recipe for {\it Config. 1} reports approximately 336 GPU hours on V100-32GB\footnote{\url{https://github.com/espnet/espnet/blob/master/egs2/librispeech/asr1/conf/tuning/train_asr_conformer10_hop_length160.yaml}}; the recipe for {\it Config. 2} reports approximately 480 GPU hours on A6000-48GB\footnote{\url{https://github.com/espnet/espnet/blob/master/egs2/librispeech/asr1/conf/tuning/train_asr_conformer7_wavlm_large.yaml}}. For {\it Config. 3}, the 1B LLM backbone's pre-training is reported as ~370k GPU hours on H100-80GB\footnote{\url{https://huggingface.co/meta-llama/Llama-3.2-1B}}, and our AR$_{\rm LLM}$ LoRA fine-tuning required approximately 106 GPU hours on A800-80GB.

\textbf{The ASR Systems with the AMD Empowered Tripartite Decoder:}
\textbf{Under {\it \textbf{Config. 1}} and {\it \textbf{Config. 2}}}, the CTC and AR components inherit frozen weights from the baseline system under {\it {Config. 1}} and {\it {Config. 2}}, respectively. The AMD employs the same architecture as the AR decoder, and is initialized with the AR decoder's parameters and subsequently fine-tuned on the LS960 training set following the same training setup as the baseline system under {\it {Config. 1}}. \textbf{Under {\it \textbf{Config. 3}}}, the LLM-based AMD (AMD$_\text{LLM}$) shares the pretrained frozen LLM weights with AR$_\text{LLM}$, where only the adapter, LoRA, and text embedding layers of AMD are randomly initialized and trained on the LS960, following the same training setup as the baseline system under {\it {Config. 3}}. The AMD fine-tuning required 161 GPU hours on A40-48GB for Config 1 and 264 GPU hours on A800-80GB for Config 2. For Config 3, both the AR$_{\rm LLM}$ and AMD$_{\rm LLM}$ LoRA fine-tuning converged in 10 epochs, requiring about 106 GPU hours and 366 GPU hours, respectively, on A800-80GB.

\subsection{Experimental Setup for the DBank}
\label{sec:exp_dbank}
\subsubsection{Dataset}
\label{sec:dbank_data}
The DBank~\cite{becker1994natural} corpus contains approximately 33 hours of neurocognitive assessment interviews recorded between 292 elderly participants (Par.) and clinical investigators (Inv.). The corpus is partitioned into a 27.2-hour training set, a 4.8-hour development (Dev) set, and a 1.1-hour evaluation (Eval) set.
The Eval set contains Cookie task recordings from 48 speakers identical to those in the ADReSS ~\cite{luz2021alzheimer} test set, while the Dev set includes these speakers' recordings from other tasks. 
The training, Dev, and Eval sets contain 688 speakers (244 elderly participants and 444 investigators), 119 speakers (43 elderly participants and 76 investigators), and 95 speakers (48 elderly participants and 47 investigators), respectively, with no speaker overlap between these sets. After silence removal ~\cite{ye2021development}, the training set contains 15.7 hours (29,682 utterances), while the Dev and Eval sets contain 2.5 hours (5,103 utterances) and 0.6 hours (928 utterances), respectively. Data augmentation via SpecAugment ~\cite{park2019specaugment} and speaker-independent speed perturbation of elderly speech and elderly speaker-dependent speed perturbation of non-aged investigators' speech ~\cite{ye2021development} expanded the training set to 58.9 hours. 
\subsubsection{Experimental Setup}
\label{sec:dbank_baseline}
\textbf{The Baseline Systems:} The Conformer ASR \textbf{baseline system under {\it \textbf{Config. 1}}} follows the same architecture as the LS960 {\it {Config. 1}} baseline system. The LS960 trained model was fine-tuned on the DBank training data for 25 epochs (Adam optimizer, learning rate=0.0025, warmup=2,000, weight decay=1e-6), with its output layer replaced to produce 100 BPE tokens derived from DBank transcripts. The final model is an average of the top 10 checkpoints based on dev set accuracy.
The \textbf{baseline system under {\it \textbf{Config. 2}}} employs a WavLM feature extractor that is initially fine-tuned on LS960, before being further fine-tuned on the DBank elderly speech data following ~\cite{hu2024self}, and kept frozen for subsequent experiments. It adopts the same architecture as the LS960 {\it {Config. 2}} baseline except for the following two hyper-parameters further adjusted for the DBank task: the convolutional downsampling module in the Conformer encoder (two-fold instead of four-fold); and the input dimensionality of Conformer encoder ($D'=1024$ rather than $D'=80$). For the {\it{Config. 2}} 
baseline, the Conformer encoder, CTC, and AR decoder are randomly initialized and trained on the DBank data in the same way as the {\it {Config. 1}} baseline system. 
For the \textbf{baseline system under {\it \textbf{Config. 3}}}, the WavLM feature extractor, Conformer encoder, and CTC are the same as those used in the baseline model under {\it {Config. 2}}, with their weights inherited and kept frozen during training. The AR$_{\rm LLM}$ decoder follows the same setup as the LS960 {\it {Config. 3}} baseline system (Sys. 9, Table \ref{tab:llama_baseline}) and is LoRA fine-tuned 
on the DBank for 10 epochs (AdamW, learning rate=0.001, warmup=1,000). The final model is an average of the top 5 checkpoints based on dev set accuracy. The DBank fine-tuning required 3 GPU hours on A40-48GB for Config 1, 4 GPU hours on A800-80GB for Config 2, and 9 GPU hours on A800-80GB for Config 3.

\textbf{The ASR Systems with the AMD Empowered Tripartite Decoder:}
Model configurations of the AMD Empowered Tripartite decoder
{\it {Config. 1-3}} are implemented following the same settings as in the LS960 task (Section \ref{sec:ls_expsetup}): The CTC and AR decoders are kept frozen while only the AMDs are either fully fine-tuned ({\it \textbf{Config. 1, 2}}) or LoRA fine-tuned ({\it \textbf{Config. 3}}), using the DBank dataset. The same training setup and hyper-parameters as the corresponding DBank baseline systems are used for each model configuration. The AMD fine-tuning required 4 GPU hours on A40-48GB for Config 1, 10 GPU hours on A800-80GB for Config 2, and 32 GPU hours on A800-80GB for Config 3.

\section{Implementation Details and Ablation Studies}
\label{sec:implementation_details}
This section discusses several implementation details and ablation studies that affect the performance of the baseline and proposed AMD-empowered system. The ablation studies conducted on LS960 include: \textbf{A.} the sampling strategy for attention mask block sizes (UNI. vs. VAR.); \textbf{B.} the selection of local top-K beam sizes ($K_1, K_2$) in the AMD beam search; and \textbf{C.} key factors for LLM integration (i.e. LoRA configurations, adapter architectures, and instruction prompts). For the DBank task, the optimal settings for the sampling strategy and LLM integration were adopted from the LS960 findings, while \textbf{D.} local top-K beam sizes were re-tuned and additional ablations on \textbf{E.} WavLM integration were performed.

\subsection{Sampling Strategies for Attention Mask Block Sizes on LS960}
\begin{table}[htbp]
\vspace{-3mm}
\setlength{\tabcolsep}{2pt}
\caption{
Ablation study of the \textbf{sampling strategies for attention mask block
sizes} (UNI. {\it vs.} VAR.) during training. Results are obtained using both \textbf{greedy search} ($K_\text{main}=1$) and \textbf{beam search} ($K_\text{main}=60$) under {\it {Config. 1}}. 
``Weights'' denotes the respective decoding weights for CTC, AR and AMD ($\lambda_1=0.3$, $\lambda_2=0.6$, $\lambda_3=0.1$) as specified in Algorithm \ref{alg:beamsearch}.
``Ave.'' denotes the average WER over \textbf{LS960 ``test-clean/other''} sets. 
``B${_{\rm TR}}$'' and ``B$_{\rm DEC}$'' denote the UNI./VAR. attention mask block size strategy used in training, and the fixed block size used during decoding, respectively.
}
\label{tab:ls960_univar}
\resizebox{\linewidth}{!}{
\begin{tabular}{c|c|c|c|c|c|ccc|c}
\hline
\multirow{2}{*}{Sys.} & \multirow{2}{*}{Encoder} & \multirow{2}{*}{Decoder} & \multirow{2}{*}{Weights}  & \multirow{2}{*}{B$_{\rm TR}$} & \multirow{2}{*}{B$_{\rm DEC}$} & \multicolumn{3}{c|}{WER} & \multirow{2}{*}{RTF} \\ \cline{7-9}
& & & & & & \multicolumn{1}{c|}{clean} & \multicolumn{1}{c|}{other} & Ave. & \\ 
\hline\hline
\multicolumn{10}{c}{Greedy Search ($K_\text{main}=1$)} \\
\hline\hline
1 &  \multirow{4}{*}{\begin{tabular}[c]{@{}c@{}}Conformer\end{tabular}} & \multirow{4}{*}{\begin{tabular}[c]{@{}c@{}}CTC+AR\\ +AMD\end{tabular}} & \multirow{4}{*}{\begin{tabular}[c]{@{}c@{}}0.3:0.6\\ :0.1\end{tabular}} & \multirow{2}{*}{UNI.} & 1 & 2.55        & 5.57        & 4.06 & 0.177 \\ \cline{1-1}
2 &  & & & & 4 & 2.69        & 5.87        & 4.28 & 0.114 \\ \cline{1-1}\cline{5-10}
3 &  & & & \multirow{2}{*}{VAR.} & 1 & 2.68        & 5.65        & 4.16 & 0.176 \\ \cline{1-1}
4 &  & & & & 4 & 2.81        & 5.97        & 4.43 & 0.113 \\ 
\hline\hline
\multicolumn{10}{c}{Beam Search ($K_\text{main}=60$)} \\
\hline\hline
5 &  \multirow{4}{*}{\begin{tabular}[c]{@{}c@{}}Conformer\end{tabular}} & \multirow{4}{*}{\begin{tabular}[c]{@{}c@{}}CTC+AR\\ +AMD\end{tabular}} & \multirow{4}{*}{\begin{tabular}[c]{@{}c@{}}0.3:0.6\\ :0.1\end{tabular}} & \multirow{2}{*}{UNI.} & 1 & 2.45 & 5.22 & 3.83 & 0.461 \\ \cline{1-1}
6 &  &  &  &  & 4 & 2.51 & 5.40 & 3.95 & 0.269 \\ \cline{1-1} \cline{5-10} 
7 &  &  &  & \multirow{2}{*}{VAR.} & 1 & 2.52 & 5.43 & 3.97 & 0.460 \\ \cline{1-1}
8 &  &  &  &  & 4 & 2.60 & 5.61 & 4.07 & 0.269 \\ \hline\hline
\end{tabular}
}
\vspace{-1mm}
\end{table}
As shown in Table \ref{tab:ls960_univar}, systems trained with UNI. mask sampling strategy consistently achieve lower WER than their VAR. counterparts at both B$_{\rm DEC}$=1 and B$_{\rm DEC}$=4, while exhibiting similar RTF values. This trend is observed in both greedy (Sys. 1 {\it vs.} 3, and Sys. 2 {\it vs.} 4, Table \ref{tab:ls960_univar}) and beam search scenario (Sys. 5 {\it vs.} 7, and Sys. 6 {\it vs.} 8, Table \ref{tab:ls960_univar}). This indicates that the \textbf{UNI.} mask sampling strategy provides more stable training process compared to the VAR. strategy, and is selected for subsequent experiments in Section \ref{sec:exp}.
\subsection{Local Top-K Hypotheses Selection on LS960} \label{sec:top_ls960}
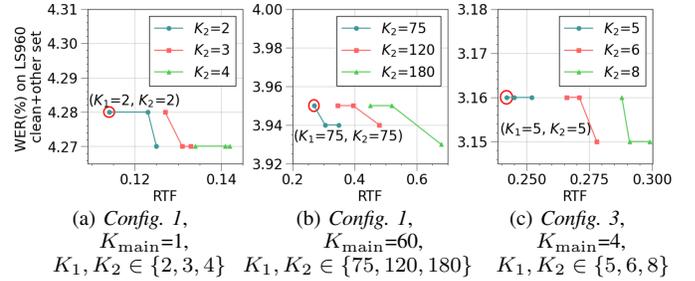
\begin{figure}[]
\centering
\begin{overpic}[width=\columnwidth]{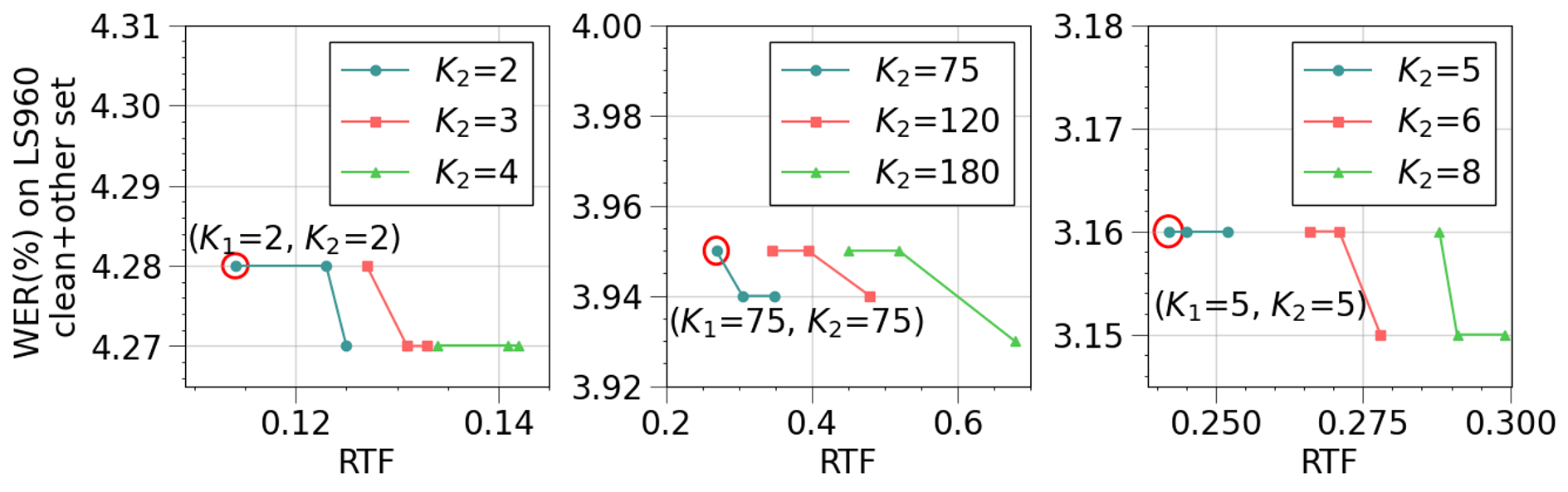}
  \put(8,-2.5){ \footnotesize (a) {\it {Config. 1}},}
  \put(12,-6){ \footnotesize {$K_{\rm main}$=1,}}
  \put(5,-9.5){ \footnotesize $K_1,K_2 \in \{2,3,4\}$}
  
  \put(43,-2.5){\footnotesize (b) {\it {Config. 1}},}
  \put(47,-6){\footnotesize $K_{\rm main}$=60,}
  \put(35,-9.5){\footnotesize $K_1,K_2 \in\{75,120,180\}$}
  
  \put(75,-2.5){\footnotesize (c) {\it {Config. 3}},}
  \put(79,-6){\footnotesize $K_{\rm main}$=4,}
  \put(73,-9.5){\footnotesize $K_1, K_2 \in \{5,6,8\}$}
\end{overpic}
\vspace{2mm}
\caption{Average WER (\%) and RTF trade-off for different $K_1$ and $K_2$ values across LS960 ``test-clean+other'' sets using (a) greedy search ($K_\text{main}$=1) under {\it {Config. 1}}; (b) beam search ($K_\text{main}$=60) under {\it {Config. 1}}, and (c) beam search ($K_\text{main}$=4) under {\it {Config. 3}}, all using fixed decoding block size B$_{\rm DEC}$=4, where $K_1$ determines the number of individual token candidates in the first pruning stage (Algorithm \ref{alg:beamsearch} line 4) and $K_2$ controls the beam width for partial hypotheses in the second pruning stage (Algorithm \ref{alg:beamsearch} line 12). Each point shows specific $(K_1, K_2)$ combinations. Red circles indicate optimal $K_1$ $K_2$ settings shown in Table \ref{tab:k1k2_summary}.}
\label{fig:k1k2}
\vspace{-5mm}
\end{figure}
For \textbf{greedy search} ($K_\text{main}=1$), experiments shown in Figure \ref{fig:k1k2}(a) for {\it {Config. 1}} demonstrate that modest increases in $K_1$ and $K_2$ relative to $K_\text{main}=1$ ($K_1=K_2=2$) yield strong performance improvements, while further increases offer negligible WER reductions ($\leq $0.01\% absolute) despite large RTF increases. Thus, $K_1=K_2=2$ is adopted for greedy search under {\it {Config. 1}}. This setting is also applied in {\it {Config. 2}} and {\it {Config. 3}} for greedy search to maintain consistency across different configurations.

For \textbf{beam search}, two specific scenarios were analyzed: {\it Config. 1-2} with $K_\text{main}=60$ (following the open-sourced ESPnet recipes, which can be found in footnotes 2 and 4), and {\it {Config. 3}} with $K_\text{main}=4$\footnote{Empirically determined based on memory-latency-performance trade-off.}. Under {\it {Config. 1}} (Figure \ref{fig:k1k2} (b)), increasing $K_1$ or $K_2$ beyond 75 yields diminishing returns in WER reduction while largely increasing computational cost. Therefore, $K_1=K_2=75$ is adopted for beam search under {\it {Config. 1}}. This same setting is applied for beam search under {\it {Config. 2}} because of the identical decoder architecture and beam size ($K_\text{main}=60$). Under {\it {Config. 3}} with $K_\text{main}=4$ (Figure \ref{fig:k1k2} (c)), varying $K_1$ and $K_2$ values from 5 to 8 produces nearly identical WER performance (ranging from 3.16\% to 3.15\%), while RTF gradually increases from 0.242 to 0.299. Therefore, $K_1=K_2=5$ is selected for {\it {Config. 3}} to optimize the WER-RTF trade-off. 

Table \ref{tab:k1k2_summary} summarizes the optimal parameter settings across {\it {Config. 1-3}} under greedy and beam search scenarios.

\begin{table}[htbp]
\small
\centering
\setlength{\tabcolsep}{3pt}
\caption{
Summary of $K_{\rm main}$, $K_1$ and $K_2$ settings on LS960.
}
\vspace{-2mm}
\label{tab:k1k2_summary}
\resizebox{0.8\columnwidth}{!}{
\begin{tabular}{c|ccc|ccc|ccc}
\hline
\multirow{2}{*}{Search Method} & \multicolumn{3}{c|}{{\it {Config. 1}}} & \multicolumn{3}{c|}{{\it {Config. 2}}} & \multicolumn{3}{c}{{\it {Config. 3}}} \\
\cline{2-10}
& $K_\text{main}$ & $K_1$ & $K_2$ & $K_\text{main}$ & $K_1$ & $K_2$ & $K_\text{main}$ & $K_1$ & $K_2$ \\
\hline\hline
Greedy search & 1 & 2 & 2 & 1 & 2 & 2 & 1 & 2 & 2 \\
Beam search & 60 & 75 & 75 & 60 & 75 & 75 & 4 & 5 & 5 \\
\hline
\end{tabular}}
\vspace{-4mm}
\end{table}
\vspace{-1mm}
\subsection{Key Factors for LLM Integration on LS960}
\begin{table}[t]
\scriptsize
\setlength\tabcolsep{1.5pt}
\caption{Performance of published LLM-based ASR results (Sys. 1-3) and baseline systems with CTC + AR$_\text{LLM}$ decoder (Sys. 4-9) on LS960 test sets. 
Systems vary in their settings of: {\bf a)} LoRA rank (Sys. 4, 5, 6); {\bf b)} adapter architectures (Sys. 6, 7, 8); and {\bf c)} whether using instruction prompts or not (Sys. 9, 8). 
}
\label{tab:llama_baseline}
\resizebox{\linewidth}{!}{
\begin{tabular}{c|c|c|c|c|c|c|cc}
\hline
\multirow{2}{*}{Sys.} & \multirow{2}{*}{Feat.} & \multirow{2}{*}{Encoder} & \multirow{2}{*}{Adapter} & \multirow{2}{*}{Decoder} & \multirow{2}{*}{Prompt} & \multirow{2}{*}{\begin{tabular}[c]{@{}c@{}}LoRA\\ Rank\end{tabular}} & \multicolumn{2}{c}{WER} \\ \cline{8-9} 
 &  &  &  &  &  &  & \multicolumn{1}{c|}{clean} & other \\ \hline\hline
\begin{tabular}[c]{@{}c@{}}1~\cite{tang2023salmonn}\end{tabular} & \multirow{7}{*}{FBank} & \begin{tabular}[c]{@{}c@{}}Whisper\\ Large\end{tabular} & Q-Former & Vicuna-13B-AR & \multirow{7}{*}{$\checkmark$} & 8 & \multicolumn{1}{c|}{2.1} & 4.9 \\ \cline{1-1} \cline{3-5} \cline{7-7}
\begin{tabular}[c]{@{}c@{}}2~\cite{chu2023qwen}\end{tabular} &  & \begin{tabular}[c]{@{}c@{}}Whisper\\ Large\end{tabular} & Linear & Qwen-7B-AR &  & - & \multicolumn{1}{c|}{2.0} & 4.2 \\ \cline{1-1} \cline{3-5} \cline{7-7}
\multirow{2}{*}{\begin{tabular}[c]{@{}c@{}}3~\cite{ma2024embarrassingly}\end{tabular}} &  & \begin{tabular}[c]{@{}c@{}}HuBERT \\ X-Large\end{tabular} & \multirow{4}{*}{Linear} & Vicuna-7B-AR &  & - & \multicolumn{1}{c|}{1.94} & 3.81 \\ \cline{3-3} \cline{5-5} \cline{7-7}
 &  & \begin{tabular}[c]{@{}c@{}}Whisper\\ Large\end{tabular} &  & \begin{tabular}[c]{@{}c@{}}TinyLlama\\ 1.1B-AR\end{tabular} &  & - & \multicolumn{1}{c|}{4.33} & 8.62 \\ \hline\hline
4 & \multirow{7}{*}{WavLM} & \multirow{7}{*}{Conformer} & \multirow{3}{*}{Linear} & \multirow{7}{*}{\begin{tabular}[c]{@{}c@{}}CTC+\\ AR$_\text{LLM}$ \\(Llama-3.2-1B-AR)\end{tabular}} & \multirow{6}{*}{$\times$} & - & \multicolumn{1}{c|}{4.55} & 6.37 \\ \cline{1-1}
5 &  &  &  &  &  & 16 & \multicolumn{1}{c|}{3.46} & 5.63 \\ \cline{1-1}
6 &  &  &  &  &  & 32 & \multicolumn{1}{c|}{3.44} & 5.62 \\ \cline{1-1} \cline{4-4} \cline{7-9} 
7 &  &  & \begin{tabular}[c]{@{}c@{}}Subsample\\ +Linear\end{tabular} &  &  & \multirow{3}{*}{16} & \multicolumn{1}{c|}{5.61} & 7.81 \\ \cline{1-1} \cline{4-4} \cline{8-9} 
8 &  &  & \multirow{2}{*}{\begin{tabular}[c]{@{}c@{}}Attention\\ +Linear\end{tabular}} &  &  &  & \multicolumn{1}{c|}{3.40} & 5.54 \\ \cline{1-1} \cline{6-6}
\textbf{9} &  &  &  &  & $\checkmark$ &  & \multicolumn{1}{c|}{\textbf{2.25}} & \textbf{4.43} \\ \hline
\end{tabular}}
\vspace{-2mm}
\end{table}

Table \ref{tab:llama_baseline} presents the performance of ASR systems with CTC + AR$_\text{LLM}$ decoder on the LS960 test sets, focusing on three critical aspects: the decoder \textbf{LoRA} configurations, where LoRA is applied to the grouped query attention components and feed-forward networks in each decoder layer, with ranks of 16 and 32 being examined; the \textbf{adapter} architectures, where three variants are considered: a simple linear network, a convolution-based sub-sampling layer with linear layers, and a self-attention layer with linear layers; and the \textbf{instruction prompts}, specifically the use of the ``\texttt{THE SPEECH IS: ... THE TRANSCRIPT IS: ...}'' template.
Several findings can be found:

First, the introduction of \textbf{LoRA} adaptation in the LLM-based AR decoder significantly improves system performance. Using a simple linear adapter, applying LoRA with rank $r=16$ to the decoder yields absolute WER reductions of 1.09\%/0.74\% (24.0\%/11.6\% relative) on ``test-clean/other'' sets respectively (Sys. 5 {\it vs.} 4, Table \ref{tab:llama_baseline}), while further increasing the rank $r$ to 32 results in marginal improvements (Sys. 6 {\it vs.} 5, Table \ref{tab:llama_baseline});

Second, among different \textbf{adapter architectures}, the ``Attention+Linear'' adapter outperforms both the simple ``Linear" adapter and the convolution-based ``Subsample+Linear" adapter. 
When using a LoRA rank of $r=16$, it produces absolute WER reductions of 0.06\% and 0.09\% (1.7\% and 1.6\% relative) on the ``test-clean/other'' sets, respectively, against the simple ``Linear" adapter (Sys. 8 {\it vs.} 5, Table \ref{tab:llama_baseline}), and also absolute reductions of 2.21\% and 2.27\% (39.4\% and 29.1\% relative) over the convolution-based ``Subsample+Linear" adapter (Sys. 8 {\it vs.} 7, Table \ref{tab:llama_baseline});

Third, the use of \textbf{instruction prompt} template is crucial for the LLMs-based decoder.
Incorporating the prompt template leads to absolute WER reductions of 1.15\% and 1.11\% (33.8\% and 20.0\% relative on the "test-clean/other" sets, respectively (Sys. 9 {\it vs.} 8, Table \ref{tab:llama_baseline}).
By using instruction prompts, competitive performance that is comparable to recently published LLM-based ASR results on the same task (Sys. 1-3, Table \ref{tab:llama_baseline}) are obtained (Sys. 9, Table \ref{tab:llama_baseline}).

Based on these findings, the ``Attention+Linear'' adapter with LoRA rank $r=16$ and the instruction prompt 
is adopted as the optimal configuration (Sys. 9, Table \ref{tab:llama_baseline}) for the systems under {\it {Config. 3}}.

\subsection{Local Top-K Hypotheses Selection on DBank}
Comparable ablation studies to LS960 task (Section~\ref{sec:top_ls960}) were conducted for the DBank dataset. The optimal parameter settings are summarized in Table \ref{tab:k1k2_summary_dbank}.
\begin{table}[htbp]
\small
\centering
\setlength{\tabcolsep}{3pt}
\caption{
Summary of $K_{\rm main}$, $K_1$, and $K_2$ settings on DBank.
}
\vspace{-2mm}
\label{tab:k1k2_summary_dbank}
\resizebox{0.8\columnwidth}{!}{
\begin{tabular}{c|ccc|ccc|ccc}
\hline
\multirow{2}{*}{Search Method} & \multicolumn{3}{c|}{{\it {Config. 1}}} & \multicolumn{3}{c|}{{\it {Config. 2}}} & \multicolumn{3}{c}{{\it {Config. 3}}} \\
\cline{2-10}
& $K_\text{main}$ & $K_1$ & $K_2$ & $K_\text{main}$ & $K_1$ & $K_2$ & $K_\text{main}$ & $K_1$ & $K_2$ \\
\hline\hline
Greedy search & 1 & 6 & 2 & 1 & 6 & 2 & 1 & 6 & 2 \\
Beam search & 60 & 75 & 65 & 60 & 75 & 65 & 4 & 24 & 8 \\
\hline
\end{tabular}}

\end{table}
\subsection{Key Factors for WavLM Integration on DBank}
Unlike LS960 task where the open-sourced model structure
was directly used as baseline system under {\it Config. 2}, ablation studies were conducted for DBank {\it Config. 2} baseline system. Based on Sys. 1, Table \ref{tab:dbank_wavlm_baseline}, which has the same structure as LS960's {\it {Config. 2}} baseline system, increasing frame rate ($\text{FR}_x$) from 12.5 Hz to 25 Hz by reducing the Conformer encoder convolutional downsampling factor from 4 to 2 leads to WER reduction by 2.35\% absolute  (9.74\% relative, Sys. 2 vs. 1, Table \ref{tab:dbank_wavlm_baseline}), while expanding encoder input dimensionality ($D'$) from 80 to 1024 further reduced WER by 0.31\% absolute (1.42\% relative, Sys. 3 vs. 2, Table \ref{tab:dbank_wavlm_baseline}). Based on these results, FR$_x=25$ Hz and $D'=1024$ are adopted as the optimal configurations for the systems under {\it {Config. 2}} (Sys. 7, Table \ref{tab:dbank_all}).
\begin{table}[t]
\scriptsize
\setlength\tabcolsep{1.5pt}
\caption{Ablation studies of the frame rate of encoder output $\boldsymbol{\mathcal{X}}$ (FR${x}$) and encoder input dimensionality ($D'$).}
\vspace{-3mm}
\label{tab:dbank_wavlm_baseline}
\resizebox{\linewidth}{!}{
\begin{tabular}{c|c|c|c|c|c|cc|cc|c}
\hline
\multirow{2}{*}{Sys.} & \multirow{2}{*}{Feats.} & \multirow{2}{*}{Encoder} & \multirow{2}{*}{$D'$} & \multirow{2}{*}{\begin{tabular}[c]{@{}c@{}}FR$_x$\\ (Hz)\end{tabular}} & \multirow{2}{*}{Decoder} & \multicolumn{2}{c|}{Eval-WER} & \multicolumn{2}{c|}{Dev-WER} & \multirow{2}{*}{Ave.} \\ 
\cline{7-10}
& & & & & & Inv. & Par. & Inv. & Par. & \\ 
\hline
1 &\multirow{3}{*}{WavLM} & \multirow{3}{*}{Conformer} & 80 & 12.5 & \multirow{3}{*}{CTC+AR} & 16.53 & 24.60 & 16.21 & 32.60 & 24.12 \\ 
2 & & & 80 & 25 & & 14.76 & 22.27 & 13.88 & 30.18 & 21.77 \\
3 & & & 1024 & 25 & & 14.76 & 21.70 & 13.78 & 29.76 & 21.46 \\
\hline
\end{tabular}}
\vspace{-3mm}
\end{table}
\section{Main Results}
\label{sec:exp}
\subsection{Main Results on LS960} This section presents the main results tested on the LS960 ``test-clean/other'' sets, comparing the performance of ASR systems with the CTC + AR decoder and with the AMD enhanced tripartite decoder under {\it {Config. 1-3}} . The detailed results are shown in Tables \ref{tab:ls960_config1}, \ref{tab:greedy_beam_wavlm}, and \ref{tab:llama_greedy_beam}.
\label{sec:ls960_mainresult}

\begin{table}[htbp]
\scriptsize
\setlength\tabcolsep{1.5pt}
\caption{Performance comparison of systems under under {\it {Config. 1}} using {\bf a)} CTC + AR decoder (Sys. 1, 6) and {\bf b)} tripartite decoder  using either greedy search (K$_{\text{main}}=1$, Sys. 2-5) or {\bf (c)} beam search (K$_{\text{main}}=60$, Sys. 7-11) on \textbf{LS960 ``test-clean/other''}
For Sys. 11, ``1-N-B'' denotes decoding with mixed block size(Section \ref{sec:4.1}), where the first N tokens are decoded in an AR manner, followed by non-AR decoding of remaining tokens with block size B.
\colorbox{yellow!20}{$\dagger$} and \colorbox{green!20}{$\ddagger$} denote that the average (Ave.) WER shows no statistically significant difference from, or achieves a statistically significant reduction compared to the corresponding baseline (Sys. 1 or Sys. 7), respectively.
\colorbox{blue!20}{$\diamond$} denotes the highlighted RTF is similar to the corresponding baselines. Other naming conventions follow those used in Table \ref{tab:ls960_univar}
}
\label{tab:ls960_config1}
\resizebox{\linewidth}{!}{
\begin{tabular}{c|c|c|c|c|ccc|c}
\hline
\multirow{2}{*}{Sys.} & \multirow{2}{*}{Encoder} & \multirow{2}{*}{Decoder} & \multirow{2}{*}{Weights} & \multirow{2}{*}{B$_{\text{DEC}}$} & \multicolumn{3}{c|}{WER} & \multirow{2}{*}{RTF} \\ \cline{6-8}
& & & & & \multicolumn{1}{c|}{clean} & \multicolumn{1}{c|}{other} & Ave. & \\ 
\hline\hline
\multicolumn{9}{c}{Greedy Search ($K_\text{main}=1$)} \\
\hline\hline
1 & \multirow{5}{*}{\begin{tabular}[c]{@{}c@{}}Conformer\end{tabular}} & CTC + AR & 0.3:0.7 & - & 2.96 & 5.80 & 4.37 & 0.148 \\ \cline{1-1} \cline{3-9} 
2 & & \multirow{4}{*}{\begin{tabular}[c]{@{}c@{}}CTC+AR\\ +AMD\end{tabular}} & \multirow{4}{*}{\begin{tabular}[c]{@{}c@{}}0.3:0.6\\ :0.1\end{tabular}} & 1 & 2.55 & 5.57 & \cellcolor{green!20}{4.06$\ddagger$} & 0.177 \\ \cline{1-1}
3 & & & & 2 & 2.61 & 5.75 & \cellcolor{green!20}{4.18$\ddagger$} & \cellcolor{blue!20}{0.149$\diamond$} \\ \cline{1-1}
4 & & & & 4 & 2.69 & 5.87 & \cellcolor{yellow!20}{4.28$\dagger$} & 0.114 \\ \cline{1-1} 
5 & & & & 8 & 2.74 & 5.92 & \cellcolor{yellow!20}{4.35$\dagger$} & 0.103 \\ 
\hline\hline
\multicolumn{9}{c}{Beam Search ($K_\text{main}=60$)} \\
\hline\hline
6 & \multirow{6}{*}{\begin{tabular}[c]{@{}c@{}}Conformer\end{tabular}} & CTC+AR & 0.3:0.7 & - & 2.43 & 5.18 & 3.79 & 0.364 \\ \cline{1-1} \cline{3-9} 
7 & & \multirow{5}{*}{\begin{tabular}[c]{@{}c@{}}CTC+AR\\ +AMD\end{tabular}} & \multirow{5}{*}{\begin{tabular}[c]{@{}c@{}}0.3:0.6\\ :0.1\end{tabular}} & 1 & 2.45 & 5.22 & \cellcolor{yellow!20}{3.83$\dagger$} & 0.461 \\ \cline{1-1}
8 & &  &  & 2 & 2.47 & 5.34 & \cellcolor{yellow!20}{3.89$\dagger$} & \cellcolor{blue!20}{0.367$\diamond$} \\ \cline{1-1}
9 & &  &  & 4 & 2.51 & 5.40 & 3.95 & 0.269 \\ \cline{1-1}
10 & &  &  & 8 & 2.63 & 5.73 & 4.18 & 0.266 \\ \cline{1-1}
11 & &  &  & 1-20-4 & 2.52 & 5.35 & \cellcolor{yellow!20}{3.93$\dagger$} & 0.279 \\ \hline
\end{tabular}}
\end{table}
\subsubsection{Performance Analysis under {\it {Config. 1}}}
 Several trends can be observed from Table \ref{tab:ls960_config1}:
\begin{enumerate}[leftmargin=8mm,labelsep=1mm]
\item[\textbf{a)}] Using a decoding block size B$_{\text{DEC}}$ of 1, the proposed AMD performs purely serial, non-parallel inference, akin to the AR decoder. 
The tripartite decoder integrating AMD outperforms the baseline CTC + AR system in greedy search. The tripartite decoder achieves an absolute averaged WER reduction of 0.31\% (7.1\% relative; Sys. 2 {\it vs.} 1, Table \ref{tab:ls960_config1})\footnote{Such performance gains may be attributed to the complementarity between the AMD and CTC + AR decoders when being combined.}, while incurs a 1.2x RTF increase due to the computational overhead of the AMD decoder (Sys. 2 {\it vs.} 1, Table \ref{tab:ls960_config1}).
\begin{figure}[htbp]
\centering
\subfloat[{\it Config. 1}]{\includegraphics[width=0.45\columnwidth]{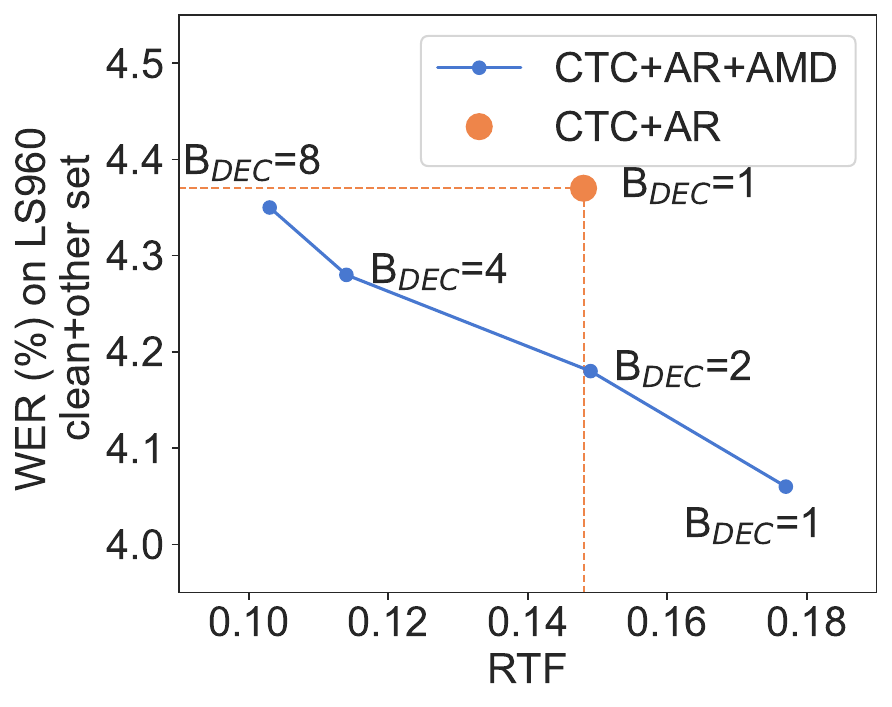}%
\label{fig:wer_rtf_config1_sub}}
\subfloat[{\it Config. 2}]{\includegraphics[width=0.45\columnwidth]{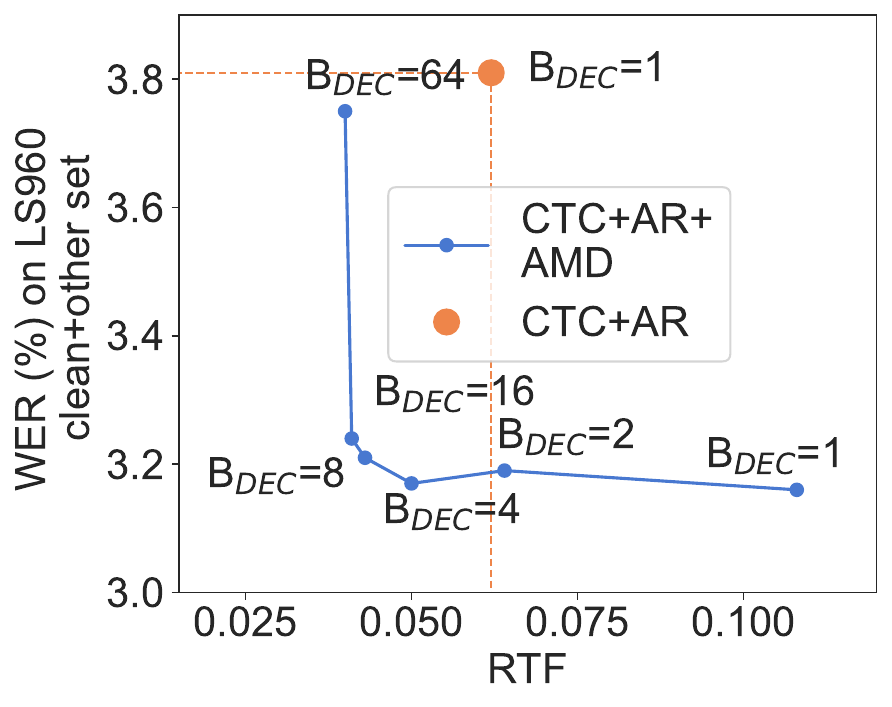}%
\label{fig:wer_rtf_config2_sub}}
\vspace{-0.7em}

\subfloat[{\it Config. 3}]{\includegraphics[width=0.45\columnwidth]{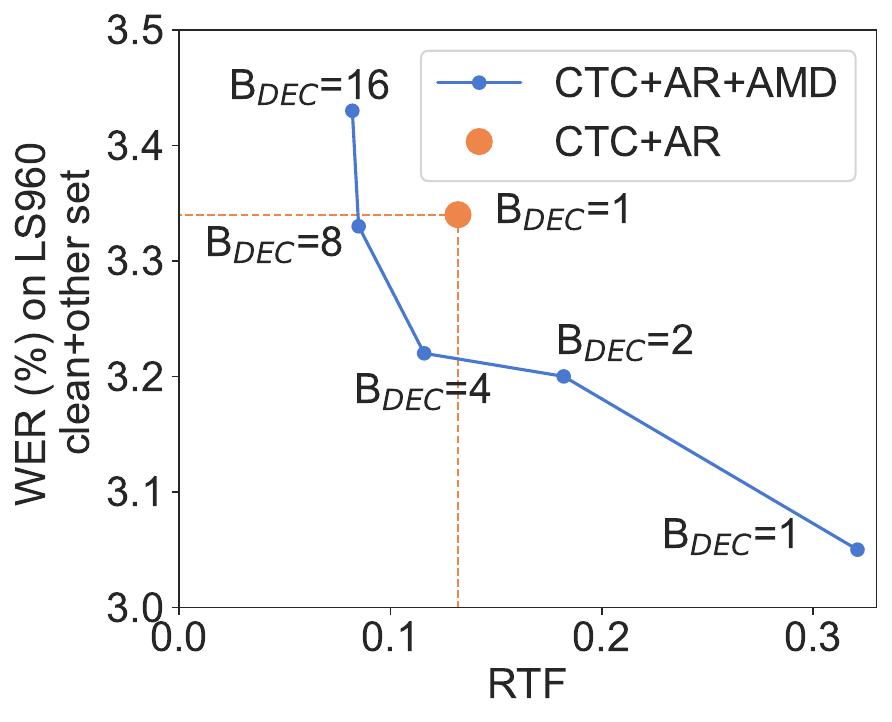}%
\label{fig:wer_rtf_config3_sub}}
\caption{Visualization of the Average WER (\%) vs. RTF trade-off on the LS960 ``test-clean+other'' sets for greedy search under (a) {\it Config. 1} (Sys. 1-5 in Table~\ref{tab:ls960_config1}); (b) {\it Config. 2} (Sys. 2-8 in Table~\ref{tab:greedy_beam_wavlm}); and (c) {\it Config. 3} (Sys. 3-8 in Table~\ref{tab:llama_greedy_beam}).}
\label{fig:wer_rtf_config}
\vspace{-4mm}
\end{figure}
\item[\textbf{b)}] As the decoding block size (B$_{\text{DEC}}$) increase from 1 to 8, the proposed tripartite decoder exhibit a clear \textbf{trade-off between WER and RTF} (as visualized in Figure~\ref{fig:wer_rtf_config}, (a)), showing an increase in average WER from 4.06\% to 4.35\% (Sys. 2-5, Table \ref{tab:ls960_config1}) alongside a decrease in RTF from 0.177 to 0.103 in greedy search.
\item[\textbf{c)}] In \textbf{greedy search}, with a decoding block size B$_{\rm DEC}$ of 8, the proposed tripartite decoder achieves a \textbf{1.44x} speedup ratio over the baseline CTC + AR system (Sys. 5 {\it vs.} 1, Table \ref{tab:ls960_config1}), with no statistically significant changes on averaged WERs. 
The tripartite decoder system (B$_{\rm DEC}$=2)  
operating with the RTF comparable to the baseline yield statistically significant WER reductions of up to \textbf{0.19\%} absolute (\textbf{4.3\%} relative; Sys. 3 {\it vs.} 1, Table \ref{tab:ls960_config1}).

\item[\textbf{d)}] In \textbf{beam search}, mixed-size decoding (Section \ref{sec:4.1})
enables the tripartite decoder to achieve up to a 1.30x speedup ratio relative to the baseline CTC + AR system (Sys. 11 {\it vs.} 6, Table \ref{tab:ls960_config1}) without statistically significant WER increase. When operating with a similar RTF, the tripartite decoder shows no significant difference in average WER compared to the CTC + AR baseline (Sys. 8 {\it vs.} 6, Table \ref{tab:ls960_config1}).
This is in contrast to the greedy search results, where systems with comparable RTF achieved significant WER reductions relative to the baseline (Sys. 3 {\it vs.} 1, Table \ref{tab:ls960_config1}). 
This performance disparity between greedy search and beam search will be analyzed in detail in Section \ref{sec:disparity_ana}.
\end{enumerate}
\begin{table}[htbp]
\vspace{-3mm}
\setlength\tabcolsep{1.5pt}
\caption{Performance comparison of systems using {\bf (a)} CTC + AR decoder under {\it {Config. 1}} (Sys. 1, 9) and {\it {Config. 2}} (Sys. 2, 10); {\bf (b)} CTC + AR + AMD tripartite decoder ({\it {Config. 2}}) using either greedy search (K$_{\text{main}}=1$, Sys. 3-8) or {\bf (c)} beam search (K$_{\text{main}}=60$, Sys. 11-16). 
FR$_{x}$ indicates the frame rate of encoder output $\boldsymbol{\mathcal{X}}$ in Eqn. (\ref{eq:cf}). 
\colorbox{yellow!20}{$\dagger$} and \colorbox{green!20}{$\ddagger$} denote that the average (Ave.) WER shows no statistically significant difference from, or achieves a statistically significant reduction compared to the corresponding baseline (Sys. 2 or Sys. 10), respectively.
\colorbox{blue!20}{$\diamond$} denotes the highlighted RTF is similar to the corresponding baselines. Other naming conventions follow those used in Table \ref{tab:ls960_univar} and \ref{tab:ls960_config1}.
}
\label{tab:greedy_beam_wavlm}
\resizebox{\linewidth}{!}{
\begin{tabular}{ccccccccccc}
\hline
\multicolumn{1}{c|}{\multirow{2}{*}{Sys.}} & \multicolumn{1}{c|}{\multirow{2}{*}{Feats.}} & \multicolumn{1}{c|}{\multirow{2}{*}{Encoder}} & \multicolumn{1}{c|}{\multirow{2}{*}{\begin{tabular}[c]{@{}c@{}}FR$_x$\\ (Hz)\end{tabular}}} & \multicolumn{1}{c|}{\multirow{2}{*}{Decoder}} & \multicolumn{1}{c|}{\multirow{2}{*}{Weights}} & \multicolumn{1}{c|}{\multirow{2}{*}{B$_{\rm DEC}$}} & \multicolumn{3}{c|}{WER} & \multirow{2}{*}{RTF} \\ \cline{8-10}
\multicolumn{1}{c|}{} & \multicolumn{1}{c|}{} & \multicolumn{1}{c|}{} & \multicolumn{1}{c|}{} & \multicolumn{1}{c|}{} & \multicolumn{1}{c|}{} & \multicolumn{1}{c|}{} & \multicolumn{1}{c|}{clean} & \multicolumn{1}{c|}{other} & \multicolumn{1}{c|}{Ave.} &  \\ \hline\hline
\multicolumn{11}{c}{Greedy Search ($K_\text{main}=1$)} \\ \hline\hline
\multicolumn{1}{c|}{1} & \multicolumn{1}{c|}{FBank} & \multicolumn{1}{c|}{\multirow{8}{*}{\begin{tabular}[c]{@{}c@{}}Conformer\end{tabular}}} & \multicolumn{1}{c|}{25} & \multicolumn{1}{c|}{CTC+AR} & \multicolumn{1}{c|}{0.3:0.7} & \multicolumn{1}{c|}{-} & 2.96 & 5.80 & \multicolumn{1}{c|}{4.37} & 0.148 \\ \cline{1-2} \cline{4-11} 
\multicolumn{1}{c|}{2} & \multicolumn{1}{c|}{\multirow{6}{*}{WavLM}} & \multicolumn{1}{c|}{} & \multicolumn{1}{c|}{\multirow{6}{*}{12.5}} & \multicolumn{1}{c|}{CTC+AR} & \multicolumn{1}{c|}{0.3:0.7} & \multicolumn{1}{c|}{-} & 2.78 & 4.85 & \multicolumn{1}{c|}{3.81} & 0.062 \\ \cline{1-1} \cline{5-11} 
\multicolumn{1}{c|}{3} & \multicolumn{1}{c|}{} & \multicolumn{1}{c|}{} & \multicolumn{1}{c|}{} & \multicolumn{1}{c|}{\multirow{5}{*}{\begin{tabular}[c]{@{}c@{}}CTC+AR\\ +AMD\end{tabular}}} & \multicolumn{1}{c|}{\multirow{5}{*}{\begin{tabular}[c]{@{}c@{}}0.3:0.6\\ :0.1\end{tabular}}} & \multicolumn{1}{c|}{1} & 2.09 & 4.22 & \multicolumn{1}{c|}{\cellcolor{green!20}{3.16$\ddagger$}} & 0.108 \\ \cline{1-1}
\multicolumn{1}{c|}{4} & \multicolumn{1}{c|}{} & \multicolumn{1}{c|}{} & \multicolumn{1}{c|}{} & \multicolumn{1}{c|}{} & \multicolumn{1}{c|}{} & \multicolumn{1}{c|}{2} & 2.09 & 4.29 & \multicolumn{1}{c|}{\cellcolor{green!20}{3.19$\ddagger$}} & \cellcolor{blue!20}{0.064$\diamond$} \\ \cline{1-1}
\multicolumn{1}{c|}{5} & \multicolumn{1}{c|}{} & \multicolumn{1}{c|}{} & \multicolumn{1}{c|}{} & \multicolumn{1}{c|}{} & \multicolumn{1}{c|}{} & \multicolumn{1}{c|}{4} & 2.10 & 4.24 & \multicolumn{1}{c|}{\cellcolor{green!20}{3.17$\ddagger$}} & 0.050 \\ \cline{1-1}
\multicolumn{1}{c|}{6} & \multicolumn{1}{c|}{} & \multicolumn{1}{c|}{} & \multicolumn{1}{c|}{} & \multicolumn{1}{c|}{} & \multicolumn{1}{c|}{} & \multicolumn{1}{c|}{8} & 2.14 & 4.27 & \multicolumn{1}{c|}{\cellcolor{green!20}{3.21$\ddagger$}} & 0.043 \\ \cline{1-1}
\multicolumn{1}{c|}{7} & \multicolumn{1}{c|}{} & \multicolumn{1}{c|}{} & \multicolumn{1}{c|}{} & \multicolumn{1}{c|}{} & \multicolumn{1}{c|}{} & \multicolumn{1}{c|}{16} & 2.15 & 4.34 & \multicolumn{1}{c|}{\cellcolor{green!20}{3.24$\ddagger$}} & 0.041 \\ 
\cline{1-1}
\multicolumn{1}{c|}{8} & \multicolumn{1}{c|}{} & \multicolumn{1}{c|}{} & \multicolumn{1}{c|}{} & \multicolumn{1}{c|}{} & \multicolumn{1}{c|}{} & \multicolumn{1}{c|}{64} & 2.64 & 4.85 & \multicolumn{1}{c|}{\cellcolor{yellow!20}{3.75$\dagger$}} & 0.040 \\ 
\hline\hline
\multicolumn{11}{c}{Beam Search ($K_\text{main}=60$)} \\ \hline\hline
\multicolumn{1}{c|}{9} & \multicolumn{1}{c|}{FBank} & \multicolumn{1}{c|}{\multirow{8}{*}{WavLM}} & \multicolumn{1}{c|}{25} & \multicolumn{1}{c|}{CTC+AR} & \multicolumn{1}{c|}{0.3:0.7} & \multicolumn{1}{c|}{-} & 2.43 & 5.18 & \multicolumn{1}{c|}{3.79} & 0.364 \\ \cline{1-2} \cline{4-11} 
\multicolumn{1}{c|}{10} & \multicolumn{1}{c|}{\multirow{7}{*}{WavLM}} & \multicolumn{1}{c|}{} & \multicolumn{1}{c|}{\multirow{7}{*}{12.5}} & \multicolumn{1}{c|}{CTC+AR} & \multicolumn{1}{c|}{0.3:0.7} & \multicolumn{1}{c|}{-} & 2.04 & 4.15 & \multicolumn{1}{c|}{3.09} & 0.268 \\ \cline{1-1} \cline{5-11} 
\multicolumn{1}{c|}{11} & \multicolumn{1}{c|}{} & \multicolumn{1}{c|}{} & \multicolumn{1}{c|}{} & \multicolumn{1}{c|}{\multirow{6}{*}{\begin{tabular}[c]{@{}c@{}}CTC+AR\\ +AMD\end{tabular}}} & \multicolumn{1}{c|}{\multirow{6}{*}{\begin{tabular}[c]{@{}c@{}}0.3:0.6\\ :0.1\end{tabular}}} & \multicolumn{1}{c|}{1} & 1.99 & 4.15 & \multicolumn{1}{c|}{\cellcolor{yellow!20}{3.07$\dagger$}} & 0.485 \\ \cline{1-1}
\multicolumn{1}{c|}{12} & \multicolumn{1}{c|}{} & \multicolumn{1}{c|}{} & \multicolumn{1}{c|}{} & \multicolumn{1}{c|}{} & \multicolumn{1}{c|}{} & \multicolumn{1}{c|}{2} & 2.08 & 4.15 & \multicolumn{1}{c|}{\cellcolor{yellow!20}{3.11$\dagger$}} & 0.298 \\ \cline{1-1}
\multicolumn{1}{c|}{13} & \multicolumn{1}{c|}{} & \multicolumn{1}{c|}{} & \multicolumn{1}{c|}{} & \multicolumn{1}{c|}{} & \multicolumn{1}{c|}{} & \multicolumn{1}{c|}{4} & 2.09 & 4.19 & \multicolumn{1}{c|}{\cellcolor{yellow!20}{3.14$\dagger$}} & 0.224 \\ \cline{1-1}
\multicolumn{1}{c|}{14} & \multicolumn{1}{c|}{} & \multicolumn{1}{c|}{} & \multicolumn{1}{c|}{} & \multicolumn{1}{c|}{} & \multicolumn{1}{c|}{} & \multicolumn{1}{c|}{8} & 2.07 & 4.20 & \multicolumn{1}{c|}{\cellcolor{yellow!20}{3.13$\dagger$}} & 0.191 \\ \cline{1-1}
\multicolumn{1}{c|}{15} & \multicolumn{1}{c|}{} & \multicolumn{1}{c|}{} & \multicolumn{1}{c|}{} & \multicolumn{1}{c|}{} & \multicolumn{1}{c|}{} & \multicolumn{1}{c|}{16} & 2.06 & 4.22 & \multicolumn{1}{c|}{\cellcolor{yellow!20}{3.13$\dagger$}} & 0.191 \\\cline{1-1}
\multicolumn{1}{c|}{16} & \multicolumn{1}{c|}{} & \multicolumn{1}{c|}{} & \multicolumn{1}{c|}{} & \multicolumn{1}{c|}{} & \multicolumn{1}{c|}{} & \multicolumn{1}{c|}{1-20-8} & 2.08 & 4.19 & \multicolumn{1}{c|}{\cellcolor{yellow!20}{3.13$\dagger$}} &  \cellcolor{blue!20}{0.262$\diamond$} \\ \hline
\end{tabular}
}
\vspace{-3mm}
\end{table}
\subsubsection{Performance Analysis under {\it {Config. 2}}} Several trends can be observed from Table \ref{tab:greedy_beam_wavlm}:
\begin{enumerate}[leftmargin=8mm,labelsep=1mm]
\item[\textbf{a)}] The \textbf{incorporation of WavLM features} significantly improves the CTC + AR system performance, with an absolute WER reduction of 0.56\% and 0.70\% (12.8\% and 18.5\% relative, Sys. 2 {\it vs.} 1, Sys. 10 {\it vs.} 9, Table \ref{tab:greedy_beam_wavlm}) using greedy search and beam search, respectively. Notably, compared to the baseline system with an encoder output frame rate of 25 Hz, WavLM-based systems operate at 12.5 Hz, further reducing the overall RTF;

\item[\textbf{b)}] In \textbf{greedy search}, a clear trade-off between WER and RTF is observed, similar to the trend in {\it Config. 1} (visualized in Figure~\ref{fig:wer_rtf_config}, (b)). Our tripartite decoder achieves up to a \textbf{1.55x} 
speedup ratio over the WavLM-enhanced CTC + AR baseline (Sys. 8 {\it vs.} 2, Table \ref{tab:greedy_beam_wavlm}) with no statistically significant changes on averaged WERs.  
When operating with an RTF comparable to the baseline (Sys. 4 {\it vs.} 2, Table \ref{tab:greedy_beam_wavlm}), the tripartite decoder system gives an absolute WER reduction of \textbf{0.62\%} (\textbf{16.3\%} relative);

\item[\textbf{c)}] In \textbf{beam search}, the tripartite decoder accelerates decoding by up to 1.40x relative to the baseline CTC + AR system (Sys. 14 {\it vs.} 10, Table \ref{tab:greedy_beam_wavlm}) without significant WER degradation. 
Moreover, mixed-size decoding allows the tripartite decoder to achieve an RTF comparable to the baseline while maintaining similar WER performance (Sys. 16 {\it vs.} 10, Table \ref{tab:greedy_beam_wavlm}), consistent with observations under {\it {Config.~1}}. 
The performance disparity between greedy search and beam search will be analyzed in detail in Section \ref{sec:disparity_ana}.
\end{enumerate}
\begin{table}[htbp]
\scriptsize
\setlength\tabcolsep{1.5pt}
\caption{Performance comparison of {\bf (a)} baseline systems under {\it {Config. 1}} (Sys. 1), {\it {Config. 2}} (Sys. 2), {\it {Config. 3}} (Sys. 3, 10) and {\bf (b)} proposed tripartite decoder under {\it {Config. 3}} using either greedy search (K$_{\text{main}}=1$, Sys. 3-9) or {\bf (c)} beam search (K$_{\text{main}}=4$, Sys. 11-16). 
\colorbox{yellow!20}{$\dagger$} and \colorbox{green!20}{$\ddagger$} denote that the average (Ave.) WER shows no statistically significant difference from, or achieves a statistically significant reduction compared to the corresponding baseline (Sys. 3 or Sys. 10), respectively.
\colorbox{blue!20}{$\diamond$} denotes the highlighted RTF is similar to the corresponding baselines. Other naming conventions follow those used in Table \ref{tab:ls960_univar} and \ref{tab:ls960_config1}.}
\label{tab:llama_greedy_beam}
\resizebox{\linewidth}{!}{
\begin{tabular}{cccccccccc}
\hline
\multicolumn{1}{c|}{\multirow{2}{*}{Sys.}} & \multicolumn{1}{c|}{\multirow{2}{*}{Feat.}} & \multicolumn{1}{c|}{\multirow{2}{*}{Encoder}} & \multicolumn{1}{c|}{\multirow{2}{*}{Decoder}} & \multicolumn{1}{c|}{\multirow{2}{*}{Weight}} & \multicolumn{1}{c|}{\multirow{2}{*}{B$_{\rm DEC}$}} & \multicolumn{3}{c|}{WER} & \multirow{2}{*}{RTF} \\ \cline{7-9}
\multicolumn{1}{c|}{} & \multicolumn{1}{c|}{} & \multicolumn{1}{c|}{} & \multicolumn{1}{c|}{} & \multicolumn{1}{c|}{} & \multicolumn{1}{c|}{} & \multicolumn{1}{c|}{clean} & \multicolumn{1}{c|}{other} & \multicolumn{1}{c|}{Ave.} &  \\ \hline\hline
\multicolumn{10}{c}{Greedy Search (K$_\text{main}$=1)} \\ \hline\hline
\multicolumn{1}{c|}{1} & \multicolumn{1}{c|}{FBank} & \multicolumn{1}{c|}{\multirow{10}{*}{Conformer}} & \multicolumn{1}{c|}{CTC+AR} & \multicolumn{1}{c|}{0.3:0.7} & \multicolumn{1}{c|}{-} & 2.96 & 5.80 & \multicolumn{1}{c|}{4.37} & 0.148 \\ \cline{1-2} \cline{4-10} 
\multicolumn{1}{c|}{2} & \multicolumn{1}{c|}{\multirow{9}{*}{WavLM}} & \multicolumn{1}{c|}{} & \multicolumn{1}{c|}{CTC+AR} & \multicolumn{1}{c|}{0.3:0.7} & \multicolumn{1}{c|}{-} & 2.78 & 4.85 & \multicolumn{1}{c|}{3.81} & 0.062 \\ \cline{1-1} \cline{4-10} 
\multicolumn{1}{c|}{3} & \multicolumn{1}{c|}{} & \multicolumn{1}{c|}{} & \multicolumn{1}{c|}{\begin{tabular}[c]{@{}c@{}}CTC+\\ AR$_\text{LLM}$\end{tabular}} & \multicolumn{1}{c|}{0.3:0.7} & \multicolumn{1}{c|}{-} & 2.25 & 4.43 & \multicolumn{1}{c|}{3.34} & 0.132 \\ \cline{1-1} \cline{4-10} 
\multicolumn{1}{c|}{4} & \multicolumn{1}{c|}{} & \multicolumn{1}{c|}{} & \multicolumn{1}{c|}{\multirow{6}{*}{\begin{tabular}[c]{@{}c@{}}CTC+\\ AR$_\text{LLM}$+ \\ AMD$_\text{LLM}$\end{tabular}}} & \multicolumn{1}{c|}{\multirow{5}{*}{\begin{tabular}[c]{@{}c@{}}0.3:0.6\\ :0.1\end{tabular}}} & \multicolumn{1}{c|}{1} & 1.98 & 4.12 & \multicolumn{1}{c|}{\cellcolor{green!20}{3.05$\ddagger$}} & 0.321 \\ \cline{1-1}
\multicolumn{1}{c|}{5} & \multicolumn{1}{c|}{} & \multicolumn{1}{c|}{} & \multicolumn{1}{c|}{} & \multicolumn{1}{c|}{} & \multicolumn{1}{c|}{2} & 2.10 & 4.31 & \multicolumn{1}{c|}{\cellcolor{green!20}{3.20$\ddagger$}} & 0.182 \\ \cline{1-1}
\multicolumn{1}{c|}{6} & \multicolumn{1}{c|}{} & \multicolumn{1}{c|}{} & \multicolumn{1}{c|}{} & \multicolumn{1}{c|}{} & \multicolumn{1}{c|}{4} & 2.11 & 4.34 & \multicolumn{1}{c|}{\cellcolor{yellow!20}{3.22$\dagger$}} & 0.116 \\ \cline{1-1}
\multicolumn{1}{c|}{7} & \multicolumn{1}{c|}{} & \multicolumn{1}{c|}{} & \multicolumn{1}{c|}{} & \multicolumn{1}{c|}{} & \multicolumn{1}{c|}{8} & 2.24 & 4.42 & \multicolumn{1}{c|}{\cellcolor{yellow!20}{3.33$\dagger$}} & 0.085 \\ \cline{1-1}
\multicolumn{1}{c|}{8} & \multicolumn{1}{c|}{} & \multicolumn{1}{c|}{} & \multicolumn{1}{c|}{} & \multicolumn{1}{c|}{} & \multicolumn{1}{c|}{16} & 2.32 & 4.55 & \multicolumn{1}{c|}{\cellcolor{yellow!20}{3.43$\dagger$}} & 0.082 \\\cline{1-1}
\multicolumn{1}{c|}{9} & \multicolumn{1}{c|}{} & \multicolumn{1}{c|}{} & \multicolumn{1}{c|}{} & \multicolumn{1}{c|}{} & \multicolumn{1}{c|}{1-10-4} & 2.11 & 4.32 & \multicolumn{1}{c|}{\cellcolor{green!20}{3.21$\ddagger$}} & \cellcolor{blue!20}{0.137$\diamond$} \\ \hline\hline
\multicolumn{10}{c}{Beam Search (K$_\text{main}$=4)} \\ \hline\hline
\multicolumn{1}{c|}{10} & \multicolumn{1}{c|}{\multirow{7}{*}{WavLM}} & \multicolumn{1}{c|}{\multirow{7}{*}{Conformer}} & \multicolumn{1}{c|}{\begin{tabular}[c]{@{}c@{}}CTC+\\ AR$_\text{LLM}$\end{tabular}} & \multicolumn{1}{c|}{0.3:0.7} & \multicolumn{1}{c|}{-} & 2.18 & 4.14 & \multicolumn{1}{c|}{3.16} & 0.370 \\ \cline{1-1} \cline{4-10} 
\multicolumn{1}{c|}{11} & \multicolumn{1}{c|}{} & \multicolumn{1}{c|}{} & \multicolumn{1}{c|}{\multirow{6}{*}{\begin{tabular}[c]{@{}c@{}}CTC+\\ AR$_\text{LLM}$+ \\ AMD$_\text{LLM}$\end{tabular}}} & \multicolumn{1}{c|}{\multirow{6}{*}{\begin{tabular}[c]{@{}c@{}}0.3:0.6\\ :0.1\end{tabular}}} & \multicolumn{1}{c|}{1} & 1.97 & 4.02 & \multicolumn{1}{c|}{\cellcolor{green!20}{2.99$\ddagger$}} & 0.836 \\ \cline{1-1}
\multicolumn{1}{c|}{12} & \multicolumn{1}{c|}{} & \multicolumn{1}{c|}{} & \multicolumn{1}{c|}{} & \multicolumn{1}{c|}{} & \multicolumn{1}{c|}{2} & 2.09 & 4.21 & \multicolumn{1}{c|}{\cellcolor{yellow!20}{3.14$\dagger$}} & 0.421 \\ \cline{1-1}
\multicolumn{1}{c|}{13} & \multicolumn{1}{c|}{} & \multicolumn{1}{c|}{} & \multicolumn{1}{c|}{} & \multicolumn{1}{c|}{} & \multicolumn{1}{c|}{4} & 2.07 & 4.25 & \multicolumn{1}{c|}{\cellcolor{yellow!20}{3.16$\dagger$}} & 0.242 \\ \cline{1-1}
\multicolumn{1}{c|}{14} & \multicolumn{1}{c|}{} & \multicolumn{1}{c|}{} & \multicolumn{1}{c|}{} & \multicolumn{1}{c|}{} & \multicolumn{1}{c|}{8} & 2.09 & 4.27 & \multicolumn{1}{c|}{\cellcolor{yellow!20}{3.17$\dagger$}} & 0.161 \\ \cline{1-1}
\multicolumn{1}{c|}{15} & \multicolumn{1}{c|}{} & \multicolumn{1}{c|}{} & \multicolumn{1}{c|}{} & \multicolumn{1}{c|}{} & \multicolumn{1}{c|}{16} & 2.09 & 4.30 & \multicolumn{1}{c|}{\cellcolor{yellow!20}{3.20$\dagger$}} & 0.160 \\  \cline{1-1}
\multicolumn{1}{c|}{16} & \multicolumn{1}{c|}{} & \multicolumn{1}{c|}{} & \multicolumn{1}{c|}{} & \multicolumn{1}{c|}{} & \multicolumn{1}{c|}{1-20-8} & 2.09 & 4.26 & \multicolumn{1}{c|}{\cellcolor{yellow!20}{3.16$\dagger$}} & \cellcolor{blue!20}{0.363$\diamond$} \\ \hline
\end{tabular}}
\vspace{-2mm}
\end{table}
\subsubsection{Performance Analysis under {\it {Config. 3}}} Table \ref{tab:llama_greedy_beam} reveals several key trends:
\begin{enumerate}[leftmargin=8mm,labelsep=1mm]
\item[\textbf{a)}] The \textbf{integration of LLM} as AR decoder significantly improves CTC + AR system performance, yielding an average WER reduction of 0.47\% (12.3\% relative) compared to the conventional AR decoder (Sys. 3 {\it vs.} 2, Table \ref{tab:llama_greedy_beam}) using greedy search.

\item[\textbf{b)}] When performing \textbf{greedy search}, a similar trade-off between WER and RTF is present, consistent with {\it Config. 1 and {\it 2}} (visualized in Figure~\ref{fig:wer_rtf_config}, (c)). The tripartite decoder achieves a speedup ratio of up to 1.61x relative to the CTC + AR baseline system (Sys. 8 {\it vs.} 3, Table \ref{tab:llama_greedy_beam}) with no statistically significant WER difference. 
When operating with an RTF comparable to the baseline, the system achieves an absolute average WER reduction of \textbf{0.13\%} (\textbf{3.8\%} relative, Sys. 9 {\it vs.} 3, Table \ref{tab:llama_greedy_beam});

\item[\textbf{c)}] In \textbf{beam search}, the tripartite decoder accelerates the decoding speed by up to \textbf{2.31x} compared to the CTC + AR baseline (Sys. 15 {\it vs.} 10, Table \ref{tab:llama_greedy_beam}) without significant WER degradation. 
When using mixed-size decoding, the tripartite decoder achieves similar RTF to the baseline while maintaining comparable WER performance (Sys. 16 {\it vs.} 10, Table \ref{tab:llama_greedy_beam}). 
Section \ref{sec:disparity_ana} presents the analysis of this disparity in performance gains over CTC + AR baselines between greedy and beam search when using tripartite decoders.
\end{enumerate}
\subsubsection{Performance Disparity between AMD Greedy and Beam Search}
\label{sec:disparity_ana}
\begin{figure}[htbp]
    \centering
    \subfloat[Lattice Density of {\it {Config. 1}}]{
        \includegraphics[width=0.45\columnwidth]{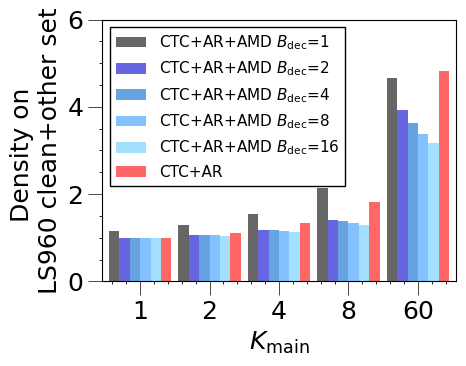}
        \label{fig:density_oraclea}
    }\hfill
    \subfloat[Oracle WER of {\it {Config. 1}}]{
        \includegraphics[width=0.47\columnwidth]{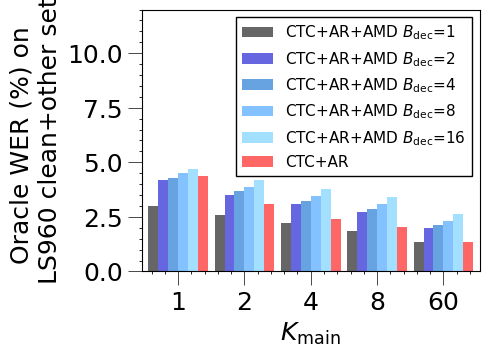}
        \label{fig:density_oracled}
    }\\[-3mm]
    \subfloat[Lattice Density of {\it {Config. 2}}]{
        \includegraphics[width=0.45\columnwidth]{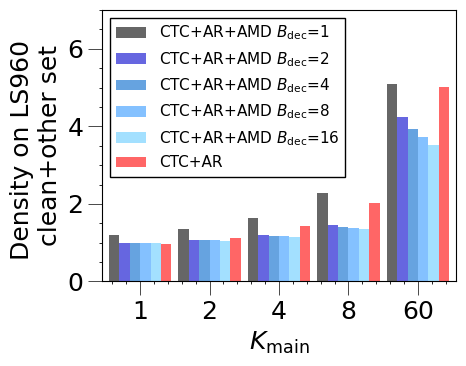}
        \label{fig:density_oracleb}
    }\hfill
    \subfloat[Oracle WER of {\it {Config. 2}}]{
        \includegraphics[width=0.47\columnwidth]{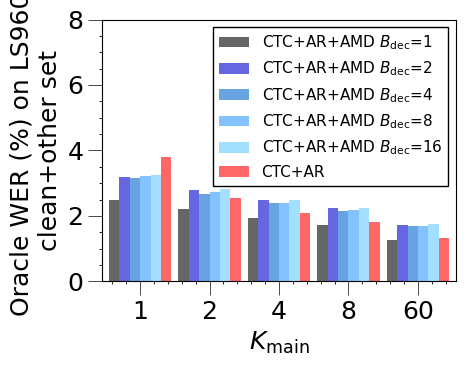}
        \label{fig:density_oraclee}
    }\\[-3mm]
    \subfloat[Lattice Density of {\it {Config. 3}}]{
        \includegraphics[width=0.45\columnwidth]{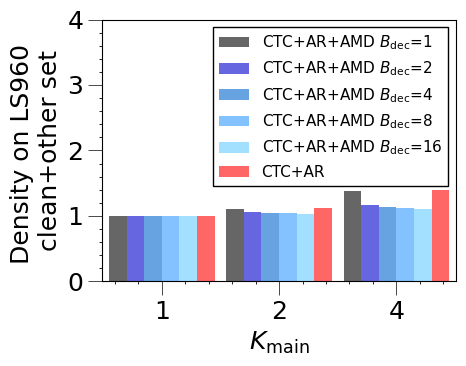}
        \label{fig:density_oraclec}
    }\hfill
    \subfloat[Oracle WER of {\it {Config. 3}}]{
        \includegraphics[width=0.45\columnwidth]{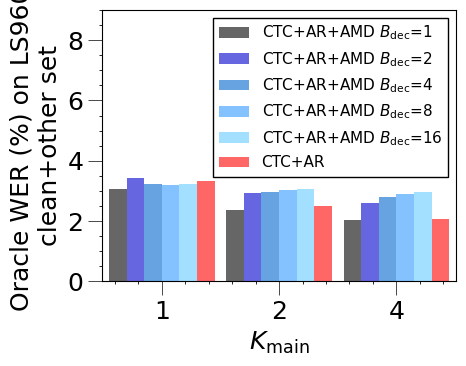}
        \label{fig:density_oraclef}
    }
    \caption{Lattice density (a, c, e) and Oracle WERs (b, d, f) computed for Conformer ASR systems using CTC + AR decoder (red bars) and using CTC + AR + AMD tripartite decoder (black and blue bars) over varying settings of $K_{\rm main}$, the main top-K hypotheses beam for the tripartite (Algorithm \ref{alg:beamsearch}) and baseline CTC + AR decoders. Results are presented for three model configurations introduced in the 1st paragraph of \ref{sec:exp}: {\it {Config. 1}} (a, b), {\it {Config. 2}} (c, d) with $K_\text{main}$ varying from 1 to 60, and {\it {Config. 3}} (e, f) with $K_\text{main}$ varying from 1 to 4. Both lattice density and oracle WERs are computed using 100-best hypotheses obtained on the LS960 ``test-clean+other'' sets.}
    \label{fig:density_oracle}
\end{figure}

In order to investigate the performance disparity when using AMD systems in greedy and beam search that previously observed (Sys. 3, 4, 9 vs. 8, 16, 16 in Tab. \ref{tab:ls960_config1}, \ref{tab:greedy_beam_wavlm} and \ref{tab:llama_greedy_beam} for {\it {Config. 1-3}} respectively)\footnote{Specifically, the AMD tripartite decoder produces statistically significant WER reductions over the CTC + AR baselines with greedy search (0.19\% absolute for Sys. 3 vs. Sys. 1 in Tab. \ref{tab:ls960_config1}, 0.62\% absolute for Sys. 4 vs. Sys. 2 in Tab. \ref{tab:greedy_beam_wavlm}, and 0.13\% absolute for Sys. 9 vs. Sys. 3 in Tab. \ref{tab:llama_greedy_beam}), while only maintaining WERs comparable to the CTC + AR baselines with beam search (Sys. 8 vs. Sys. 6 in Tab. \ref{tab:ls960_config1}, Sys. 16 vs. Sys. 10 in Tab. \ref{tab:greedy_beam_wavlm}, and Sys. 16 vs. Sys. 10 in Tab. \ref{tab:llama_greedy_beam}).}, their corresponding lattice density and oracle WER measures are analyzed in this section.
Lattice density measures the average number of distinct tokens in the 100-best list for each ground truth token. Oracle WER indicates the lowest possible WER achievable by selecting the optimal hypothesis in the 100-best list. 
Two trends can be observed in Figure \ref{fig:density_oracle}:
\begin{enumerate}[leftmargin=8mm,labelsep=1mm]
\item[\textbf{a)}] \textbf{For greedy search} (K$_{\text{main}}$=1), the tripartite decoder produces lattices with higher (when B$_{\text{DEC}}$=1) or comparable (when B$_{\text{DEC}}>$1) density, when compared to those produced by the CTC + AR counterparts across {\it {Config. 1-3}} (black \& blue {\it vs.} red bars in the leftmost bar chart of Figure \ref{fig:density_oracle}, (a), (c), (e)). This is achieved while maintaining lower (B$_{\text{DEC}}$=1) or comparable (B$_{\text{DEC}}>$1) oracle WERs (black \& blue {\it vs.} red bars in the leftmost bar chart of Figure \ref{fig:density_oracle}, (b), (d), (f)). This indicates that the AMD tripartite decoder produces fewer search errors than the CTC + AR baselines under greedy search.
\item[\textbf{b)}] \textbf{For beam search} (K$_{\text{main}}>1$), we observe different trends. {\bf When B$_{\text{DEC}}$=1, AMD performs serial, non-parallel prediction} akin to the AR decoder. In this case, 
the AMD tripartite decoder produces higher or comparable lattice density measures, and lower oracle WERs when compared to the CTC + AR baselines (black {\it vs.} red bars in Figure \ref{fig:density_oracle}, (a), (c), (e) for lattice density, and Figure \ref{fig:density_oracle}, (b), (d), (f) for oracle WER, from second to rightmost bar chart groups). These trends are similar to those found in the greedy search scenario in \textbf{a)}. In contrast, {\bf when B$_{\text{DEC}}=2,4,8,16$, AMD performs parallel NAR prediction}. In these cases, the AMD tripartite decoder produces lower or comparable lattice density measures, and higher oracle WERs when compared to the CTC + AR baselines (blue bars from dark to light {\it vs.} red bars in Figure \ref{fig:density_oracle}, (a), (c), (e) for lattice density, and Figure \ref{fig:density_oracle}, (b), (d), (f) for oracle WER, from second to rightmost bar chart groups). These led to reduced performance gains from AMD over the CTC + AR baselines in beam search compared to those obtained in greedy search (Sys. 3, 4, 9 vs. 8, 16, 16 in Tab. \ref{tab:ls960_config1}, \ref{tab:greedy_beam_wavlm} and \ref{tab:llama_greedy_beam} for {\it{Config. 1-3}}).
\end{enumerate}
\subsubsection{Analysis of Mixed Block-Size Decoding}
\label{sec:mixed_block_analysis}
To empirically validate the mixed block-size decoding technique described in Section \ref{sec:4.1}, we conducted a further analysis for greedy search under {\it {Config. 3}}. The mixed block-size decoding strategy offers a more flexible balance between performance and efficiency compared to the fixed block-size decoding, as shown in Table \ref{tab:ablation_mixed_llm}:
\begin{enumerate}[leftmargin=8mm,labelsep=1mm]
    \item[\textbf{a)}] The strategy allows for configurations that prioritize accuracy. For instance, the system with B$_{\rm DEC}$=4 and N=10 achieves a statistically significant absolute WER reduction of 0.13\% (3.9\% relative) compared to the CTC+AR$_\text{LLM}$ baseline, while operating at a similar RTF (0.137 vs. 0.132).
    \item[\textbf{b)}] Conversely, it also enables configurations that prioritize speed without performance degradation. The system with B$_{\rm DEC}$=16 and N=20 achieves a 1.28x speedup over the baseline (RTF of 0.103 vs. 0.132) while maintaining a comparable (and slightly improved) WER.
\end{enumerate}
\begin{table}[h!]
\centering
\caption{Ablation study on the mixed block size decoding strategy under {\it Config. 3}, grouped by Parallel Block Size (B$_{\rm DEC}$). ``N'' denotes the Switch Point, the number of initial tokens decoded autoregressively; N=0 indicates the Fixed Block Size strategy.}
\label{tab:ablation_mixed_llm}
\begin{tabular}{c|c|c|c}
\hline\hline
B$_{\rm DEC}$ & N & WER (\%) & RTF \\ 
\hline\hline
\multicolumn{2}{c|}{CTC+AR$_\text{LLM}$} & 3.34 & 0.132 \\ 
\hline
\multirow{4}{*}{4} & 0 & 3.22 & 0.116 \\
& 10 & 3.21 & 0.137 \\
& 20 & 3.20 & 0.163 \\
\hline
\multirow{4}{*}{8} & 0 & 3.33 & 0.085 \\
& 10 & 3.27 & 0.107 \\
& 20 & 3.22 & 0.124 \\
\hline
\multirow{4}{*}{16} & 0 & 3.43 & 0.082 \\
& 10 & 3.39 & 0.092 \\
& 20 & 3.32 & 0.103 \\
\hline\hline
\end{tabular}
\end{table}

\subsubsection{Analysis of AR and AMD Attention}
\begin{figure}[htb]
    \centering 
    \subfloat[AR Cross Attention\label{fig:ar_src_attn}]{
        \includegraphics[width=0.45\linewidth]{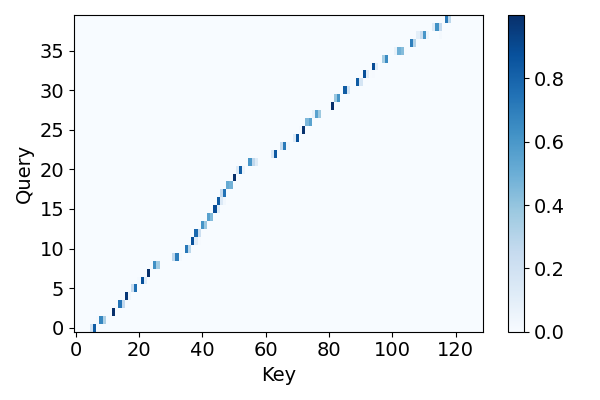}
    }
    \hfill
    \subfloat[AR Self Attention\label{fig:ar_self_attn}]{
        \includegraphics[width=0.45\linewidth]{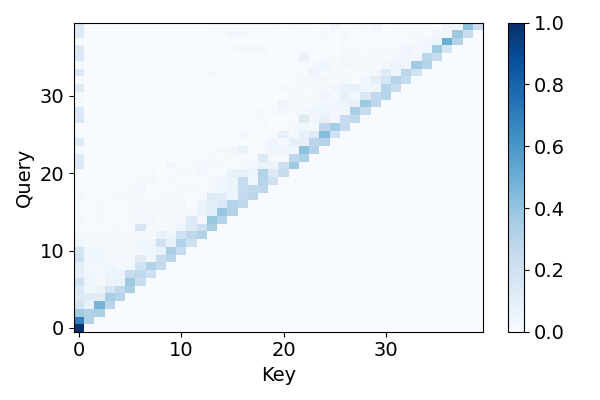}
    }
    \\ 
    \subfloat[AMD Cross Attention\label{fig:amd_src_attn}]{
        \includegraphics[width=0.45\linewidth]{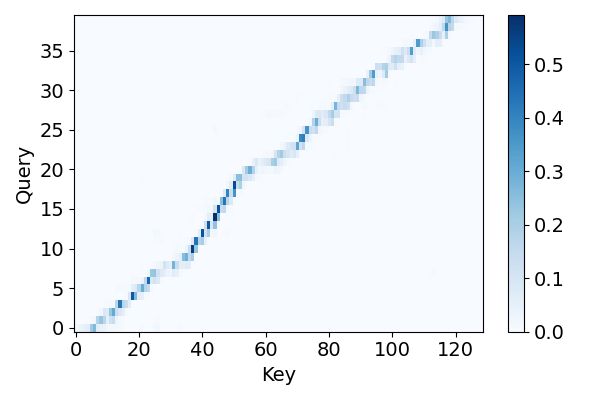}
    }
    \hfill 
    \subfloat[AMD Self Attention\label{fig:amd_self_attn}]{
        \includegraphics[width=0.45\linewidth]{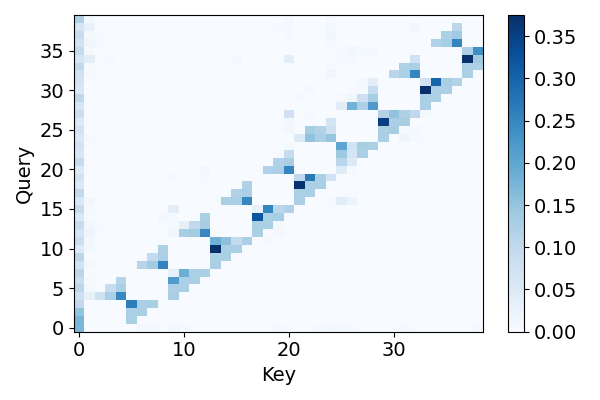}
    }
    \caption{Visualization of self and cross-attention patterns under {\it Config. 2} for (a)-(b) the AR decoder and (c)-(d) the AMD. (a) and (c) display the cross-attention between decoder output tokens and speech embeddings, while (b) and (d) display the self-attention maps of the decoder output tokens. The utterance is from the LS960 ``test-clean'' set, and the attention map is extracted from the last layer and averaged across all attention heads.}
    \label{fig:attention_comparison}
    \vspace{-3mm}
\end{figure}

To provide mechanistic insight of the AMD, we compares the attention patterns of the AR decoder and the AMD, examining both self-attention and cross-attention. As shown in Figure~\ref{fig:attention_comparison}, two observations can be made:
\begin{enumerate}[leftmargin=8mm,labelsep=1mm]
    \item[\textbf{a)}] Both the AR decoder and AMD attend to encoder frames in a roughly monotonic alignment, and the AMD exhibit broader attention corresponding to the block ((c) vs. (a)).
    \item[\textbf{b)}] By comparing (b) and (d), the AR decoder attends primarily to the immediately preceding token, reflecting its strict left-to-right, token-level dependency. In contrast, the AMD decoder attends to both the preceding left context tokens and the right context tokens provided by the CTC greedy result, demonstrating its block-level processing which incorporates future context.
\end{enumerate}
\subsection{Main Results on DBank task}
\label{sec:exp_dbank}
Table \ref{tab:dbank_all} presents the performance comparison between baseline systems and the proposed tripartite decoder across {\it {Config. 1-3}} on the DBank Eval and Dev sets.

\begin{table}[t]
\scriptsize
\setlength\tabcolsep{1.5pt}
\caption{Performance comparison of {\bf (a)} baseline systems under {\it {Config. 1}} (Sys. 1, 21), {\it {Config. 2}} (Sys. 7, 27), and {\it {Config. 3}} (Sys. 14, 33); and {\it {Config. 3}} (Sys. 14, 33); and {\bf (b)} proposed tripartite decoder integrating AMD under {\it {Config. 1}} (Sys. 2-6, 22-26), {\it {Config. 2}} (Sys. 8-13, 28-32), and {\it {Config. 3}} (Sys. 15-20, 34-38). Results are reported for both greedy search (Sys. 1-20) and beam search (Sys. 21-38). ``Ave.'' denotes the average WER across \textbf{DBank Eval and Dev} sets.
\colorbox{yellow!20}{$\dagger$} and \colorbox{green!20}{$\ddagger$} denote that the average (Ave.) WER shows no statistically significant difference from, or achieves a statistically significant reduction over the comparable baseline.
\colorbox{blue!20}{$\diamond$} denotes the highlighted RTF is similar to the comparable baselines. 
Other naming conventions follow those used in Table \ref{tab:ls960_univar} and \ref{tab:ls960_config1}.}
\vspace{-3mm}
\label{tab:dbank_all}
\resizebox{\linewidth}{!}{
\begin{tabular}{ccccccccccccc}
\hline
\multicolumn{1}{c|}{\multirow{2}{*}{Sys.}} & \multicolumn{1}{c|}{\multirow{2}{*}{Config.}} & \multicolumn{1}{c|}{\multirow{2}{*}{Feats.}} & \multicolumn{1}{c|}{\multirow{2}{*}{Encoder}} & \multicolumn{1}{c|}{\multirow{2}{*}{Decoder}} & \multicolumn{1}{c|}{\multirow{2}{*}{Weights}} & \multicolumn{1}{c|}{\multirow{2}{*}{B$_{\rm DEC}$}} & \multicolumn{2}{c|}{Eval-WER} & \multicolumn{2}{c|}{Dev-WER} & \multicolumn{1}{c|}{\multirow{2}{*}{Ave.}} & \multirow{2}{*}{RTF} \\ \cline{8-11}
\multicolumn{1}{c|}{} & \multicolumn{1}{c|}{} & \multicolumn{1}{c|}{} & \multicolumn{1}{c|}{} & \multicolumn{1}{c|}{} & \multicolumn{1}{c|}{} & \multicolumn{1}{c|}{} & Inv. & \multicolumn{1}{c|}{Par.} & Inv. & \multicolumn{1}{c|}{Par.} & \multicolumn{1}{c|}{} &  \\ \hline\hline
\multicolumn{13}{c}{Greedy Search (K$_\text{main}=1$)} \\ \hline\hline
\multicolumn{1}{c|}{1} & \multicolumn{1}{c|}{\multirow{6}{*}{1}}  & \multicolumn{1}{c|}{\multirow{6}{*}{FBank}} & \multicolumn{1}{c|}{\multirow{21}{*}{Conformer}} & \multicolumn{1}{c|}{CTC+AR} & \multicolumn{1}{c|}{0.3:0.7} & \multicolumn{1}{c|}{-} & 15.42 & \multicolumn{1}{c|}{24.83} & 15.71 & \multicolumn{1}{c|}{35.69} & \multicolumn{1}{c|}{25.14} & 0.112 \\ \cline{1-1} \cline{5-13} 
\multicolumn{1}{c|}{2} & \multicolumn{1}{c|}{}  & \multicolumn{1}{c|}{} & \multicolumn{1}{c|}{} & \multicolumn{1}{c|}{\multirow{5}{*}{\begin{tabular}[c]{@{}c@{}}CTC+AR\\ +AMD\end{tabular}}} & \multicolumn{1}{c|}{\multirow{5}{*}{\begin{tabular}[c]{@{}c@{}}0.3:0.6\\ :0.1\end{tabular}}} & \multicolumn{1}{c|}{1} & 15.87 & \multicolumn{1}{c|}{24.03} & 15.6 & \multicolumn{1}{c|}{34.61} & \multicolumn{1}{c|}{{\cellcolor{green!20}{24.55}$\ddagger$}} & 0.167 \\ \cline{1-1}
\multicolumn{1}{c|}{3} & \multicolumn{1}{c|}{} & \multicolumn{1}{c|}{} & \multicolumn{1}{c|}{} & \multicolumn{1}{c|}{} & \multicolumn{1}{c|}{} & \multicolumn{1}{c|}{2} & 15.87 & \multicolumn{1}{c|}{24.06} & 15.5 & \multicolumn{1}{c|}{35.03} & \multicolumn{1}{c|}{{\cellcolor{green!20}{24.68}$\ddagger$}} & \cellcolor{blue!20}{0.108$\diamond$} \\ \cline{1-1}
\multicolumn{1}{c|}{4} & \multicolumn{1}{c|}{} & \multicolumn{1}{c|}{} & \multicolumn{1}{c|}{} & \multicolumn{1}{c|}{} & \multicolumn{1}{c|}{} & \multicolumn{1}{c|}{4} & 15.87 & \multicolumn{1}{c|}{25.42} & 15.91 & \multicolumn{1}{c|}{35.97} & \multicolumn{1}{c|}{{\cellcolor{yellow!20}{25.44}$\dagger$}} & 0.081 \\ \cline{1-1}
\multicolumn{1}{c|}{5} & \multicolumn{1}{c|}{} & \multicolumn{1}{c|}{} & \multicolumn{1}{c|}{} & \multicolumn{1}{c|}{} & \multicolumn{1}{c|}{} & \multicolumn{1}{c|}{8} & 15.98 & \multicolumn{1}{c|}{26.76} & 16.57 & \multicolumn{1}{c|}{37.61} & \multicolumn{1}{c|}{26.59} & 0.069 \\ \cline{1-1}
\multicolumn{1}{c|}{6} & \multicolumn{1}{c|}{} & \multicolumn{1}{c|}{} & \multicolumn{1}{c|}{} & \multicolumn{1}{c|}{} & \multicolumn{1}{c|}{} & \multicolumn{1}{c|}{16} & 15.42 & \multicolumn{1}{c|}{27.81} & 17.21 & \multicolumn{1}{c|}{37.81} & \multicolumn{1}{c|}{27.09} & 0.066 \\ \cline{1-3} \cline{5-13} 
\multicolumn{1}{c|}{7} & \multicolumn{1}{c|}{\multirow{6}{*}{2}}  & \multicolumn{1}{c|}{\multirow{14}{*}{WavLM}} & \multicolumn{1}{c|}{} & \multicolumn{1}{c|}{CTC+AR} & \multicolumn{1}{c|}{0.3:0.7} & \multicolumn{1}{c|}{-} & 14.76 & \multicolumn{1}{c|}{21.70} & 13.78 & \multicolumn{1}{c|}{29.76} & \multicolumn{1}{c|}{21.46} & 0.133 \\ \cline{1-1} \cline{5-13} 
\multicolumn{1}{c|}{8} & \multicolumn{1}{c|}{} & \multicolumn{1}{c|}{} & \multicolumn{1}{c|}{} & \multicolumn{1}{c|}{\multirow{6}{*}{\begin{tabular}[c]{@{}c@{}}CTC+AR\\ +AMD\end{tabular}}} & \multicolumn{1}{c|}{\multirow{6}{*}{\begin{tabular}[c]{@{}c@{}}0.4:0.5\\ :0.1\end{tabular}}} & \multicolumn{1}{c|}{1} & 14.20 & \multicolumn{1}{c|}{19.77} & 13.31 & \multicolumn{1}{c|}{28.05} & \multicolumn{1}{c|}{{\cellcolor{green!20}{20.26}$\ddagger$}} & 0.227 \\ \cline{1-1}
\multicolumn{1}{c|}{9} & \multicolumn{1}{c|}{} & \multicolumn{1}{c|}{} & \multicolumn{1}{c|}{} & \multicolumn{1}{c|}{} & \multicolumn{1}{c|}{} & \multicolumn{1}{c|}{2} & 15.20 & \multicolumn{1}{c|}{21.39} & 13.76 & \multicolumn{1}{c|}{29.49} & \multicolumn{1}{c|}{{\cellcolor{yellow!20}{21.31}$\dagger$}} & 0.128 \\ \cline{1-1}
\multicolumn{1}{c|}{10} & \multicolumn{1}{c|}{} & \multicolumn{1}{c|}{} & \multicolumn{1}{c|}{} & \multicolumn{1}{c|}{} & \multicolumn{1}{c|}{} & \multicolumn{1}{c|}{4} & 14.98 & \multicolumn{1}{c|}{21.52} & 14.18 & \multicolumn{1}{c|}{29.92} & \multicolumn{1}{c|}{{\cellcolor{yellow!20}{21.67}$\dagger$}} & 0.093 \\ \cline{1-1}
\multicolumn{1}{c|}{11} & \multicolumn{1}{c|}{} & \multicolumn{1}{c|}{} & \multicolumn{1}{c|}{} & \multicolumn{1}{c|}{} & \multicolumn{1}{c|}{} & \multicolumn{1}{c|}{8} & 14.65 & \multicolumn{1}{c|}{21.62} & 14.27 & \multicolumn{1}{c|}{30.04} & \multicolumn{1}{c|}{{\cellcolor{yellow!20}{21.76}$\dagger$}} & 0.081 \\ \cline{1-1}
\multicolumn{1}{c|}{12} & \multicolumn{1}{c|}{} & \multicolumn{1}{c|}{} & \multicolumn{1}{c|}{} & \multicolumn{1}{c|}{} & \multicolumn{1}{c|}{} & \multicolumn{1}{c|}{16} & 14.87 & \multicolumn{1}{c|}{21.89} & 14.63 & \multicolumn{1}{c|}{30.26} & \multicolumn{1}{c|}{22.04} & 0.077 \\ \cline{1-1} \cline{8-13} 
\multicolumn{1}{c|}{13} & \multicolumn{1}{c|}{} & \multicolumn{1}{c|}{} & \multicolumn{1}{c|}{} & \multicolumn{1}{c|}{} & \multicolumn{1}{c|}{} & \multicolumn{1}{c|}{1-15-8} & 14.53 & \multicolumn{1}{c|}{20.95} & 13.64 & \multicolumn{1}{c|}{29.26} & \multicolumn{1}{c|}{{\cellcolor{green!20}{21.08}$\ddagger$}} & \cellcolor{blue!20}{0.137$\diamond$} \\ \cline{1-2} \cline{5-13} 
\multicolumn{1}{c|}{14} & \multicolumn{1}{c|}{\multirow{6}{*}{3}}  & \multicolumn{1}{c|}{} & \multicolumn{1}{c|}{} & \multicolumn{1}{c|}{\begin{tabular}[c]{@{}c@{}}CTC+\\ AR$_\text{LLM}$\end{tabular}} & \multicolumn{1}{c|}{0.3:0.7} & \multicolumn{1}{c|}{-} & 14.53 & \multicolumn{1}{c|}{19.14} & 13.82 & \multicolumn{1}{c|}{29.00} & \multicolumn{1}{c|}{20.75} & 0.203 \\ \cline{1-1} \cline{5-13} 
\multicolumn{1}{c|}{15} & \multicolumn{1}{c|}{} & \multicolumn{1}{c|}{} & \multicolumn{1}{c|}{} & \multicolumn{1}{c|}{\multirow{6}{*}{\begin{tabular}[c]{@{}c@{}}CTC+\\ AR$_\text{LLM}$+\\ AMD$_\text{LLM}$\end{tabular}}} & \multicolumn{1}{c|}{\multirow{6}{*}{\begin{tabular}[c]{@{}c@{}}0.3:0.5\\ :0.2\end{tabular}}} & \multicolumn{1}{c|}{1} & 14.20 & \multicolumn{1}{c|}{18.49} & 13.4 & \multicolumn{1}{c|}{28.01} & \multicolumn{1}{c|}{{\cellcolor{green!20}{20.06}$\ddagger$}} & 0.732 \\ \cline{1-1}
\multicolumn{1}{c|}{16} & \multicolumn{1}{c|}{} & \multicolumn{1}{c|}{} & \multicolumn{1}{c|}{} & \multicolumn{1}{c|}{} & \multicolumn{1}{c|}{} & \multicolumn{1}{c|}{2} & 14.65 & \multicolumn{1}{c|}{18.81} & 13.61 & \multicolumn{1}{c|}{28.47} & \multicolumn{1}{c|}{{\cellcolor{green!20}{20.40}$\ddagger$}} & 0.306 \\ \cline{1-1}
\multicolumn{1}{c|}{17} & \multicolumn{1}{c|}{} & \multicolumn{1}{c|}{} & \multicolumn{1}{c|}{} & \multicolumn{1}{c|}{} & \multicolumn{1}{c|}{} & \multicolumn{1}{c|}{4} & 14.09 & \multicolumn{1}{c|}{19.02} & 13.66 & \multicolumn{1}{c|}{28.32} & \multicolumn{1}{c|}{{\cellcolor{green!20}{20.38}$\ddagger$}} & 0.181 \\ \cline{1-1}
\multicolumn{1}{c|}{18} & \multicolumn{1}{c|}{} & \multicolumn{1}{c|}{} & \multicolumn{1}{c|}{} & \multicolumn{1}{c|}{} & \multicolumn{1}{c|}{} & \multicolumn{1}{c|}{8} & 14.87 & \multicolumn{1}{c|}{19.42} & 14.13 & \multicolumn{1}{c|}{29.00} & \multicolumn{1}{c|}{{\cellcolor{yellow!20}{20.93}$\dagger$}} & 0.126 \\ \cline{1-1}
\multicolumn{1}{c|}{19} & \multicolumn{1}{c|}{} & \multicolumn{1}{c|}{} & \multicolumn{1}{c|}{} & \multicolumn{1}{c|}{} & \multicolumn{1}{c|}{} & \multicolumn{1}{c|}{16} & 18.64 & \multicolumn{1}{c|}{20.82} & 17.45 & \multicolumn{1}{c|}{31.39} & \multicolumn{1}{c|}{23.57} & 0.111 \\ \cline{1-1}
\multicolumn{1}{c|}{20} & \multicolumn{1}{c|}{} & \multicolumn{1}{c|}{} & \multicolumn{1}{c|}{} & \multicolumn{1}{c|}{} & \multicolumn{1}{c|}{} & \multicolumn{1}{c|}{1-2-4} & 14.42 & \multicolumn{1}{c|}{18.85} & 13.68 & \multicolumn{1}{c|}{28.25} & \multicolumn{1}{c|}{{\cellcolor{green!20}{20.34}$\ddagger$}} & \cellcolor{blue!20}{0.214$\diamond$} \\ \hline\hline
\multicolumn{13}{c}{Beam Search (K$_\text{main}=60$)} \\ \hline\hline
\multicolumn{1}{c|}{21} & \multicolumn{1}{c|}{\multirow{6}{*}{1}} & \multicolumn{1}{c|}{\multirow{6}{*}{FBank}} & \multicolumn{1}{c|}{\multirow{12}{*}{Conformer}} & \multicolumn{1}{c|}{CTC+AR} & \multicolumn{1}{c|}{0.3:0.7} & \multicolumn{1}{c|}{-} & 13.65 & \multicolumn{1}{c|}{22.96} & 15.15 & \multicolumn{1}{c|}{33.50} & \multicolumn{1}{c|}{23.68} & 0.486 \\ \cline{1-1} \cline{5-13} 
\multicolumn{1}{c|}{22} & \multicolumn{1}{c|}{} & \multicolumn{1}{c|}{} & \multicolumn{1}{c|}{} & \multicolumn{1}{c|}{\multirow{5}{*}{\begin{tabular}[c]{@{}c@{}}CTC+AR\\ +AMD\end{tabular}}} & \multicolumn{1}{c|}{\multirow{5}{*}{\begin{tabular}[c]{@{}c@{}}0.3:0.6\\ :0.1\end{tabular}}} & \multicolumn{1}{c|}{1} & 14.87 & \multicolumn{1}{c|}{22.48} & 14.87 & \multicolumn{1}{c|}{32.98} & \multicolumn{1}{c|}{{\cellcolor{yellow!20}{23.32}$\dagger$}} & 0.896 \\ \cline{1-1}
\multicolumn{1}{c|}{23} & \multicolumn{1}{c|}{} & \multicolumn{1}{c|}{} & \multicolumn{1}{c|}{} & \multicolumn{1}{c|}{} & \multicolumn{1}{c|}{} & \multicolumn{1}{c|}{2} & 15.76 & \multicolumn{1}{c|}{22.73} & 14.72 & \multicolumn{1}{c|}{33.18} & \multicolumn{1}{c|}{{\cellcolor{yellow!20}{23.40}$\dagger$}} & \cellcolor{blue!20}{0.492$\diamond$} \\ \cline{1-1}
\multicolumn{1}{c|}{24} & \multicolumn{1}{c|}{} & \multicolumn{1}{c|}{} & \multicolumn{1}{c|}{} & \multicolumn{1}{c|}{} & \multicolumn{1}{c|}{} & \multicolumn{1}{c|}{4} & 15.31 & \multicolumn{1}{c|}{23.47} & 14.93 & \multicolumn{1}{c|}{33.54} & \multicolumn{1}{c|}{{\cellcolor{yellow!20}{23.74}$\dagger$}} & 0.461 \\ \cline{1-1}
\multicolumn{1}{c|}{25} & \multicolumn{1}{c|}{} & \multicolumn{1}{c|}{} & \multicolumn{1}{c|}{} & \multicolumn{1}{c|}{} & \multicolumn{1}{c|}{} & \multicolumn{1}{c|}{8} & 15.42 & \multicolumn{1}{c|}{24.01} & 15.19 & \multicolumn{1}{c|}{33.91} & \multicolumn{1}{c|}{24.09} & 0.379 \\ \cline{1-1}
\multicolumn{1}{c|}{26} & \multicolumn{1}{c|}{} & \multicolumn{1}{c|}{} & \multicolumn{1}{c|}{} & \multicolumn{1}{c|}{} & \multicolumn{1}{c|}{} & \multicolumn{1}{c|}{16} & 15.42 & \multicolumn{1}{c|}{24.52} & 15.39 & \multicolumn{1}{c|}{34.25} & \multicolumn{1}{c|}{24.39} & 0.383 \\ \cline{1-3} \cline{5-13} 
\multicolumn{1}{c|}{27} & \multicolumn{1}{c|}{\multirow{5}{*}{2}} & \multicolumn{1}{c|}{\multirow{6}{*}{WavLM}} & \multicolumn{1}{c|}{} & \multicolumn{1}{c|}{CTC+AR} & \multicolumn{1}{c|}{0.3:0.7} & \multicolumn{1}{c|}{-} & 13.98 & \multicolumn{1}{c|}{19.08} & 13.30 & \multicolumn{1}{c|}{27.80} & \multicolumn{1}{c|}{20.03} & 0.528 \\ \cline{1-1} \cline{5-13} 
\multicolumn{1}{c|}{28} &\multicolumn{1}{c|}{} & \multicolumn{1}{c|}{} & \multicolumn{1}{c|}{} & \multicolumn{1}{c|}{\multirow{5}{*}{\begin{tabular}[c]{@{}c@{}}CTC+AR\\ +AMD\end{tabular}}} & \multicolumn{1}{c|}{\multirow{5}{*}{\begin{tabular}[c]{@{}c@{}}0.3:0.6\\ :0.1\end{tabular}}} & \multicolumn{1}{c|}{1} & 13.87 & \multicolumn{1}{c|}{18.64} & 13.10 & \multicolumn{1}{c|}{27.47} & \multicolumn{1}{c|}{{\cellcolor{yellow!20}{19.75}$\dagger$}} & 0.834 \\ \cline{1-1}
\multicolumn{1}{c|}{29} & \multicolumn{1}{c|}{} & \multicolumn{1}{c|}{} & \multicolumn{1}{c|}{} & \multicolumn{1}{c|}{} & \multicolumn{1}{c|}{} & \multicolumn{1}{c|}{2} & 13.98 & \multicolumn{1}{c|}{19.12} & 13.23 & \multicolumn{1}{c|}{27.94} & \multicolumn{1}{c|}{{\cellcolor{yellow!20}{20.07}$\dagger$}} & \cellcolor{blue!20}{0.547$\diamond$} \\ \cline{1-1}
\multicolumn{1}{c|}{30} & \multicolumn{1}{c|}{} & \multicolumn{1}{c|}{} & \multicolumn{1}{c|}{} & \multicolumn{1}{c|}{} & \multicolumn{1}{c|}{} & \multicolumn{1}{c|}{4} & 13.76 & \multicolumn{1}{c|}{19.48} & 13.36 & \multicolumn{1}{c|}{28.07} & \multicolumn{1}{c|}{{\cellcolor{yellow!20}{20.23}$\dagger$}} & 0.395 \\ \cline{1-1}
\multicolumn{1}{c|}{31} & \multicolumn{1}{c|}{} & \multicolumn{1}{c|}{} & \multicolumn{1}{c|}{} & \multicolumn{1}{c|}{} & \multicolumn{1}{c|}{} & \multicolumn{1}{c|}{8} & 13.65 & \multicolumn{1}{c|}{19.88} & 13.49 & \multicolumn{1}{c|}{28.17} & \multicolumn{1}{c|}{20.38} & 0.338 \\ \cline{1-1}
\multicolumn{1}{c|}{32} & \multicolumn{1}{c|}{} & \multicolumn{1}{c|}{} & \multicolumn{1}{c|}{} & \multicolumn{1}{c|}{} & \multicolumn{1}{c|}{} & \multicolumn{1}{c|}{16} & 13.87 & \multicolumn{1}{c|}{19.96} & 13.67 & \multicolumn{1}{c|}{28.30} & \multicolumn{1}{c|}{20.52} & 0.339 \\ \hline\hline
\multicolumn{13}{c}{Beam Search (K$_\text{main}=4$)} \\ \hline\hline
\multicolumn{1}{c|}{33} & \multicolumn{1}{c|}{\multirow{6}{*}{3}}  & \multicolumn{1}{c|}{\multirow{6}{*}{WavLM}} & \multicolumn{1}{c|}{\multirow{6}{*}{Conformer}} & \multicolumn{1}{c|}{\begin{tabular}[c]{@{}c@{}}CTC+\\ AR$_\text{LLM}$\end{tabular}} & \multicolumn{1}{c|}{0.3:0.7} & \multicolumn{1}{c|}{-} & 13.87 & \multicolumn{1}{c|}{18.34} & 13.37 & \multicolumn{1}{c|}{27.55} & \multicolumn{1}{c|}{19.84} & 0.649 \\ \cline{1-1} \cline{5-13} 
\multicolumn{1}{c|}{34} & \multicolumn{1}{c|}{}& \multicolumn{1}{c|}{} & \multicolumn{1}{c|}{} & \multicolumn{1}{c|}{\multirow{5}{*}{\begin{tabular}[c]{@{}c@{}}CTC+\\ AR$_\text{LLM}$+\\ AMD$_\text{LLM}$\end{tabular}}} & \multicolumn{1}{c|}{\multirow{5}{*}{\begin{tabular}[c]{@{}c@{}}0.4:0.4\\ :0.2\end{tabular}}} & \multicolumn{1}{c|}{1} & 13.87 & \multicolumn{1}{c|}{18.16} & 13.23 & \multicolumn{1}{c|}{27.62} & \multicolumn{1}{c|}{{\cellcolor{yellow!20}{19.77}$\dagger$}} & 1.385 \\ \cline{1-1}
\multicolumn{1}{c|}{35} & \multicolumn{1}{c|}{} & \multicolumn{1}{c|}{} & \multicolumn{1}{c|}{} & \multicolumn{1}{c|}{} & \multicolumn{1}{c|}{} & \multicolumn{1}{c|}{2} & 14.09 & \multicolumn{1}{c|}{18.37} & 13.44 & \multicolumn{1}{c|}{27.88} & \multicolumn{1}{c|}{{\cellcolor{yellow!20}{20.00}$\dagger$}} & \cellcolor{blue!20}{0.696$\diamond$} \\ \cline{1-1}
\multicolumn{1}{c|}{36} & \multicolumn{1}{c|}{} & \multicolumn{1}{c|}{} & \multicolumn{1}{c|}{} & \multicolumn{1}{c|}{} & \multicolumn{1}{c|}{} & \multicolumn{1}{c|}{4} & 14.09 & \multicolumn{1}{c|}{18.53} & 13.65 & \multicolumn{1}{c|}{27.91} & \multicolumn{1}{c|}{{\cellcolor{yellow!20}{20.13}$\dagger$}} & 0.421 \\ \cline{1-1}
\multicolumn{1}{c|}{37} & \multicolumn{1}{c|}{} & \multicolumn{1}{c|}{} & \multicolumn{1}{c|}{} & \multicolumn{1}{c|}{} & \multicolumn{1}{c|}{} & \multicolumn{1}{c|}{8} & 14.31 & \multicolumn{1}{c|}{18.81} & 13.67 & \multicolumn{1}{c|}{28.11} & \multicolumn{1}{c|}{20.27} & 0.331 \\ \cline{1-1}
\multicolumn{1}{c|}{38} & \multicolumn{1}{c|}{} & \multicolumn{1}{c|}{} & \multicolumn{1}{c|}{} & \multicolumn{1}{c|}{} & \multicolumn{1}{c|}{} & \multicolumn{1}{c|}{16} & 18.53 & \multicolumn{1}{c|}{19.94} & 17.10 & \multicolumn{1}{c|}{30.51} & \multicolumn{1}{c|}{22.93} & 0.332 \\ \hline
\end{tabular}}
\end{table}
\label{sec:dbank_mainresult}
\subsubsection{Performance Analysis of \textbf{greedy search}} Several trends can be observed when performing greedy search (Sys. 1 - 20, Table \ref{tab:dbank_all}) under {\it {Config. 1-3}}:
\begin{enumerate}[leftmargin=8mm,labelsep=1mm]
    \item[\textbf{a)}] The baseline systems with CTC + AR decoder demonstrate progressive performance improvements across {\it {Config. 1-3}}. The integration of WavLM features shows WER reduction of 3.68\% absolute (14.6\% relative) (Sys. 7 {\it vs.} 1, Table \ref{tab:dbank_all}) compared to the baseline system under {\it {Config. 1}}, and the integration of AR$_{\rm LLM}$ decoder yields a further WER reduction of 0.71\% absolute (3.3\% relative) (Sys. 14 {\it vs.} 7, Table \ref{tab:dbank_all});
    \item[{\bf b)}] When performing purely serial, non-parallel inference (B$_\text{DEC}=1$), the tripartite decoder integrating AMD consistently outperforms the baselines across {\it {Config.1-3}}. The tripartite decoder achieves statistically significant average WER reductions of 0.75\%, 1.20\% and 0.29\% absolute (3.0\%, 5.6\% and 1.4\% relative) under {\it {Config. 1-3}} respectively (Sys. 2 {\it vs.} 1, Sys. 8 {\it vs.} 7, Sys. 15 {\it vs.} 14, Table \ref{tab:dbank_all}), while incurs RTF increases of 1.42x, 1.71x, and 1.33x due to the computational overhead of the AMD decoder.
    \item[\textbf{c)}] As the B$_{\text{DEC}}$ increases from 1 to 16, the proposed tripartite decoder exhibits a clear \textbf{trade-off between WER and RTF}, showing an increase in average WER from 24.55\% to 27.09\% alongside a decrease in RTF from 0.167 to 0.066 under {\it {Config. 1}}. The same patterns are observed under {\it {Config. 2}} (Sys. 12 - 8, Table \ref{tab:dbank_all}) and {\it {Config. 3}} (Sys. 19 - 15, Table \ref{tab:dbank_all}).
    \item[{\bf e)}] The tripartite decoder achieves \textbf{consistent speedup} over the corresponding CTC + AR baselines without statistically significant WER increase. The tripartite decoder achieves speedup ratios of up to \textbf{1.38x}, \textbf{1.64x} and \textbf{1.61x} with B$_{\rm DEC}$ of 4, 8, and 8 under {\it {Config. 1-3}} respectively (Sys. 4 {\it vs.} 1, Sys. 11 {\it vs.} 7, Sys. 18 {\it vs.} 14, Table \ref{tab:dbank_all});
    \item[{\bf f)}] When operating at comparable RTFs to CTC + AR baseline systems, the tripartite decoder \textbf{consistently yields statistically significant WER reductions} over the baselines. Absolute averaged WER reductions of \textbf{0.46\%}, \textbf{0.38\%} and \textbf{0.41\%} (1.8\%, 1.8\% and 2.0\% relative) are achieved under {\it {Config. 1-3}} respectively, using B$_{\rm DEC}$ values of 2, 1-15-8 and 1-2-4 (Sys. 3 {\it vs.} 1, Sys. 13 {\it vs.} 7, Sys. 20 {\it vs.} 14, Table \ref{tab:dbank_all});
\end{enumerate}

\subsubsection{Performance Analysis of \textbf{beam search}} With beam search (Sys. 21 - 38, Table \ref{tab:dbank_all}), the following trends can be observed under {\it {Config. 1-3}}:
\begin{enumerate}[leftmargin=8mm,labelsep=1mm]
    \item[{\bf a)}] When using B$_{\text{DEC}} = 2$  under {\it {Config. 1-3}}, the tripartite decoder achieves similar RTFs to the baselines while maintaining comparable WER performance (Sys. 23 {\it vs.} 21, Sys. 29 {\it vs.} 27, Sys. 35 {\it vs.} 33, Table \ref{tab:dbank_all}). This disparity in performance gains over CTC + AR baselines between greedy and beam search when using tripartite decoders is consistent with observations on LS960 task.
    \item[{\bf b)}] With B$_\text{DEC}=4$, the proposed tripartite decoder accelerates decoding by up to 1.22x, 1.32x and 1.54x over the corresponding CTC + AR baselines under {\it {Config. 1-3}} respectively, without statistically significant WER increase (Sys. 24 {\it vs.} 21, Sys. 30 {\it vs.} 27, Sys. 36 {\it vs.} 33, Table \ref{tab:dbank_all}).
\end{enumerate}
\vspace{-4mm}
\section{Conclusion}
\label{sec:conclusion}
NAR approaches aim primarily to achieve significant decoding speedup while the maintaining recognition accuracy that is comparable to AR baselines. This paper proposed a novel NAR block-based attention mask decoder (AMD) that effectively improves decoding efficiency while maintaining ASR accuracy, and also offers flexibility in balancing performance-efficiency trade-offs for ASR systems. The proposed AMD decoder performs parallel inference within contiguous blocks of output labels while maintaining monotonic left-to-right prediction between blocks. A one-pass beam search algorithm was designed to dynamically combine CTC, AR, and AMD probabilities during inference, eliminating the need for multi-pass decoding frameworks commonly used in prior NAR approaches. Mixed-size attention-masking blocks were further developed to facilitate cold-start monotonic inference for initial tokens before switching to parallel label prediction for the remaining sequence. Experiments on the normal speech LS960 and DBank elderly speech corpus demonstrate the effectiveness of the proposed tripartite decoder across three model configurations: \textbf{a)} Conformer encoder-decoder ASR systems; \textbf{b)} their integration with WavLM features; and \textbf{c)} their further integration with an LLM-based decoder. When evaluated on the LS960 task, the AMD empowered tripartite decoder achieves decoding speedup ratios of up to \textbf{1.44x}, \textbf{1.55x}, and \textbf{2.31x} under the three model configurations respectively, without statistically significant WER increase. When operating with RTFs comparable to the CTC + AR baselines, the tripartite decoder yields statistically significant absolute averaged WER reductions of \textbf{0.19\%}, \textbf{0.62\%} and \textbf{0.13\%} (\textbf{4.3\%}, \textbf{16.3\%}, and \textbf{3.8\%} relative) across the three model configurations. Similar trends were observed on the DBank task.

While the proposed AMD is theoretically applicable to larger LLM backbones due to its model-agnostic design and the use of scalable components like LoRA, practical computational resource constraints limited the scope of our LLM experiments to a 1B parameter model. Nevertheless, these experiments provide valuable insights for resource-constrained scenarios and demonstrate the feasibility of efficiently integrating AMD within an LLM using parameter-sharing and LoRA fine-tuning. Future work could explore scaling to larger models as resources permit.

\section{Acknowledgements}
\label{ssec:a}
This research is supported by Hong Kong RGC GRF grant No. 14200021, 14200220, 14200324, TRS grant No. T45-407/19N, Innovation \& Technology Fund grant No. ITS/218/21, Research Project of Institute of Software, Chinese Academy of Sciences (ISCAS-ZD-202401, ISCAS-JCMS-202306), Youth Innovation Promotion Association CAS Grant (2023119)
\bibliographystyle{IEEEtran}
\bibliography{nar_ref}

\vspace{-1.0cm}
\begin{IEEEbiography}
[{\includegraphics[width=1in,height=1.25in,clip,keepaspectratio]{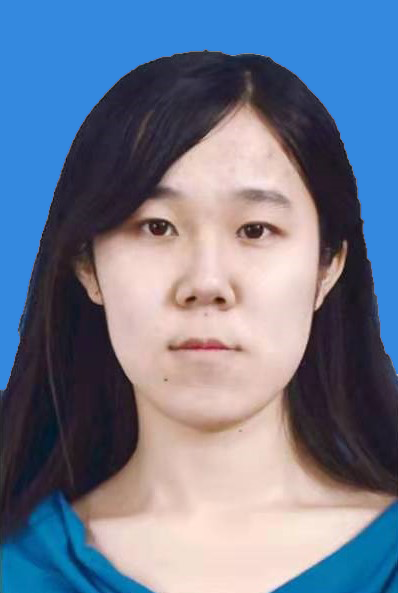}}]{Tianzi Wang}
 recieved the M.S. degree in electrical and computer engineering from Johns Hopkins University, Baltimore, MD, USA in 2020. She is currently working toward the Ph.D. degree with The Chinese University of Hong Kong, Hong Kong. Her research interests include architectural adaptation and disordered speech recognition.
\end{IEEEbiography}

\vspace{-1.0cm}
\begin{IEEEbiography}
[{\includegraphics[width=1in,height=1.25in,clip,keepaspectratio]{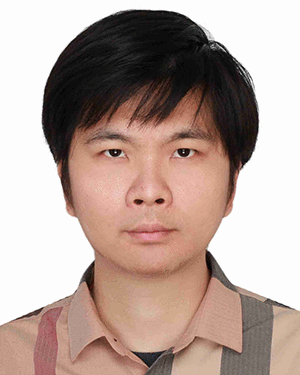}}]{Xurong Xie} received the B.S. degree in mathematics from Sun Yat-sen University, Guangzhou, China, the M.S. degree in machine learning from University College London, London, U.K., and the Ph.D. degree in electronic engineering from The Chinese University of Hong Kong, Hong Kong. He is currently an Associate Professor with the Institute of Software, Chinese Academy of Sciences, Beijing, China. His research interests include disorder speech processing, adaptation techniques on speech and language modeling, human-computer interaction, and computational neuroscience. He was the recipient of the Best Student Paper Award at IEEE ICASSP 2019.
\end{IEEEbiography}

\vspace{-1.0cm}
\begin{IEEEbiography}
[{\includegraphics[width=1in,height=1.25in,clip,keepaspectratio]{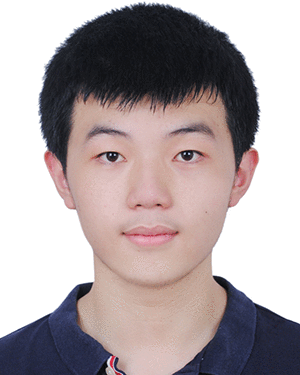}}]{Zengrui Jin} received the Ph.D. degree from the Department of Systems Engineering and Engineering Management, The Chinese University of Hong Kong, under the supervision of Prof. Xunying Liu, following his undergraduate studies with the Dalian University of Technology. His Ph.D. thesis focuses on adversarial learning and reinforcement learning based data augmentation for pathological speech recognition systems. He is currently a Postdoctoral Researcher with Tsinghua University.
\end{IEEEbiography}

\vspace{-1.0cm}
\begin{IEEEbiography}
[{\includegraphics[width=1in,height=1.25in,clip,keepaspectratio]{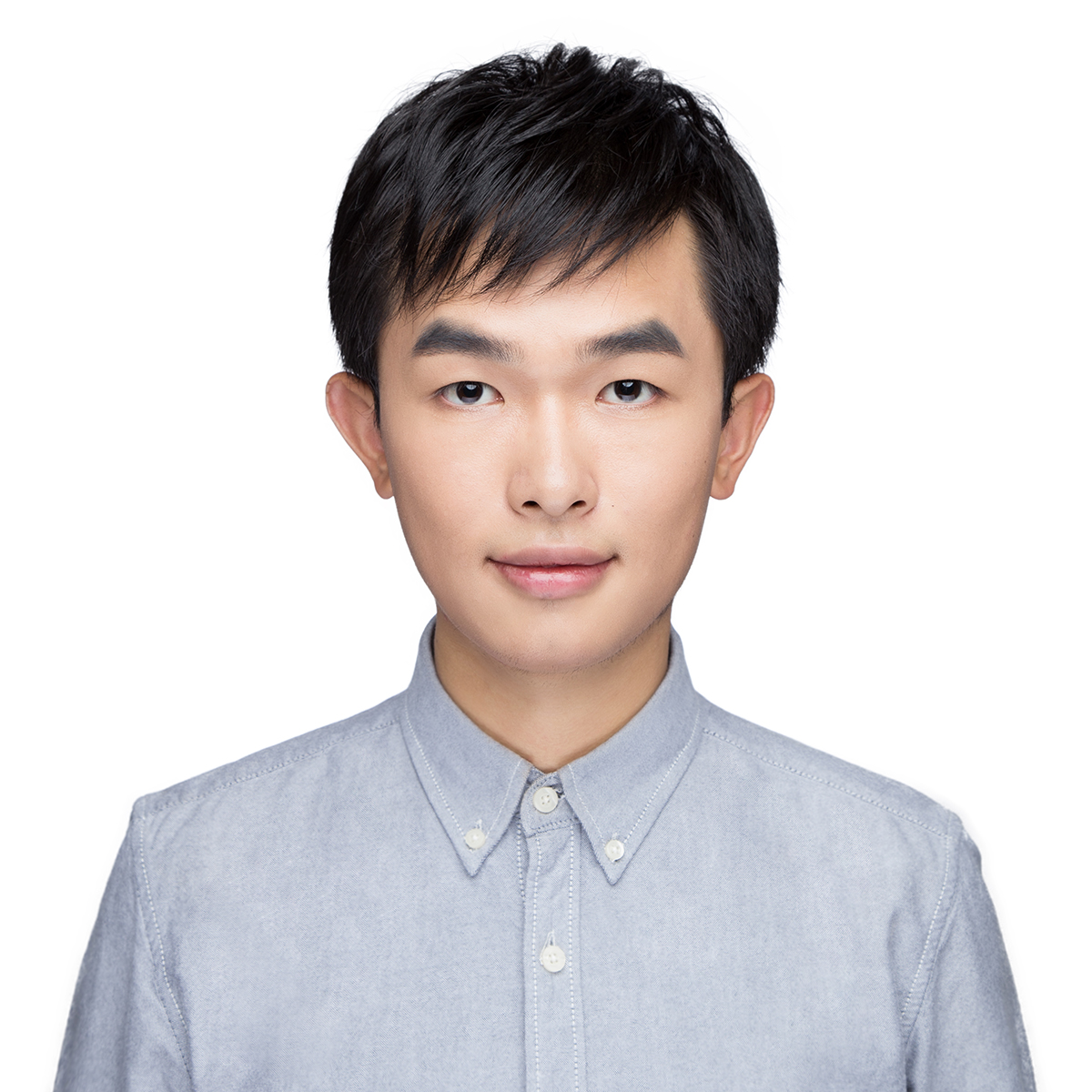}}]{Mengzhe Geng} received the B.Sc. degree in mathematics \& information engineering and the Ph.D. degree in systems engineering \& engineering management from The Chinese University of Hong Kong, Hong Kong, in 2019 and 2023, respectively. He is currently an assistant research officer at the National Research Council Canada. His research focuses on developing inclusive speech technologies for healthcare and low-resource languages, as well as advancing equitable AI solutions.
\end{IEEEbiography}

\vspace{-1.0cm}
\begin{IEEEbiography}
[{\includegraphics[width=1in,height=1.25in,clip,keepaspectratio]{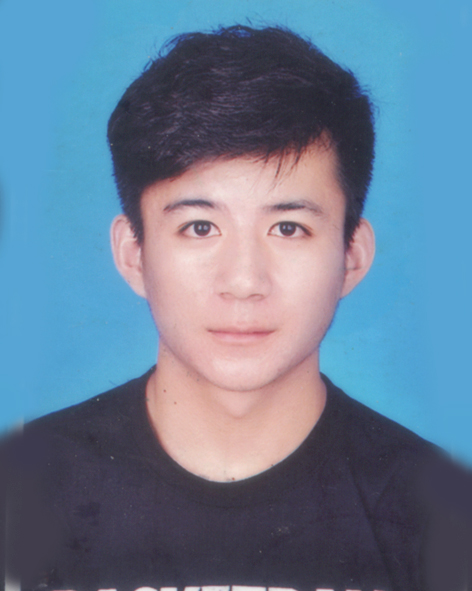}}]{Jiajun Deng}
 received the Bachelor's degree from Central China Normal University, Wuhan, China, in 2017 and the Master's degree from University of Science and Technology of China, Hefei, China, in 2020. He is currently working toward the Ph.D. degree with The Chinese University of Hong Kong, Hong Kong. His current research interests include speaker and domain adaptation in speech recognition.
\end{IEEEbiography}

\vspace{-1.0cm}
\begin{IEEEbiography}
[{\includegraphics[width=1in,height=1.25in,clip,keepaspectratio]{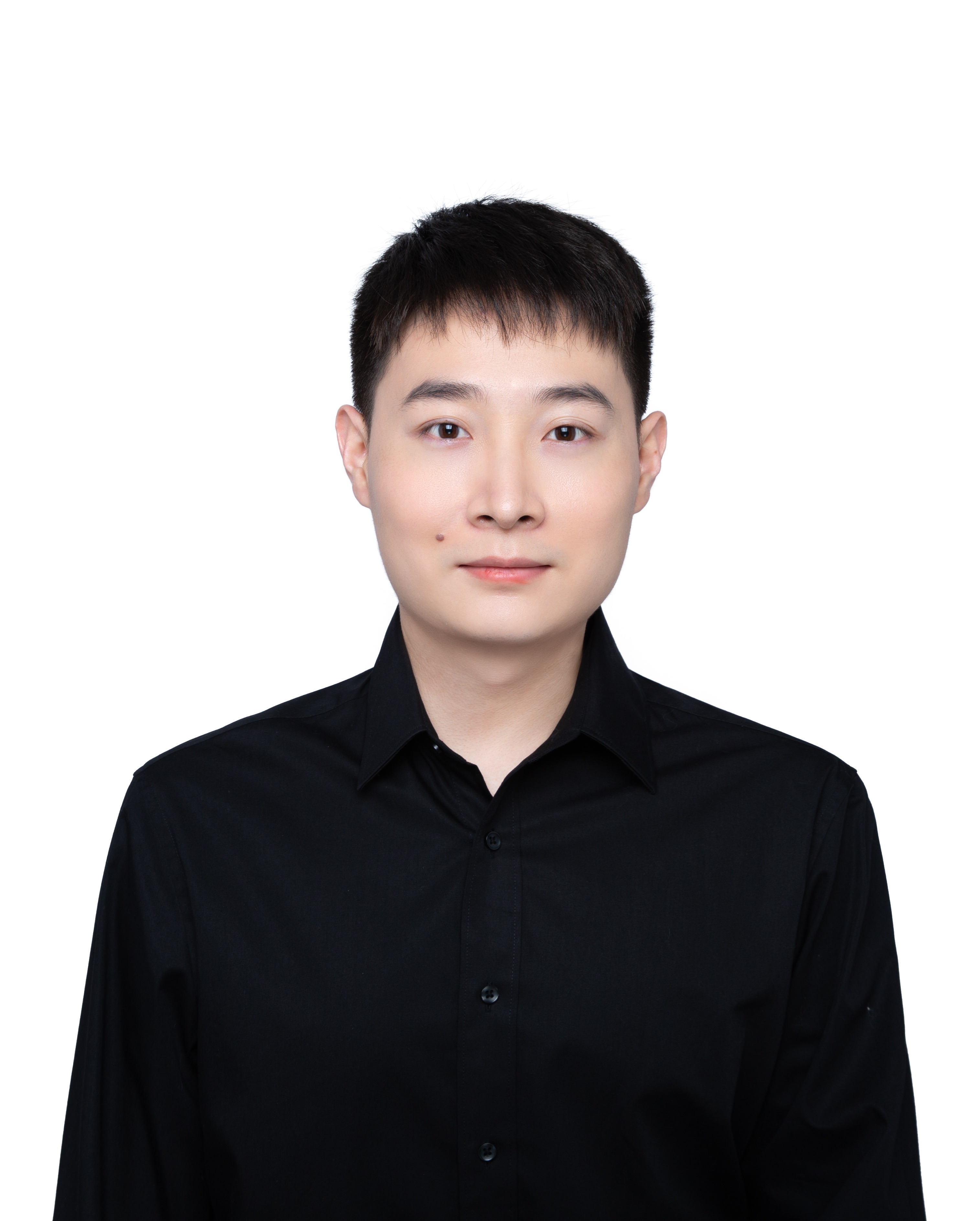}}]{Zhaoping Li} 
received his B.S. and M.S. degrees in Electrical Engineering from Northwestern Polytechnical University, Xi'an, China. He is currently a Ph.D. candidate at the Chinese University of Hong Kong, where his research focuses on machine learning and speech and language processing. His work specializes in developing lightweight and efficient deep learning systems.
\end{IEEEbiography}

\vspace{-1.0cm}
\begin{IEEEbiography}
[{\includegraphics[width=1in,height=1.25in,clip,keepaspectratio]{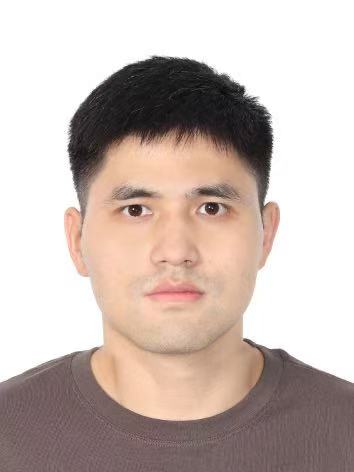}}]{Shoukang Hu} 
received the B.E. degree in Mechanical and Electrical Engineering from University of Electronic Science and Technology of China, Chengdu, China, in 2017. and the Ph.D. degree in system engineering and engineering management from The Chinese University of Hong Kong, Hong Kong, China. He was a Research Fellow in Nanyang Technological University, Singapore, and a research scientist at Sony AI, Japan. He is currently a senior researcher at Microsoft Research Asia. His current research interests include Automatic Speech Recognition, 3D Vision and Automated Deep Learning.
\end{IEEEbiography}

\vspace{-1.0cm}
\begin{IEEEbiography}
[{\includegraphics[width=1in,height=1.25in,clip,keepaspectratio]{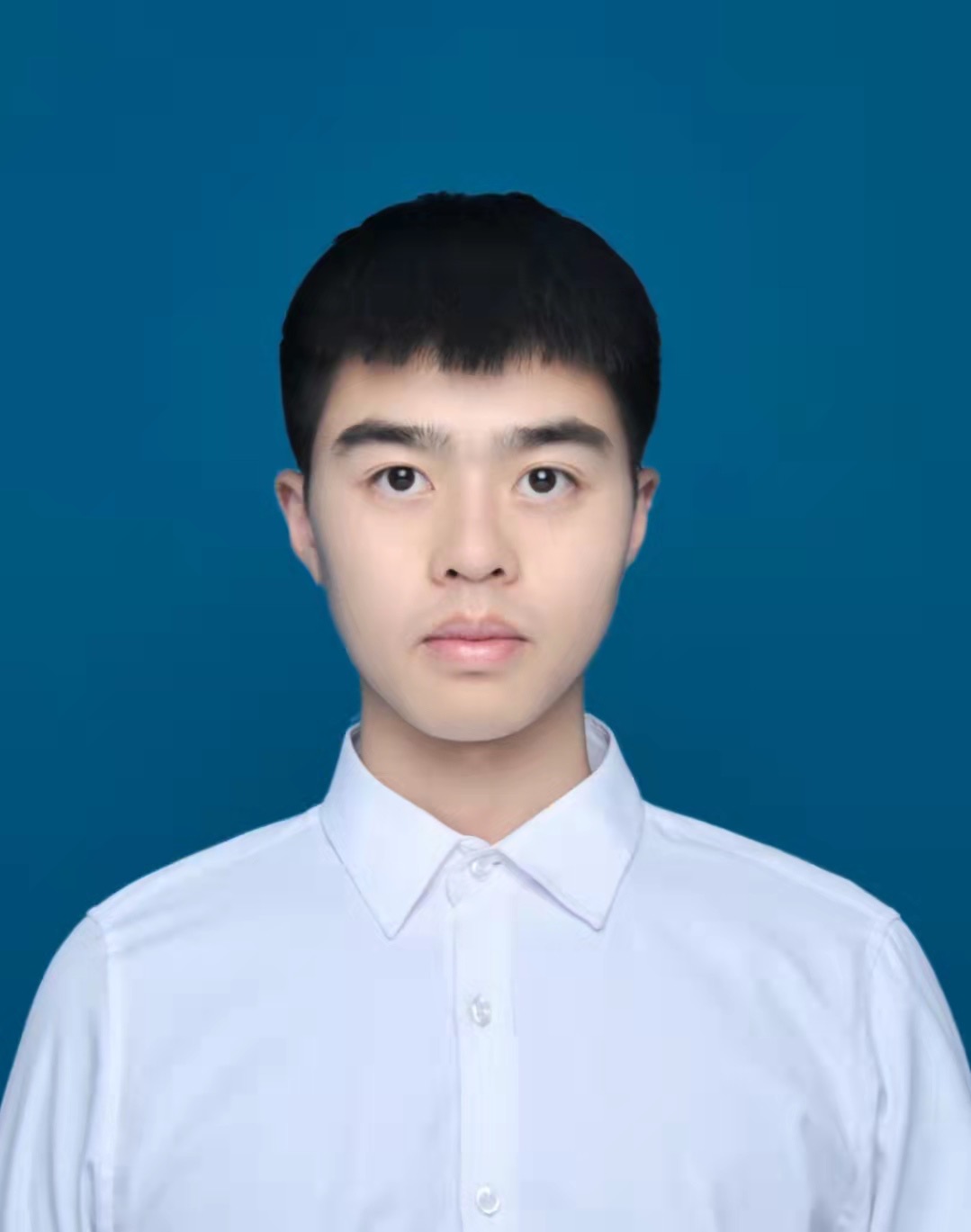}}]{Shujie Hu}
received the B.E. degree in computer science and technology from Sichuan University, Chengdu, China, in 2021. He is currently working toward the Ph.D. degree with The Chinese University of Hong Kong, Hong Kong. His current research interests include elderly and disordered speech recognition, and multimodal speech recognition.
\end{IEEEbiography}

\vspace{-1.0cm}
\begin{IEEEbiography}
[{\includegraphics[width=1in,height=1.25in,clip,keepaspectratio]{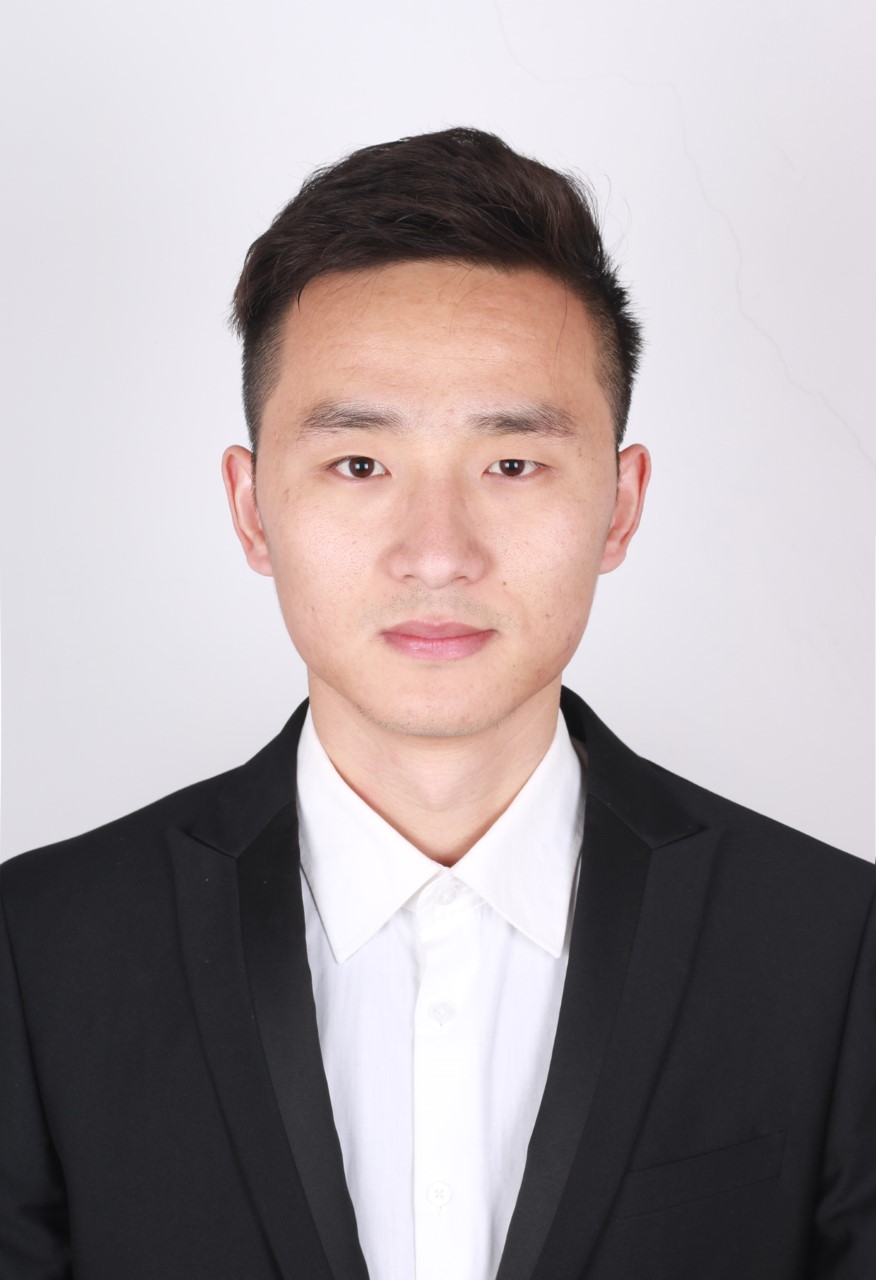}}]{Guinan Li}
received the B.E. degree in automation from Nanjing University of Aeronautics and Astronautics, Nanjing, China, in 2015 and the M.S. degree in control science and engineering from Zhejiang University, Hangzhou, China, in 2018. He is currently working toward the Ph.D. degree with The Chinese University of Hong Kong, Hong Kong. His current research interests include multi-modal speech separation, dereverberation and automatic speech recognition.
\end{IEEEbiography}

\vspace{-1.0cm}
\begin{IEEEbiography}
[{\includegraphics[width=1in,height=1.25in,clip,keepaspectratio]{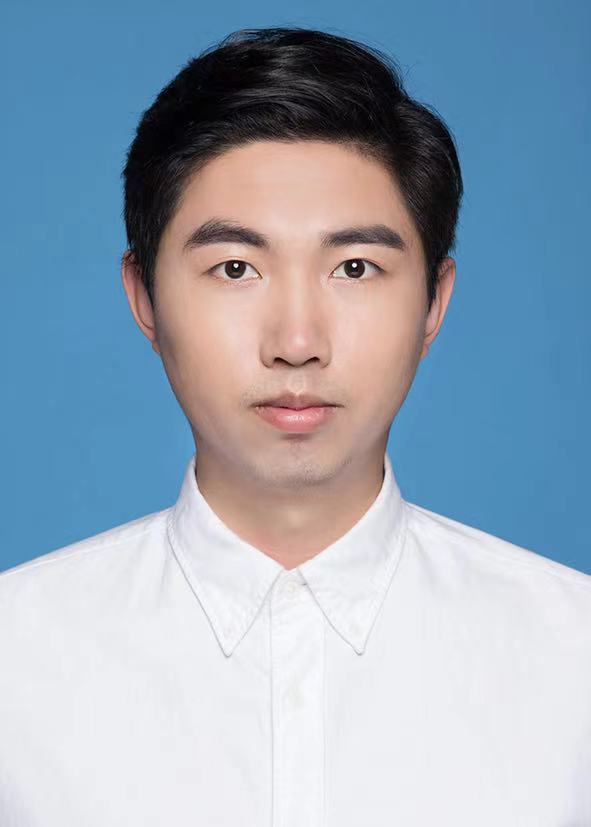}}]{Mingyu Cui}
reveived the B.E. degree in Computer Science and Software Engineering from SouthEast University in 2019. Currently, he is a Ph.D. student at the Chinese University of Hong Kong. His current research interests include long context ASR.
\end{IEEEbiography}

\vspace{-1.0cm}
\begin{IEEEbiography}
[{\includegraphics[width=1in,height=1.25in,clip,keepaspectratio]{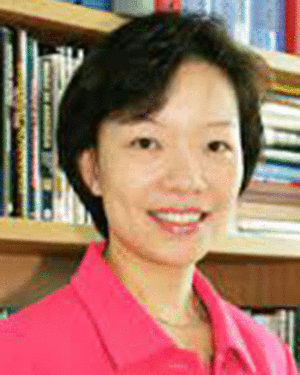}}]{Helen Meng}
(Fellow, IEEE) received the B.S., M.S., and Ph.D. degrees in electrical engineering from the Massachusetts Institute of Technology, Cambridge, MA, USA. In 1998, she joined the Chinese University of Hong Kong, Hong Kong, where she is currently the Chair Professor with the Department of Systems Engineering and Engineering Management. She was the former Department Chairman and Associate Dean of Research with the Faculty of Engineering. Her research interests include human–computer interaction via multimodal and multilingual spoken language systems, spoken dialog systems, computer-aided pronunciation training, speech processing in assistive technologies, health-related applications, and Big Data decision analytics. She was the Editor-in-Chief of the IEEE Transactions on Audio, Speech and Language Processing from 2009 to 2011. She was the recipient of the IEEE Signal Processing Society Leo L. Beranek Meritorious Service Award in 2019. She was also on the Elected Board Member of the International Speech Communication Association and an International Advisory Board Member. She is a Fellow of ISCA, HKCS, and HKIE.
\end{IEEEbiography}

\vspace{-1.0cm}
\begin{IEEEbiography}[{\includegraphics[width=1in,height=1.25in,clip,keepaspectratio]{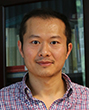}}]{Xunying Liu} (Member, IEEE) received his Ph.D. degree in speech recognition and M.Phil. degree in computer speech and language processing both from University of Cambridge, prior to his undergraduate study Shanghai Jiao Tong University. He had been a Senior Research Associate at the Machine Intelligence Laboratory of the Cambridge University Engineering Department, and from 2016 has been an Associate Professor in the Department of Systems Engineering and Engineering Management, the Chinese University of Hong Kong. He and his students received a number of best paper awards and nominations, including a Best Paper Award at ISCA Interspeech2010 for the paper titled ``Language Model Cross Adaptation for LVCSR System Combination'', and a Best Paper Award at IEEE ICASSP2019 for their paper titled ``BLHUC: Bayesian Learning of Hidden Unit Contributions for Deep Neural Network Speaker Adaptation''. His current research interests include machine learning, large vocabulary continuous speech recognition, statistical language modelling, noise robust speech recognition, audio-visual speech recognition, computer aided language learning, speech synthesis and assistive technology. He is a Member of ISCA.
\end{IEEEbiography}
\end{document}